\documentclass[footinbib,a4paper,aps,superscriptaddress,reprint,twocolumn,preprintnumbers,amsmath,amssymb,nobalancelastpage,10pt]{revtex4-2}

\usepackage{amsmath}
\usepackage{verbatim}
\usepackage{graphicx}
\usepackage{color}
\usepackage{braket}
\usepackage{yfonts}
\usepackage{soul}
\usepackage[normalem]{ulem}
\usepackage{dsfont}
\usepackage[colorlinks=true,citecolor=blue,linkcolor=blue,urlcolor=blue]{hyperref}

\begin{document}

\title{Fermionic quantum processing with programmable neutral atom arrays}

\author{D. Gonz\'alez-Cuadra}
\affiliation{Institute for Theoretical Physics, University of Innsbruck, 6020 Innsbruck, Austria}
\affiliation{Institute for Quantum Optics and Quantum Information of the Austrian Academy of Sciences,
6020 Innsbruck, Austria}
\author{D. Bluvstein}
\affiliation{Department of Physics, Harvard University, Cambridge, MA 02138, USA}
\author{M. Kalinowski}
\affiliation{Department of Physics, Harvard University, Cambridge, MA 02138, USA}
\author{R. Kaubruegger}
\affiliation{Institute for Theoretical Physics, University of Innsbruck, 6020 Innsbruck, Austria}
\affiliation{Institute for Quantum Optics and Quantum Information of the Austrian Academy of Sciences,
6020 Innsbruck, Austria}
\author{N. Maskara}
\affiliation{Department of Physics, Harvard University, Cambridge, MA 02138, USA}
\author{P. Naldesi}
\affiliation{Institute for Theoretical Physics, University of Innsbruck, 6020 Innsbruck, Austria}
\affiliation{Institute for Quantum Optics and Quantum Information of the Austrian Academy of Sciences,
6020 Innsbruck, Austria}
\author{T. V. Zache}
\affiliation{Institute for Theoretical Physics, University of Innsbruck, 6020 Innsbruck, Austria}
\affiliation{Institute for Quantum Optics and Quantum Information of the Austrian Academy of Sciences,
6020 Innsbruck, Austria}
\author{A. M. Kaufman}
\affiliation{Department of Physics, University of Colorado, Boulder, Colorado 80309, USA}
\affiliation{JILA, University of Colorado and National Institute of Standards and Technology, Boulder, Colorado 80309, USA}
\author{M. D. Lukin}
\affiliation{Department of Physics, Harvard University, Cambridge, MA 02138, USA}
\author{H. Pichler}
\affiliation{Institute for Theoretical Physics, University of Innsbruck, 6020 Innsbruck, Austria}
\affiliation{Institute for Quantum Optics and Quantum Information of the Austrian Academy of Sciences,
6020 Innsbruck, Austria}
\author{B. Vermersch}
\affiliation{Université Grenoble Alpes, CNRS, Laboratoire de Physique et Modélisation des Milieux Condensés (LPMMC), Grenoble 38000, France}
\author{Jun Ye}
\affiliation{Department of Physics, University of Colorado, Boulder, Colorado 80309, USA}
\affiliation{JILA, University of Colorado and National Institute of Standards and Technology, Boulder, Colorado 80309, USA}
\author{P. Zoller}
\affiliation{Institute for Theoretical Physics, University of Innsbruck, 6020 Innsbruck, Austria}
\affiliation{Institute for Quantum Optics and Quantum Information of the Austrian Academy of Sciences,
6020 Innsbruck, Austria}

\begin{abstract}
Simulating the properties of many-body fermionic systems is an outstanding computational challenge relevant to material science, quantum chemistry, and particle physics. Although qubit-based quantum computers can potentially tackle this problem more efficiently than classical devices, encoding non-local fermionic statistics introduces an overhead in the required resources, limiting their applicability on near-term architectures. In this work, we present a fermionic quantum processor, where fermionic models are locally encoded in a fermionic register and simulated in a hardware-efficient manner using fermionic gates. We consider in particular fermionic atoms in programmable tweezer arrays and develop different protocols to implement non-local tunneling gates, guaranteeing Fermi statistics at the hardware level. We use this gate set, together with Rydberg-mediated interaction gates, to find efficient circuit decompositions for digital and variational quantum simulation algorithms, illustrated here for molecular energy estimation. Finally, we consider a combined fermion-qubit architecture, where both the motional and internal degrees of freedom of the atoms are harnessed to efficiently implement quantum phase estimation, as well as to simulate lattice gauge theory dynamics.
\end{abstract}

\maketitle

\section{Introduction}
The study of strongly correlated fermionic systems lies at the core of some of the most interesting problems in modern physics.
These include profound questions regarding the inner workings of the universe, such as the physics of quark-gluon plasmas \cite{Berges_2021}, as well as technologically pressing challenges in material science and quantum chemistry, from high-temperature superconductivity \cite{Varma_2020} to nitrogen fixation \cite{Reiher_2017}. The defining feature of fermionic many-body systems is the fundamental indistinguishability of their constituents, which dictates the anti-symmetry of the wavefunction and the corresponding quantum statistics. Importantly, this indistinguishabilty gives rise to the so-called sign problem, which severely limits the applicability of many numerical approaches, such as Monte Carlo methods, highlighting the innate difficulty of solving fermionic many-body problems on classical computers~\cite{Troyer_2005}.

One of the most promising alternatives to address these problems is provided by quantum computers~\cite{Feynman_1982}.
Traditionally, quantum computing involves distinguishable spin-1/2 particles, where quantum information is stored in superposition states of qubit registers, and processed by the action of quantum gates. The quantum statistics of fermions needs to be encoded in such qubit-based devices on a \textit{software} level, which incurs overhead in circuit depths~\cite{Abrams_1997, Ortiz_2001, Bravyi_2002, Whitfield_2011} or qubit numbers~\cite{Ball_2005, Verstraete_2005, Whitfield_2016}.
This presents a substantial challenge for current experiments where noise limits gate and readout fidelities. Although this approach has been applied to simple fermionic  models from quantum chemistry~\cite{Peruzzo2014, OMalley_2016, Kandala2017, Hempel2018, Arute2020}, condensed-matter~\cite{Lanyon_2011, Salathe_2015, Barends_2015, Choi_2017, Zhang_2017, Satzinger_2021, Kyprianidis_2021, Randall_2021, Mi_2022, Zhang_2022_2} and particle physics~\cite{Martinez_2016, Klco_2018, Kokail_2019}, using quantum processors based on superconducting circuits or trapped ions, experimental studies have so far been restricted to small system sizes.

Neutral atom systems provide a route to bypass this issue and construct quantum devices where fermionic statistics are built-in on a \textit{hardware} level. The natural indistinguishability of atoms, which come as bosons or fermions, is for instance leveraged in celebrated analog quantum simulations of Hubbard models in optical lattices \cite{Jaksch_2005, Lewenstein_2007, Gross2017}. Recently, optical tweezers have emerged as powerful tools to trap and manipulate neutral atoms with an unprecedented level of programmability and scalability~\cite{Wilk_2010, Saffman_2010, Saffman_2016, Levine_2019, Graham_2019, Madjarov_2020, Henriet_2020, Bluvstein_2021, morgado2021quantum}. So far, these systems have, however, mainly been used to realize spin models with distinguishable constituents~\cite{Labuhn_2016, Bernien_2017, Keesling_2019, Leseleuc_2019, Ebadi_2021, Scholl_2021, Semeghini_2021, Scholl_2022}, where each atomic position is pinned to a specific tweezer, internal electronic or nuclear spin states are used to represent qubit states, and interactions between these qubits are implemented using highly-excited Rydberg states.

In this work, we envision the next-generation of such tweezer setups, where not only the internal but also the external degrees of freedom are coherently controlled, and fully integrated in the quantum processing architecture. This is a crucial prerequisite for capitalizing on the indistinguishability of (fermionic) atoms, which requires the possibility for their center-of-mass wave functions to overlap, e.g., by coherently delocalizing atoms across tweezers. Remarkably, this motional control has already been demonstrated in pioneering proof-of-principle experiments with tweezer pairs and double-well potentials~\cite{Kaufman2014, Murmann_2015, Bergschneider_2019, Becher_2020, Spar_2022, Young_2022, Yan_2022}. Below, we describe a blueprint of the elements required for such a \textit{fermionic quantum processor}~\cite{Bravyi_2002}, where both quantum hardware and software are co-designed to efficiently simulate fermionic models. 
More precisely, we describe protocols for the basic set of fermionic quantum gates, including, apart from Rydberg-mediated interacting gates, digital tunneling gates (or `fermionic beam splitters') implemented through \textsc{Merge} or \textsc{Shuttle} protocols.
While the present work focuses on a setup with   tweezer arrays, we note alternative setups involving optical lattices, as developed originally in the context of bosonic atoms \cite{Jaksch_1999, Mandel_2003_0, Mandel_2003, Daley_2008, belmechri2013microwave, Robens_2015, Lam_2021, Zhang_2022}. We exemplify our proposal for concrete atomic systems, and discuss the requirements and experimental challenges for their physical implementation. Furthermore, we provide illustrative examples of application of such a fermionic quantum processor in the context of digital quantum simulation for quantum chemistry and for lattice gauge theories (LGT), where we use this fermionic gate set to find efficient circuit decompositions and demonstrate considerable depth reductions compared to traditional qubit-based approaches.

\begin{figure}[t]
    \centering
    \includegraphics[width=0.9\linewidth]{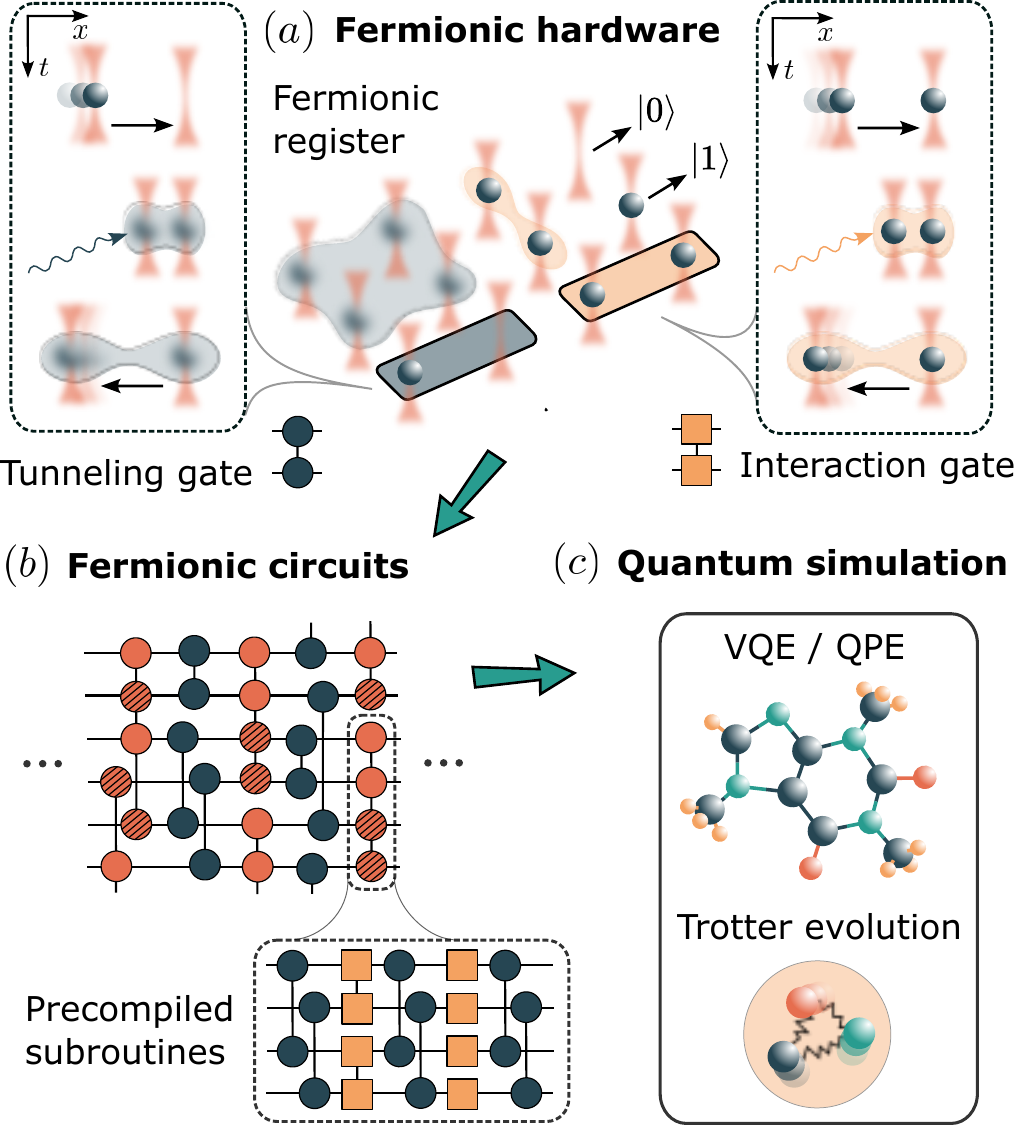}
    \caption{{\bf Fermionic quantum processor:} (a) we consider a fermionic register based on fermionic atoms trapped in optical tweezers, where quantum information is encoded in the atomic occupation and processed using fermionic gates. The latter includes tunneling processes, delocalizing atoms between different tweezers, as well as interaction gates, based on the Rydberg blockade mechanism. (b) We use these gates to construct fermionic quantum circuits, where certain subroutines are first precompiled to minimize circuit depths. (c) Fermionic circuits are particularly suited for quantum simulation of fermionic models, avoiding non-local overheads. Here, we consider the ground-state energy estimation of molecules using VQE and QPE, as well as Trotter time evolution of LGTs.
    }
   \label{fig:fig1}
\end{figure}

\section{Hardwired Fermi statistics}
We consider fermionic atoms in an array of microtraps that represent the fermionic quantum register [Fig.~\ref{fig:fig1}(a)]. We write $c^{\vphantom{\dagger}}_{j,\sigma}$ ($c_{j,\sigma}^{\dagger}$) for the annihilation (creation) operator of atoms on lattice site $j$, which we assume to be prepared in the trap ground state, and $\sigma$ labeling internal atomic states [qudits~\cite{Ringbauer_2022}]. The quantum state of the quantum register comprising $N$ atoms occupying $L$ microtraps will thus be a quantum superposition of all possible configurations. To be concrete, we will illustrate the quantum gate set, including in particular digital tunneling gates, for spinless fermions, i.e., dropping the index $\sigma$ for the moment. In this case, the state of the quantum register is given by a superposition of Fock states $\ket{n_{1},\ldots,n_{L}}$, where $n_{j}=0,1$ is the atomic occupation, and $\sum_{j=1}^{L}n_{j}=N$. In the last section, we consider again spinful fermions and work with a combined fermion-qubit register to encode more general models, which are not purely fermionic, such as LGTs.

\begin{figure*}[t]
    \centering
    \includegraphics[width=0.9\linewidth]{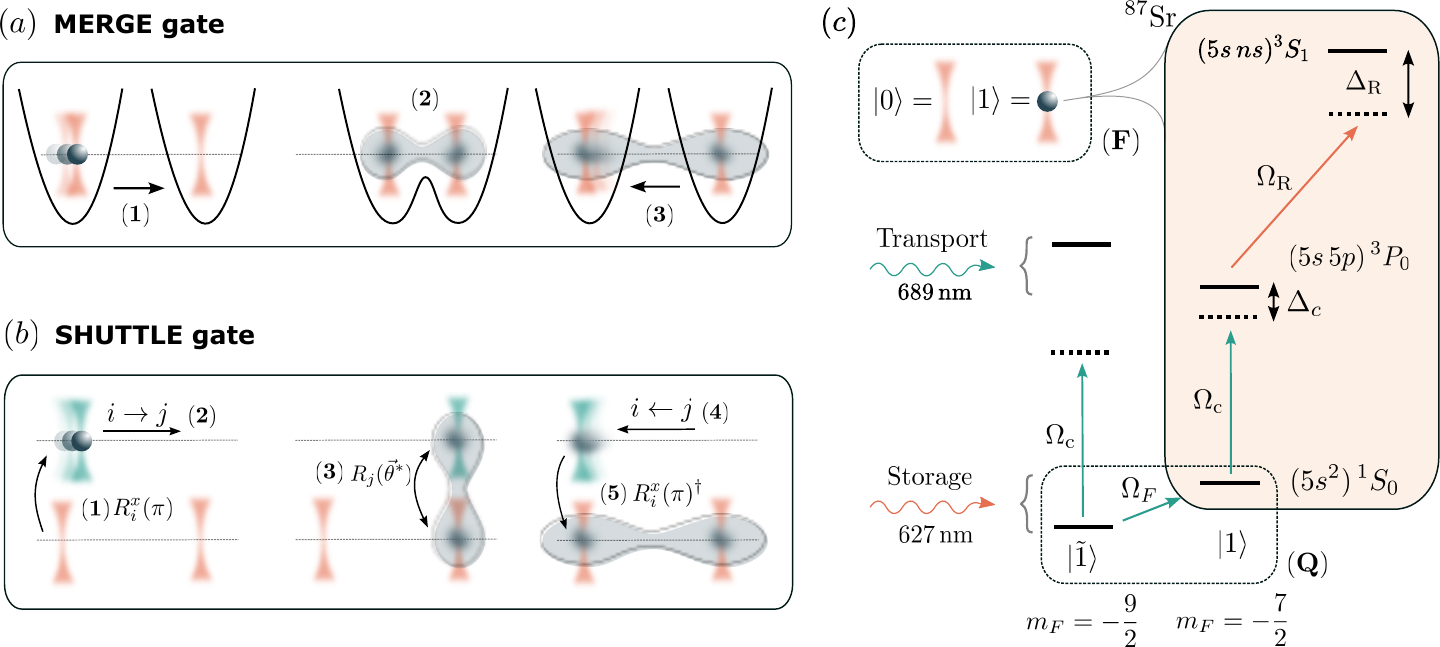}
    \caption{{\bf Fermion-qubit register and tunneling gates:} (a) and (b) show the sequence of laser pulses and tweezer moves used to implement a \textsc{Merge} and a \textsc{Shuttle} gate between a pair of sites $(i,j)$, respectively. While the illustration shows the case of a single initially localized atom, we emphasize that the protocols also apply to situations where both tweezers contain contributions from many-body superpositions of several atoms delocalized over the whole system. (c) Level structure of ${}^{87}{\rm Sr}$. A fermion register ({\bf F}) is built by encoding quantum information into the presence/absence of an atom trapped by a given storage tweezer (red) in one of the hyperfine states of the ground-state manifold ${}^1 S_0$. The latter is laser-coupled to the meta-stable excited state ${}^3 P_0$, with Rabi frequency $\Omega_c$ and detuning $\Delta_c$, trapped by a second transport tweezer (green). Interactions between pairs of atoms are turned on by exciting the atom to the Rydberg state ${}^3 S_1$, using a Rabi frequency $\Omega_{\rm R}$, where $\Delta_{\rm R}$ is the corresponding detuning. Other hyperfine levels, energy resolved using a magnetic field and coupled through a microwave frequency $\Omega_F$, serve as a qubit register ({\bf Q}).
    }
    \label{fig:fig2}
\end{figure*}

In the context of quantum simulation (QS) of many-body systems~\cite{Feynman_1982}, we are interested in particle-number conserving unitaries acting on this register~\cite{Lloyd_1996}, which can be constructed using the gate set 
\begin{equation}
\label{eq:BK}
\mathcal{BK} = \left\{e^{i\pi/4n_i},\,e^{i\pi n_in_j},\,e^{i\pi/4\left(c^\dagger_i c^{\vphantom{\dagger}}_j + \text{H.c.}\right)}\right\},
\end{equation}
as shown by Bravyi and Kitaev~\cite{Bravyi_2002}. 
As we demonstrate in the next section, the circuit depth required to simulate fermionic models can be considerably shortened by considering instead the more general set \mbox{$\mathcal{G} = \left\{\mathcal{U}^{\rm (int)}_{i,j}(\theta),\,\mathcal{U}^{\rm (t)}_{i,j}(\vec{\theta})\right\}$}, where 
\begin{align}
\mathcal{U}^{(\rm int)}_{i, j}(\theta) &\equiv e^{-i\theta \, n_i n_j} \;,\\
\mathcal{U}^{\rm (t)}_{i,j}(\vec{\theta}) &\equiv e^{-i\left[\frac{\theta_1}{2}\left(e^{-i\theta_2}c^\dagger_i c^{\vphantom{\dagger}}_j + \text{H.c.}\right) + \frac{\theta_3}{2}\left(n_i - n_j\right)\right]}
\end{align}
are generalized interaction (int) and tunneling (t) gates, respectively [Fig.~\ref{fig:fig1}(a)], and $\vec{\theta} = (\theta_1,\theta_2,\theta_3)$.
In the context of qubit-based quantum computation, where single-qubit rotations together with one entangling interaction gate are sufficient to achieve universality~\cite{Lloyd_1995}, fermionic degrees of freedom first need to be encoded into qubits using, e.g., a Jordan-Wigner (JW) transformation~\cite{McArdle_2020}. Tunneling gates, required to simulate many-body fermionic systems, can be implemented in the case of a JW encoding using $\mathcal{O}(L)$ entangling gates~\cite{McArdle_2020}.
Fermionic atoms trapped in the motional ground state of optical tweezers offer the unique possibility to avoid this overhead by implementing the gate set $\mathcal{G}$ directly. Specifically, the tunneling gate, $\mathcal{U}^{\rm (t)}_{i,j}(\vec{\theta})$, can be realized using different approaches that  exploit the capability of dynamically rearranging the tweezer positions, two of which we discuss now in more detail.

The \textsc{Merge} approach realizes the tunneling gate by temporarily bringing the two tweezers $i$ and $j$ so close together that atoms can tunnel between the corresponding lowest vibrational states [see Fig.~\ref{fig:fig2}(a)], as it has been demonstrated in recent experiments~\cite{Kaufman2014, Spar_2022, Young_2022, Yan_2022}. More generally, one may fully merge (and subsequently separate) the tweezer pair using custom designed merging and splitting protocols. The gate parameters $\vec{\theta}$ can in either case be completely controlled by the tweezer parameters and the details of the merging protocol, such as the tweezer depths, the time-dependent distance of the tweezer minima, and the duration of this coupling process. In practice, these can be determined via optimal control techniques for given target tunnel parameters $\vec{\theta}$, and allow gate execution on timescales set by the inverse trapping frequency of the tweezers. This approach is therefore natural for light 
atoms, such as Lithium, where relatively small trap depths are sufficient for large trap frequencies.

The \textsc{Shuttle} approach offers an alternative way to realize the tunnel gate, which is based on the capability to realize state-dependent optical potentials \cite{Daley_2008,ye2008quantum}, and is thus naturally suited for alkaline-earth atoms. We therefore illustrate this idea for the specific example of ${}^{87}{\rm Sr}$, a fermionic isotope of strontium with nuclear spin $I = 9/2$ below and in Fig~\ref{fig:fig2}. The central idea is to use two sets of tweezers: a set of static \emph{storage} traps, whose occupations define the fermionic register, and a set of \emph{transport} traps, which serve as a ``shuttle'' for atoms. Crucially, the wavelength of the storage and transport tweezers is chosen such that they trap two different internal states of an atom, respectively. For instance, for ${}^{87}{\rm Sr}$, one can trap the state ${}^1 S_0$~\cite{Cooper_2018} in the storage tweezers, and independently trap the clock state ${}^3 P_0$~\cite{Daley_2008, Heinz_2020} in transport tweezers, reminiscent of the collisional entangling quantum gate with spin-dependent lattices for bosonic atoms \cite{Jaksch_1999, Mandel_2003, Zhang_2022}. 
Here, we extend these ideas and design a fermionic shuttle that implements the tunneling gate $\mathcal{U}^{\rm (t)}_{i,j}(\vec{\theta})$. Importantly, when a storage and transport tweezer overlap and their potential shapes match, atoms can be coherently transferred or coherently split between the two tweezers simply by using laser pulses that implement internal rotations $R_i(\vec{\theta}) = e^{-i\left[\frac{\theta_1}{2}\left({\rm cos}\,\theta_2 X+ {\rm sin}\,\theta_2 Y\right) + \frac{\theta_3}{2} Z\right]}$, where $X$, $Y$ and $Z$ are Pauli matrices acting on the atomic subspace spanned by $^1S_0$ and $^3P_0$.
Using this mechanism, one can construct the full tunneling gate, $\mathcal{U}^{\rm (t)}_{i,j}(\vec{\theta})$, as follows [see also Fig.~\ref{fig:fig2}(b)]: {\bf (1)} we first bring a transport tweezer to a storage site $i$ and perform a $\pi$-pulse rotation, i.e. \mbox{$R^x_i(\pi) \equiv R_i(\pi, 0, 0)$}, {\bf (2)} after which we move the transport tweezer to site $j$. {\bf (3)} We then perform a second pulse $R_j(\vec{\theta}^*)$, with $\vec{\theta}^* = (\theta_1,\theta_2 + \frac{\pi}{2}, \theta_3)$, {\bf (4)}, bring the transport tweezer back to site $i$, and {\bf (5)} finally undo the initial $\pi$-pulse. 
We note that both the \textsc{Shuttle} and the \textsc{Merge} approach can be fully parallelized. We comment on potential error sources for both approaches in a separate section below.

In these setups, the interaction gate $\mathcal{U}^{\rm (int)}_{i,j}(\theta)$ is essentially equivalent to a standard qubit entangling gate that has 
already been implemented for alkaline~\cite{Wilk_2010, Levine_2019, Graham_2019, Bluvstein_2021} as well as alkaline-earth atoms~\cite{Madjarov_2020, Schine2022} using the Rydberg blockade mechanism. To realize it, we first rearrange the tweezers hosting the modes $i$ and $j$ and bring them to a distance that lies within the Rydberg blockade radius. We then drive the atoms with a laser that couples the atoms internal state to a Rydberg state. Owing to the Rydberg blockade mechanism, properly chosen laser pulses result in a unitary $e^{i\phi_{01} \left(n_i + n_j\right) + i \phi_{11}n_i n_j}$~\cite{Levine_2019}. The phases $\phi_{01/11}$ depend on the shape of the Rydberg laser pulse~\cite{Levine_2019} and choosing $\phi_{11} = 2\phi_{01} - \theta$ provides the desired interaction gate up to the single-particle phase shift $\phi_{01}$.
We note that resonantly coupling the atom to the untrapped Rydberg state could create excitations in the trap, which should be suppressed to preserve the atom's indistinguishability. We note that these effects can be reduced by trapping the atoms also in the Rydberg state~\cite{Wilson_2022}. Alternatively, one can work in the dressed-Rydberg regime, where decay from the Rydberg state is suppressed and interactions become independent of distance, thus avoiding the repulsive forces between atoms. Below, we will discuss these together with other experimental challenges, including several strategies to mitigate the dominant error sources.

\section{Fermionic quantum circuits}
Now we employ the set of fermionic gates $\mathcal{G}$ to construct the quantum circuits required for QS of fermionic systems. Let us first focus on purely fermionic Hamiltonians, which we generalize below to fermion-boson models relevant to high-energy physics. To be explicit, we consider the particle-number-conserving fermionic Hamiltonian:

\begin{align}
\label{eq:QC_ham}
\mathcal{H}& = 
\sum_{ij}   
h_{ij}^{(1)} \:
c^\dagger_i c^{\vphantom{\dagger}}_j + 
\sum_{ijkl} 
h_{ijkl}^{(2)} \:
c^\dagger_i c^\dagger_j c^{\vphantom{\dagger}}_k c^{\vphantom{\dagger}}_l \;
\end{align}
with complex parameters $h_{ij}^{(1)}$ and $h_{ijkl}^{(2)}$.
This Hamiltonian is frequently used both in condensed matter~\cite{Jaksch_2005} and quantum chemistry~\cite{McArdle_2020}, where the indices can denote either the position of electrons in a solid-state crystal or the orbitals of a molecule, respectively. Many QS algorithms use as subroutines unitary operations obtained from exponentiating each term in the Hamiltonian. Apart from the tunneling and interaction gates introduced above, these include density-dependent tunneling (dt) as well as pair-tunneling (pt) processes,
\begin{align}
\mathcal{U}^{\rm (dt)}_{ijk}(\theta_1, \theta_2) &\equiv e^{-i\, \theta_1 (e^{-i\theta_2} c_{i}^{\dagger}n_jc^{\vphantom{\dagger}}_k + \text{H.c.})}, \label{density_t}\\
\mathcal{U}^{\rm (pt)}_{i,j,k,l}(\theta_1, \theta_2) &\equiv\ e^{-i\, \theta_1 \left(e^{-i\theta_2}c_{i}^{\dagger}c_{j}^{\dagger}c^{\vphantom{\dagger}}_k c^{\vphantom{\dagger}}_l + \text{H.c.} \right)}. \label{pair_t}
\end{align}

\begin{figure}[t]
    \centering
    \includegraphics[width=1.0\linewidth]{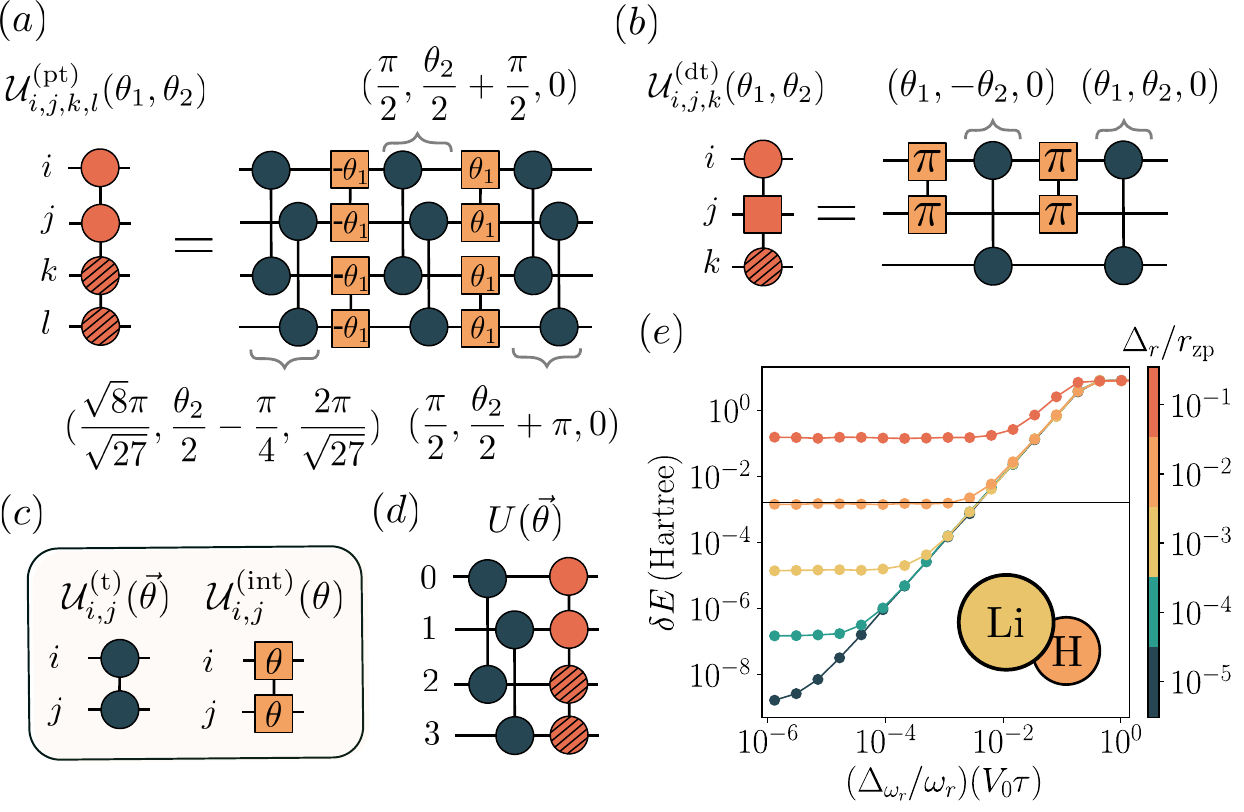}
    \caption{{\bf Fermionic subroutines and variational circuits: } (a) and (b) show the circuit decomposition of the pair-tunneling and density-dependent tunneling gates, respectively, in terms of basic fermionic gates (c). (d) Variational circuit used to prepare the ground-state of the LiH molecule (e) Average energy difference $\overline{\delta E}$ in the presence of fluctuations in the trapping frequency and tweezer positions, characterized by standard deviations $\Delta_{\omega_r}$ and $\Delta_{r}$, respectively. $V_0$, $\omega_r$ and $r_{zp}$ denote the depth, radial frequency and zero-point fluctuations of the harmonic trap, and $\tau$ is the transfer pulse time (Methods). The dotted line signals chemical accuracy, $\delta E^* \approx 1.59$ mHa.}
    \label{fig:fig3}
\end{figure}

For both \eqref{density_t} and \eqref{pair_t}, we derived a specific decomposition in terms of the gate set $\mathcal{G}$, which are shown in Figs.~\ref{fig:fig3}(a) and (b), respectively. Both decompositions have a constant circuit depth of 5 gates and are exact and optimal.
This provides a clear advantage in terms of depth with respect to the circuit obtained using e.g. JW transformations~\cite{McArdle_2020}, as well as with approximate decompositions based on the universal $\mathcal{BK}$ set~[\ref{eq:BK}]. Moreover, note that the all-to-all connectivity of atom arrays plays a crucial point in the case of quantum chemistry, where the presence of non-local terms in \eqref{eq:QC_ham} introduces further overhead for architectures with only nearest-neighbor gates~\cite{Kandala_2017}.

These fermionic quantum circuits can then be used as precompiled subroutines, carrying over the polynomial saving of resources to the full QS algorithm. For instance, the real-time evolution under Hamiltonian~\eqref{eq:QC_ham} can be simulated by a first-order Trotter expansion~\cite{Trotter_1959},
\begin{equation}
e^{-i\mathcal{H}t} \approx \left[\prod_{i,j}\mathcal{U}^{\rm (t)}_{i,j}(h^{(1)}_{i,j}\delta t,0,0)\prod_{i,j,k,l}\mathcal{U}^{\rm (pt)}_{i,j,k,l}(h^{(2)}_{i,j,k,l}\delta t,0)\right]^{t/\delta t},
\end{equation}
where $\delta t$ is the Trotter time and the local gates can be applied in parallel across the system. 
Another application are hybrid quantum-classical algorithms such as the variational quantum eigensolver (VQE)~\cite{McClean_2016}, used in particular to find the ground-state energy of molecules using near-term devices, where $\mathcal{U}^{\rm (pt)}_{i,j,k,l}(\theta, \frac{\pi}{2})$ allows us to construct variational states that are both hardware-efficient~\cite{Kandala_2017} and so-called ``chemically-inspired''~\cite{McArdle_2020}. This includes the disentangled unitary coupled cluster ansatz~\cite{Anand_2022}, $\ket{\psi(\vec{\theta})} = U(\vec \theta)\ket{\psi_{\rm HF}}$,  with 
\begin{equation}
U(\vec \theta) = \prod_{i,\alpha}\mathcal{U}^{\rm (t)}_{i,\alpha}(\theta^{(1)}_{i\alpha}, \frac{\pi}{2}, 0)\prod_{i,j,\alpha,\beta}\mathcal{U}^{\rm (pt)}_{i,j,\alpha,\beta}(\theta^{(2)}_{ij\alpha\beta}, \frac{\pi}{2}),
\end{equation}
where the products run over occupied $(i, j)$ and virtual $(\alpha, \beta)$ modes with respect to the initial Hartree-Fock product state $\ket{\psi_{\rm HF}}$. The energy functional $E(\vec{\theta}) = \bra{\psi(\vec{\theta})}\mathcal{H}\ket{\vec{\psi}(\vec{\theta})}$, which is then classically minimized, can be also efficiently constructed in our fermionic processor, by first applying randomized tunneling gates followed by measurements in the occupation basis~\cite{Naldesi2022, Low_2022}.

We illustrate our approach for the LiH molecule. We calculate the energy difference $\delta E \equiv E_{\rm exp}(\vec{\theta}^*) - E_0$, where $\vec{\theta}^*$ are the optimal variational parameters that minimize $E(\vec{\theta})$, $E_{\rm exp}$ is obtained by evaluating the latter in the presence of experimental errors and $E_0$ is the exact ground-state energy~\footnote{We consider the LiH molecule at a fixed interatomic distance of $1.45$ {\AA} in the Born-Oppenheimer approximation, with two electrons and four active orbitals. The Hamiltonian parameters $h^{(1)}_{i,j}$ and $h^{(2)}_{i,j,k,l}$ are calculated using OpenFermion~\cite{McClean_2020}.}. The variational circuit $U(\vec \theta)$ is depicted in Fig.~\ref{fig:fig3}(d), and Fig.~\ref{fig:fig3}(e) shows the average $\overline{\delta E}$ for random fluctuations in the trapping frequency and tweezer positions (Methods), providing an estimate on the required precision to reach chemical accuracy. Note that $\overline{\delta E}$ can be reduced by lowering the tweezer depth, as well as by increasing the pulse intensity to reduce the implementation time, and the full protocol including tweezer transport can be further improved using optimal control~\cite{DeChiara_2008, Pagano_2022}. 

A detailed study for larger molecules using more advanced algorithms such as ADAPT-VQE~\cite{Grimsley_2019, qubitadapt_vqe, tetris_vqe}, where the advantage of fermionic processing compared to qubit-based protocols is expected to grow with the number of orbitals, will be included in a separate work~\cite{Naldesi_2023}.

\begin{figure}[t]
    \centering
    \includegraphics[width=0.9\linewidth]{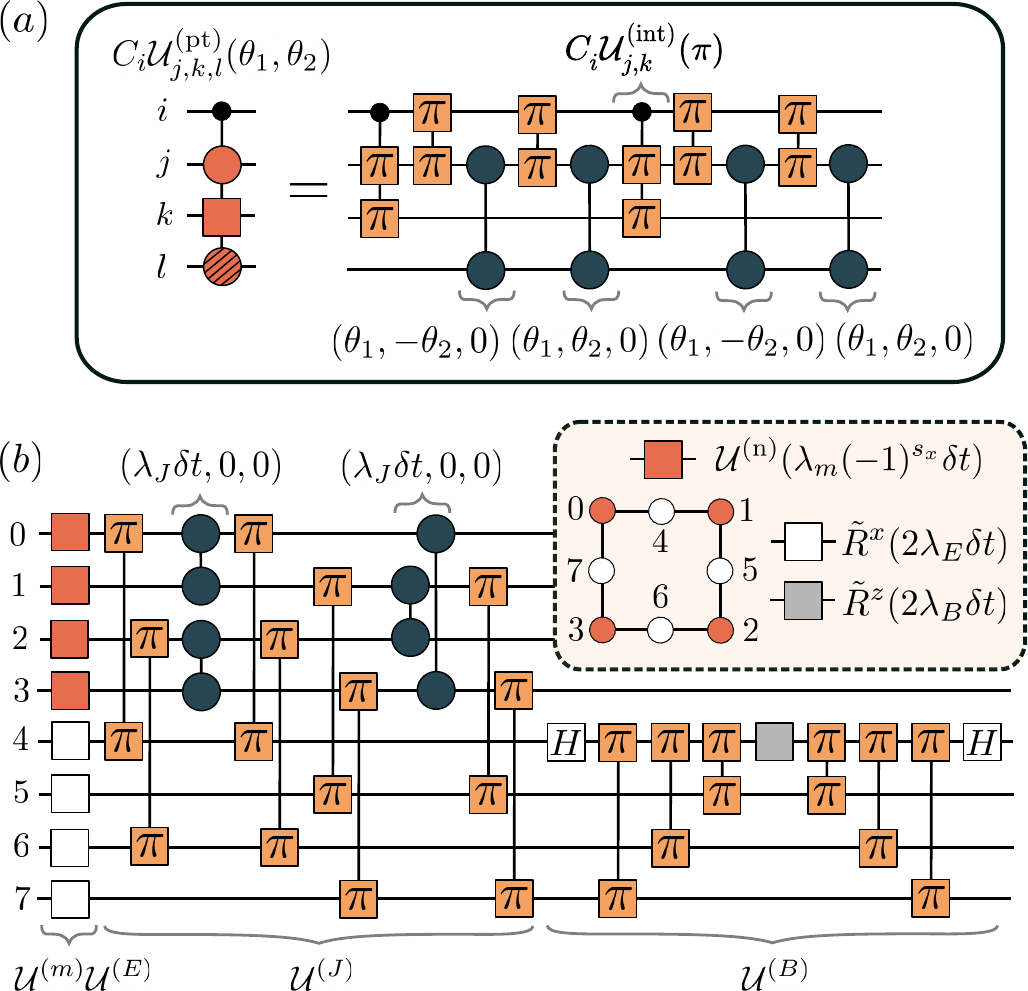}
    \caption{{\bf Fermion-qubit quantum circuits:} (a) Decomposition of a density-dependent fermionic hopping controlled by a qubit ancilla in terms of fermionic gates and the three body fermion-qubit gate $C_i\mathcal{U}^{\rm (int)}_{j,k}(\pi)$. (b) Trotter step required to time-evolve one plaquette under the $\mathbb{Z}_2$ LGT Hamiltonian~\eqref{eq:LGT_H}, where atoms $0$ to $4$ and $5$ to $7$ encode local matter and gauge fields, respectively. The unitary circuits $\mathcal{U}^{(p)}$, with $p\in \{m, E, J, B\}$, implement the exponential of each term in the Hamiltonian, and the single-qubit rotations are given by $\tilde{R}^x(\theta) = e^{-i\theta/2 X}$ and $\tilde{R}^z(\theta) = e^{-i\theta/2 Z}$, acting on the $\{\ket{\tilde{1}},\ket{1}\}$ subspace, while $\mathcal{U}^{\rm (n)}(\theta) = e^{-i\theta n}$ acts on $\{\ket{0},\ket{1}\}$.
    }
    \label{fig:fig4}
\end{figure}

\section{Fermion-qubit architecture}
We now consider spinful fermionic atoms, combining the fermionic register and fermion gates introduced above with a more standard qubit-based architecture. This allows us to encode both qubit ancillas and fermionic modes locally, leading to efficient implementations of more advance QS algorithms such as quantum phase estimation (QPE)\cite{Kitaev_1995}, as well as to simulate boson-fermion models such as LGTs in a hardware-efficient manner.

To be specific, we illustrate the architecture using again Sr as an example. Qubit ancillas can be readily included by considering two hyperfine levels of $^1S_0$ [Fig.~\ref{fig:fig2}(a)], where one level, denoted $\ket{\tilde{1}}$, is decoupled from the $^3P_0$ clock manifold by a magnetic field, and therefore behaves similar to $\ket{0}$ for gates in $\mathcal{G}$. Rotations $\tilde{R}$ between the states $\ket{\tilde{1}}$ and $\ket{1}$  can be implemented through microwave frequencies [Fig.~\ref{fig:fig2}(a)], enlarging the gate set to $\tilde{\mathcal{G}}$, including both $\mathcal{G}$ and $\tilde{R}$. In the fermion-qubit register, the same type of atoms can either encode a qubit ancilla or a fermionic mode. In the case of QPE, the energy of $\mathcal{H}$ can be estimated by applying a Trotter time evolution under the terms in $\mathcal{H}$ controlled by an ancilla, with a precision that grows with the number of ancillas~\cite{Kitaev_1995}. As an example, we show in Fig.~\ref{fig:fig4}(a) a fermion-qubit circuit associated to a controlled density-dependent hopping process, and other controlled operations can be decomposed similarly in terms of our fermionic gate set. We note that this decomposition requires the three-body gate $C_i\mathcal{U}^{\rm (int)}_{j,k}(\pi) = \ket{\tilde{1}}\bra{\tilde{1}}\otimes\mathbb{I} + \ket{1}\bra{1}\otimes\mathcal{U}^{\rm (int)}_{j,k}(\pi)$, which can be directly implemented using the Rydberg blockade mechanism~\cite{Levine_2019}.

Finally, the fermion-qubit architecture allows us to go beyond purely fermionic models, and consider for instance LGTs, where fermions are coupled to dynamical (bosonic) gauge fields. Consider for simplicity a $\mathbb{Z}_2$ LGT described by the Hamiltonian
\begin{equation}
\label{eq:LGT_H}
\begin{aligned}
&\mathcal{H} = \lambda_E\sum_{\langle x, y \rangle} \sigma^x_{\langle x, y \rangle} + \lambda_B\sum_\square \sigma^z_\square\\
 &+\lambda_J\sum_{\langle x, y \rangle}\left(c^\dagger_{x}\sigma^z_{\langle x, y \rangle}c^{\vphantom{\dagger}}_{y} + \text{H.c.}\right) + \lambda_m \sum_x (-1)^{s_x}n_{x},
\end{aligned}
\end{equation}
where fermion and spin operators, representing matter and gauge degrees of freedom, respectively, act on the sites $x$ and links $\langle x, y \rangle$ of a $D$-dimensional lattice. The first row in \eqref{eq:LGT_H} contains the pure-gauge dynamics, including four-body plaquette operators $\sigma^z_\square$ acting with $\sigma^z_{\langle x, y \rangle}$ on each link around a plaquette $\square$ [Fig.~\ref{fig:fig4}(b)], while the second one includes gauge-matter interactions. The Hamiltonian is invariant under local $\mathbb{Z}_2$  transformation, i.e. $\mathcal{H} = \mathcal{V}^\dagger_x \mathcal{H} \mathcal{V}_x$ $\forall x$, with $\mathcal{V}_x = (-1)^{n_x} \prod_{y}\sigma^x_{\langle x, y \rangle}$. 

Apart from serving as simplified models to study fermionic confinement~\cite{Borla_2020, Gonzalez_2020}, $\mathbb{Z}_2$ LGTs emerge in condensed-matter systems~\cite{Fradkin_2013}, displaying strongly-correlated phenomena such as high-Tc superconductivity~\cite{Baskaran_1988}, topological order~\cite{Wen_2017, Assaad_2016, Gonzalez_2020, Borla_2022} and unconventional dynamics~\cite{Iadecola_2020, Sai_2022, Halimeh_2022}. For $D>1$, the model presents a sign problem away from half-filling and can not be solved efficiently using classical methods. The corresponding real-time dynamics can be efficiently simulated using quantum devices~\cite{Wiese_2013, Zohar_2015, Dalmonte_2016, Banuls_2020, Aidelsburger_2022, Klco_2022} through e.g. a first-order Trotter expansion. In our fermion-qubit architecture, the gate set $\tilde{\mathcal{G}}$ allows us to construct Trotter steps [shown in Fig.~\ref{fig:fig4}(b)] with a constant circuit depth~\cite{Zohar_2017, Lamm_2019, Gonzalez_2022}, thanks to the local structure of~\eqref{eq:LGT_H}, the possibility to parallelize fermionic gates and the lack of JW strings. The above protocol can be generalized to non-abelian gauge fields, required to address the full Standard Model of particle physics, by further extending it to a fermion-qudit architecture~\cite{Gonzalez_2022}, which will be presented in detail in a separate work~\cite{Gonzalez_2023}.

\section{Experimental challenges and considerations}

We finish by discussing in more detail some of the experimental challenges that should be overcome to build a fermionic quantum processor, including the main error sources for the gates discussed above, as well as different strategies to minimize them. The experimental setup for our proposed fermionic quantum processor is similar to existing reconfigurable tweezer platforms with high-fidelity Rydberg gates~\cite{Bluvstein_2021,Schine2022}; however, the main new challenge will be coherently controlling the motional degrees of freedom. This introduces new pathways for decoherence, primarily coming from leakage out of the 
fermionic register (i.e., heating to motional excited states) and dephasing from inhomogeneity between different tweezer sites.

Leakage, or heating, out of the motional ground state can arise from state preparation errors, scattering by the tweezer, pulsing the traps off for the Rydberg gate, and from moving the atoms; we argue here that all of these effects can be greatly suppressed. 
First, heating from tweezer scattering is negligible for sufficiently large detunings.
By utilizing tight radial and axial confinement, 3D motional ground state preparation has been realized at levels of $\approx 95\%$~\cite{Kaufman2014,Young_2022}. However, by ``spilling out'' all motional excited states, we can convert the motional ground state occupancy to nearly 100\% provided we can nondestructively check for atom presence, which can be done directly using an ancilla atom and a high-fidelity Rydberg gate to check for atom presence~\cite{serwane2011deterministic, Cong_2022}. 

Consider for example the 50 kHz trap depth for the Lithium tweezers, relevant for the \textsc{Merge} gate, used in~\cite{Spar_2022,Yan_2022}, which would give 0.03 Hz scattering rate and an even smaller heating rate due to the Lamb-Dicke parameter. Pulsing the trap off during the Rydberg gate could cause heating, but the probability of transitioning to higher motional states will be roughly $(\omega t)^2 / 4$, where $\omega = 2\pi \times 15$ kHz is the trap frequency~\cite{Spar_2022,Yan_2022} and $t \approx 100$ ns is the implementation time required for 99.9\% fidelity gates. This probability then evaluates to $< 10^{-4}$ per gate, and so does not contribute significantly compared to a 99.9\% gate fidelity. Finally, moving the atoms can cause heating and, as a conservative estimate, we utilize the heating rates calculated in~\cite{Bluvstein_2021}. If the atoms are placed relatively close together, at a distance of several micrometers, then for the move to be 99.9-99.99\% fidelity each move has to be $\approx 500 ~\mu$s. This can be significantly speed-up using optimal control~\cite{Lam.2021}.

In general, we expect that trap inhomogeneities will be the dominant source of dephasing for degrees of freedom coherently encoded in motional states.
With a Rydberg gate fidelity of $\sim 99.9\%$ we would like to perform $\sim 1000$ operations, where in general the tweezer geometry is reconfigured between each round. If it takes $\sim 500~ \mu$s to move atoms between gates, that sets the total operation time of $\sim 500$ ms. For  the example of Lithium-6, a trap depth of 50 kHz and a reasonable standard deviation between tweezers of $0.2$\% leads to a coherence time of $T^{*}_2 \approx 2$ ms, which is only enough time for $\approx$ 4 moves. Note that this is for shallow trap depths of Lithium-6, a light atom, and that this effect will be even more exacerbated for heavier atoms like strontium, where larger trap depths (laser intensities) are required for the same trap frequencies.

\begin{figure}[t]
    \centering
    \includegraphics{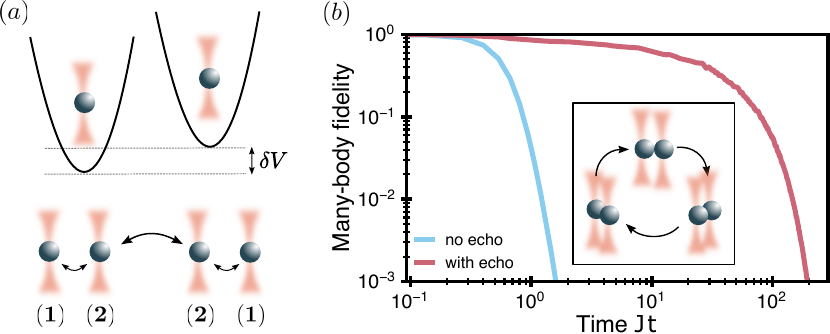}
    \caption{\textbf{Robustness to trap inhomogeneity.} (a) Spatial variations in the light intensity lead to relative dephasing between tweezers ($\delta V$). Performing a positional echo, either by repeatedly switching the positions of the two tweezers or including a full swap on top of a tunneling gate, allows us to cancel out this static source of noise. 
    (b) The achievable physical time in digital simulation is extended by two orders of magnitude when utilizing the echo procedure. The simulation was performed for a simple hopping Hamiltonian in a 1D configuration, where the many-body echo is especially simple (inset), with a hundred sites and a Gaussian phase disorder with $\sigma_\theta$=0.035 applied between tunneling events (see main text and Methods).}
    \label{fig:5}
\end{figure}

Whenever there is static inhomogeneity in a system, such as positional trap-depth dependence, the natural approach is to perform a motional echo procedure~\cite{Hahn_1950}. 
We can permute around the various tweezer positions
so that all atoms acquire the same average phase due to the positional trap depth dependence.
First, consider two tweezers with unequal trap intensities and thus unequal energies at the bottom of the trap, as depicted in Fig.~\ref{fig:5}(a). Due to the different energies, the atoms at these sites experience dephasing. However, this effect can be canceled out by repeatedly swapping the tweezer positions after each application of the tunneling gate (see Methods for details). In Fig.~\ref{fig:5}(b), we illustrate this idea by simulating the
Floquet evolution of a 1D chain with nearest-neighbor hopping (Methods). The results show that performing the echo procedure allows us to extend the useful simulation time by two orders of magnitude. In practice, the inhomogeneities might not be exactly static, in which case the echo needs to be applied at a rate faster than that timescale. We also note that there could potentially be other methods to design robust sequences of tunnel gates. For example, circuits could be precompiled in order to minimize the number of necessary swap operations while taking into account the specific distribution of spatial coherence. 

In addition, trap inhomogeneities could be further reduced by storing the atoms in state-dependent optical lattices in combination with tweezers, which could be used to move the atoms to the desired position in order to apply the corresponding gates~\cite{Jaksch_1999, Mandel_2003_0, Mandel_2003, Daley_2008, belmechri2013microwave, Robens_2015, Lam_2021, Zhang_2022}. In this architecture, the \textsc{Merge} gate could be implemented using superlattices, by first placing the atoms in double wells~\cite{Nascimbene_2012}, while the \textsc{Shuttle} gate could be implemented by using the lattice and tweezers as storage and transport potentials, respectively.

Finally, for the \textsc{Shuttle} gate based on alkaline-earth atoms like strontium as discussed above, additional error sources include phase shifts and losses during step {\bf (3)} of the protocol due to elastic and inelastic collisions, respectively, between atoms in the ${}^3 P_0$ and ${}^1 S_0$ states. The former are of the order of a few kHz for $^{87}$Sr~\cite{Zhang_2014, Goban_2018}, and should be removed by an appropriate calibration or measured and compensated by additional phase shift gates. Inelastic collisions are even smaller~\cite{Bishof_2011} than typical Rabi frequencies and can therefore be safely neglected. 
Finally, we note that the optical potentials for the storage and transport tweezers need to match to properly transfer atoms between them, and fluctuations in the tweezer location and laser intensity will lead to imperfect transfer processes (Methods). The effect of such fluctuations is taken into account in the variational preparation of a simple molecule, illustrated in Figure~\ref{fig:fig3}. 

\section{Conclusions and outlook}
We presented a fermionic quantum processor based on fermionic atoms trapped in tweezer arrays, and showed how to locally encode and quantum simulate fermionic models in a hardware-efficient manner. We illustrated  the advantages of our approach with respect to qubit-based devices for VQE and QPE in quantum chemistry, as well as for Trotter time evolution of gauge theories. We note that the proposed hardware can be used to run more advance and potentially new fermionic quantum algorithms, further optimizing the required resources and facilitating the reach of quantum advantage in the near term. The fermionic gate set could also be extended to include particle non-conserving processes, using e.g. atom reservoirs, allowing to implement error correction protocols, which we plan to investigate in the future.

\section*{Methods}

\subsection*{Error model for VQE}
We exemplify the fermionic quantum simulation of the LiH molecule using VQE, where we consider in particular the tunneling gate implemented using the \textsc{Shuttle}
 protocol in strontium. We consider optical tweezers generated by laser fields with the following intensity profile in the radial $r$ and longitudinal $z$ directions,
\begin{equation}
I(r, z) = I_0\left(\frac{w_0}{w_z}\right)e^{-2(r / w_z)^2},
\end{equation}
with $w_z = w_0\sqrt{1 + (z / z_R)^2}$, where $w_0$ and $z_R = \pi w_0^2/\lambda$ are the waist and Rayleigh length of the tweezer, respectively, and $\lambda$ is the laser (trapping) wavelength. For a properly chosen wavelength, the latter gives rise to an AC Stark shift on a neutral atom, leading to an optical potential $V(r, z) = -{\rm Re}(\alpha) / (2\epsilon_0 c) I(r, z)$, where $\alpha$ is the atom polarizability for the corresponding energy level, $\epsilon_0$ is the permittivity of free space, and $c$ is the speed of light. The trapping potential $V(r, z)$ can be approximated around its minimum using a second-order Taylor expansion, leading to the following harmonic potential,
\begin{equation}
\label{eq:harmonic_V}
V(r, z) = V_0 + \frac{1}{2} m \omega_r^2 r^2 + \frac{1}{2} m \omega_z^2 z^2,
\end{equation}
where $V_0 \equiv V(0, 0)$, $m$ is the mass of the atom and the frequencies are given by $\omega_r = \sqrt{4V_0/(mw_0^2)}$ and $\omega_z = \sqrt{2V_0/(mz_R^2)}$. For typical experimental parameters these frequencies are related by $\omega_r /\omega_z \approx 10$.  Here we consider random fluctuations both in the frequencies $\omega_r$ and $\omega_z$ and in the relative position between tweezers, given by $\delta r$ and $\delta z$, following all these parameters independent Gaussian distributions.

These experimental fluctuations introduce errors in the fermionic quantum gates, since they lead to imperfect rotations between the storage (S) and transport (P) tweezers. In particular, both the Rabi coupling and the detuning of a give pulse are modified due to the imperfect overlap between the wavefunctions, $\Omega(\delta r, \delta z) = \Omega_0 f(\delta r, \delta z)$ and $\Delta(\delta r, \delta z) = \Delta_0 f(\delta r, \delta z)$, with
\begin{equation}
f(\delta r, \delta z) = \int \! {\rm d} r {\rm d} z \, 2\pi r \, \psi^*_S(r, z) \psi^{\vphantom{*}}_P(r +\delta r, z + \delta z),
\end{equation}
where we consider Gaussian wavefunctions for the ground state of the harmonic potential \eqref{eq:harmonic_V},
\begin{equation}
\psi_{S/P}(r, z) = \frac{e^{- (r / 2r_{\rm zp})^2}}{(2 \pi r_{\rm zp})^{1/4}} \frac{e^{- (z / 2z_{\rm zp})^2}}{(2 \pi z_{\rm zp})^{1/4}},\end{equation}
with zero-point fluctuations given by $r_{\rm zp} = 1 / \sqrt{2m\omega_r}$ and $z_{\rm zp} = 1 / \sqrt{2m\omega_z}$. Moreover, fluctuations in the trap frequency lead to a non-zero difference between the trap depths, $\delta V_0 = V^S_0 - V^P_0$, introducing an extra unwanted detuning, $\tilde{\Delta} = \Delta + \delta V_0$, giving rise to an extra angle $\delta V_0 \tau$ for $Z$ rotations, where $\tau$ is the pulse time.

\subsection*{Motional echo scheme}

The positional trap-depth dependence leading to a static dephasing can be mitigated by an appropriate echo procedure. Due to the positional nature of this noise, the natural approach is to ensure that each fermionic register spends equal time at each tweezer site and, on average, experiences the same disorder pattern. The key insight is that, once the evolution is digital, the atom $i$ does not need to always reside in the tweezer site $i$. That is, we can permute the atoms around the various tweezer positions either by moving them around or introducing swap operations, which effectively change which atom resides in which tweezer. 

As described in the text, the dephasing between the two tweezers coming from different trap depths can be canceled out by repeatedly swapping the tweezer positions after each application of the tunneling gate. In a many-body context, the atoms need to be shuffled in an ergodic fashion such that they can spend equal time at every tweezer site; since this process is deterministic it is easy to keep track of the orbital labels in the classical experimental software.

This approach is particularly simple in a 1D chain with periodic boundary conditions [Fig.~\ref{fig:5}(b) inset]. 
Here, the two sublattices of tweezers, labeled with even and odd numbers, will sequentially translate by one site in space after every tunneling gate. This results in each orbital spending the same amount of time at each position and, moreover, experiencing the same history of trap depths as the neighboring site, up to boundary conditions, which further improves the echo performance. More concretely, during the $t$-th timestep, the $i$-th atom resides in the $\sigma_t(i)$-th tweezer. Since the inhomogeniety is (to first order) static, but varies from tweezer to tweezer, this can be modelled via a time-dependent disorder Hamiltonian $D_t = \sum_x h_{\sigma_t(x)} \hat{n}_x$.
The effective dynamics is captured by alternating disorder evolution and target evolution. Moving into an interaction picture with respect to the disorder, and assuming each moving step takes time $\tau$, the hopping terms $c_i^\dagger c_j$ become dressed as
\begin{align}
    (c_i^\dagger c_j)(t) &= e^{-i \tau \sum_{t' \leq t} (h_{\sigma_{t'}(j)} - h_{\sigma_{t'}(i)})} c_i^\dagger c_j,
\end{align}
where $j = i+1$,
For the cyclic-shift strategy depicted in the inset of Fig.~\ref{fig:5}(b) we have $\sigma_t(x) = (x - t) \mod L$, where $L$ is the length of the chain. Here the noise between the two sites becomes time-correlated (between Floquet rounds), $\sigma_{t+1}(i+1) = \sigma_t(i)$ and the relative accumulated disorder is
\begin{align}
    \sum_{t' \leq t} (h_{\sigma_{t'}(i+1)} - h_{\sigma_{t'}(i)}) = h_{\sigma_0(i+1)} - h_{\sigma_t(i)},
\end{align}
which implies that the phase noise on the hopping term $t_{ij}$ remains bounded for all time. Compare this to the situation when no echo is performed where the relative phase grows linearly in time. We note the special nature of this echo procedure, which is intrinsically linked to spatial positions of the traps and cannot be performed by applying global operations, in contrast to the typical echo sequences present in spin systems.
A similar effect occurs in the disorder cancellation strategy using swaps, since if two sites $x_1$ and $x_2$ interact at time-step $t$, then $\sigma_t(x_1) = \sigma_{t+1}(x_2)$ and $\sigma_t(x_2) = \sigma_{t+1}(x_1)$.

The results shown in Fig.~\ref{fig:5}(b) were obtained by simulating 100 tweezers with 20 atoms in a 1D ring configuration evolving under a nearest-neighbor tunneling Hamiltonian with the hopping rate $J=1$ and the time step $\tau=0.13$, using free-fermion methods. The Hamiltonian evolution is split into two parallel Floquet rounds and a Gaussian phase disorder with $\sigma_\theta = 0.035$ is applied between each round.

\acknowledgements

We thank I.~Bloch for comments on the manuscript, and discussion of the advantages of hybrid tweezer-optical lattice platforms. 
Work at Innsbruck was supported by
the US Air Force Office of Scientific Research (AFOSR) via IOE Grant No. FA9550-19-1-7044 LASCEM, the European Union’s Horizon 2020 research and innovation program under Grant Agreement No. 101079862 (PASQuanS2), by the Simons Collaboration on Ultra-Quantum Matter, which is a grant from the Simons Foundation (651440, P.Z.), and by  ERC Starting grant QARA (101041435, H.P.).
Work at JILA is supported by NSF QLCI OMA-2016244 and NIST.
B.V.~and P.N.~acknowledge funding from the Austrian Science Foundation (FWF, P 32597 N). 
Work at Grenoble is supported by the French National Research Agency via the JCJC project QRand (ANR-20-CE47-0005), and via the research programs EPIQ (ANR-22-PETQ-0007, Plan France 2030), and QUBITAF (ANR-22-PETQ-0004, Plan France 2030). Work at Harvard was supported by the US Department of Energy (DE-SC0021013 and DOE Quantum Systems Accelerator Center, contract number 7568717), the National Science Foundation, and the Center for Ultracold Atoms. D.B.~acknowledges support from the NSF Graduate Research
Fellowship Program (grant DGE1745303) and The Fannie and John Hertz Foundation. N.M. acknowledges support from the Department of Energy Computational Science Graduate Fellowship under Award Number DE-SC0021110.

\bibliography{bibliography}

\begin{thebibliography}{117}%
\makeatletter
\providecommand \@ifxundefined [1]{%
 \@ifx{#1\undefined}
}%
\providecommand \@ifnum [1]{%
 \ifnum #1\expandafter \@firstoftwo
 \else \expandafter \@secondoftwo
 \fi
}%
\providecommand \@ifx [1]{%
 \ifx #1\expandafter \@firstoftwo
 \else \expandafter \@secondoftwo
 \fi
}%
\providecommand \natexlab [1]{#1}%
\providecommand \enquote  [1]{``#1''}%
\providecommand \bibnamefont  [1]{#1}%
\providecommand \bibfnamefont [1]{#1}%
\providecommand \citenamefont [1]{#1}%
\providecommand \href@noop [0]{\@secondoftwo}%
\providecommand \href [0]{\begingroup \@sanitize@url \@href}%
\providecommand \@href[1]{\@@startlink{#1}\@@href}%
\providecommand \@@href[1]{\endgroup#1\@@endlink}%
\providecommand \@sanitize@url [0]{\catcode `\\12\catcode `\$12\catcode
  `\&12\catcode `\#12\catcode `\^12\catcode `\_12\catcode `\%12\relax}%
\providecommand \@@startlink[1]{}%
\providecommand \@@endlink[0]{}%
\providecommand \url  [0]{\begingroup\@sanitize@url \@url }%
\providecommand \@url [1]{\endgroup\@href {#1}{\urlprefix }}%
\providecommand \urlprefix  [0]{URL }%
\providecommand \Eprint [0]{\href }%
\providecommand \doibase [0]{https://doi.org/}%
\providecommand \selectlanguage [0]{\@gobble}%
\providecommand \bibinfo  [0]{\@secondoftwo}%
\providecommand \bibfield  [0]{\@secondoftwo}%
\providecommand \translation [1]{[#1]}%
\providecommand \BibitemOpen [0]{}%
\providecommand \bibitemStop [0]{}%
\providecommand \bibitemNoStop [0]{.\EOS\space}%
\providecommand \EOS [0]{\spacefactor3000\relax}%
\providecommand \BibitemShut  [1]{\csname bibitem#1\endcsname}%
\let\auto@bib@innerbib\@empty
\bibitem [{\citenamefont {Berges}\ \emph {et~al.}(2021)\citenamefont {Berges},
  \citenamefont {Heller}, \citenamefont {Mazeliauskas},\ and\ \citenamefont
  {Venugopalan}}]{Berges_2021}%
  \BibitemOpen
  \bibfield  {author} {\bibinfo {author} {\bibfnamefont {J.}~\bibnamefont
  {Berges}}, \bibinfo {author} {\bibfnamefont {M.~P.}\ \bibnamefont {Heller}},
  \bibinfo {author} {\bibfnamefont {A.}~\bibnamefont {Mazeliauskas}},\ and\
  \bibinfo {author} {\bibfnamefont {R.}~\bibnamefont {Venugopalan}},\
  }\bibfield  {title} {\bibinfo {title} {Qcd thermalization: Ab initio
  approaches and interdisciplinary connections},\ }\href@noop {} {\bibfield
  {journal} {\bibinfo  {journal} {Reviews of Modern Physics}\ }\textbf
  {\bibinfo {volume} {93}},\ \bibinfo {pages} {035003} (\bibinfo {year}
  {2021})}\BibitemShut {NoStop}%
\bibitem [{\citenamefont {Varma}(2020)}]{Varma_2020}%
  \BibitemOpen
  \bibfield  {author} {\bibinfo {author} {\bibfnamefont {C.~M.}\ \bibnamefont
  {Varma}},\ }\bibfield  {title} {\bibinfo {title} {Colloquium: Linear in
  temperature resistivity and associated mysteries including high temperature
  superconductivity},\ }\href {https://doi.org/10.1103/RevModPhys.92.031001}
  {\bibfield  {journal} {\bibinfo  {journal} {Rev. Mod. Phys.}\ }\textbf
  {\bibinfo {volume} {92}},\ \bibinfo {pages} {031001} (\bibinfo {year}
  {2020})}\BibitemShut {NoStop}%
\bibitem [{\citenamefont {Reiher}\ \emph {et~al.}(2017)\citenamefont {Reiher},
  \citenamefont {Wiebe}, \citenamefont {Svore}, \citenamefont {Wecker},\ and\
  \citenamefont {Troyer}}]{Reiher_2017}%
  \BibitemOpen
  \bibfield  {author} {\bibinfo {author} {\bibfnamefont {M.}~\bibnamefont
  {Reiher}}, \bibinfo {author} {\bibfnamefont {N.}~\bibnamefont {Wiebe}},
  \bibinfo {author} {\bibfnamefont {K.~M.}\ \bibnamefont {Svore}}, \bibinfo
  {author} {\bibfnamefont {D.}~\bibnamefont {Wecker}},\ and\ \bibinfo {author}
  {\bibfnamefont {M.}~\bibnamefont {Troyer}},\ }\bibfield  {title} {\bibinfo
  {title} {Elucidating reaction mechanisms on quantum computers},\ }\href
  {https://doi.org/10.1073/pnas.1619152114} {\bibfield  {journal} {\bibinfo
  {journal} {Proceedings of the National Academy of Sciences}\ }\textbf
  {\bibinfo {volume} {114}},\ \bibinfo {pages} {7555} (\bibinfo {year}
  {2017})}\BibitemShut {NoStop}%
\bibitem [{\citenamefont {Troyer}\ and\ \citenamefont
  {Wiese}(2005)}]{Troyer_2005}%
  \BibitemOpen
  \bibfield  {author} {\bibinfo {author} {\bibfnamefont {M.}~\bibnamefont
  {Troyer}}\ and\ \bibinfo {author} {\bibfnamefont {U.-J.}\ \bibnamefont
  {Wiese}},\ }\bibfield  {title} {\bibinfo {title} {Computational complexity
  and fundamental limitations to fermionic quantum monte carlo simulations},\
  }\href {https://doi.org/10.1103/PhysRevLett.94.170201} {\bibfield  {journal}
  {\bibinfo  {journal} {Phys. Rev. Lett.}\ }\textbf {\bibinfo {volume} {94}},\
  \bibinfo {pages} {170201} (\bibinfo {year} {2005})}\BibitemShut {NoStop}%
\bibitem [{\citenamefont {Feynman}(1981)}]{Feynman_1982}%
  \BibitemOpen
  \bibfield  {author} {\bibinfo {author} {\bibfnamefont {R.~P.}\ \bibnamefont
  {Feynman}},\ }\bibfield  {title} {\bibinfo {title} {Simulating physics with
  computers},\ }\href {https://doi.org/10.1007/BF02650179} {\bibfield
  {journal} {\bibinfo  {journal} {International Journal of Theoretical
  Physics}\ }\textbf {\bibinfo {volume} {21}},\ \bibinfo {pages} {467}
  (\bibinfo {year} {1981})}\BibitemShut {NoStop}%
\bibitem [{\citenamefont {Abrams}\ and\ \citenamefont
  {Lloyd}(1997)}]{Abrams_1997}%
  \BibitemOpen
  \bibfield  {author} {\bibinfo {author} {\bibfnamefont {D.~S.}\ \bibnamefont
  {Abrams}}\ and\ \bibinfo {author} {\bibfnamefont {S.}~\bibnamefont {Lloyd}},\
  }\bibfield  {title} {\bibinfo {title} {Simulation of many-body fermi systems
  on a universal quantum computer},\ }\href
  {https://doi.org/10.1103/PhysRevLett.79.2586} {\bibfield  {journal} {\bibinfo
   {journal} {Phys. Rev. Lett.}\ }\textbf {\bibinfo {volume} {79}},\ \bibinfo
  {pages} {2586} (\bibinfo {year} {1997})}\BibitemShut {NoStop}%
\bibitem [{\citenamefont {Ortiz}\ \emph {et~al.}(2001)\citenamefont {Ortiz},
  \citenamefont {Gubernatis}, \citenamefont {Knill},\ and\ \citenamefont
  {Laflamme}}]{Ortiz_2001}%
  \BibitemOpen
  \bibfield  {author} {\bibinfo {author} {\bibfnamefont {G.}~\bibnamefont
  {Ortiz}}, \bibinfo {author} {\bibfnamefont {J.~E.}\ \bibnamefont
  {Gubernatis}}, \bibinfo {author} {\bibfnamefont {E.}~\bibnamefont {Knill}},\
  and\ \bibinfo {author} {\bibfnamefont {R.}~\bibnamefont {Laflamme}},\
  }\bibfield  {title} {\bibinfo {title} {Quantum algorithms for fermionic
  simulations},\ }\href {https://doi.org/10.1103/PhysRevA.64.022319} {\bibfield
   {journal} {\bibinfo  {journal} {Phys. Rev. A}\ }\textbf {\bibinfo {volume}
  {64}},\ \bibinfo {pages} {022319} (\bibinfo {year} {2001})}\BibitemShut
  {NoStop}%
\bibitem [{\citenamefont {Bravyi}\ and\ \citenamefont
  {Kitaev}(2002)}]{Bravyi_2002}%
  \BibitemOpen
  \bibfield  {author} {\bibinfo {author} {\bibfnamefont {S.~B.}\ \bibnamefont
  {Bravyi}}\ and\ \bibinfo {author} {\bibfnamefont {A.~Y.}\ \bibnamefont
  {Kitaev}},\ }\bibfield  {title} {\bibinfo {title} {Fermionic quantum
  computation},\ }\href
  {https://doi.org/https://doi.org/10.1006/aphy.2002.6254} {\bibfield
  {journal} {\bibinfo  {journal} {Annals of Physics}\ }\textbf {\bibinfo
  {volume} {298}},\ \bibinfo {pages} {210} (\bibinfo {year}
  {2002})}\BibitemShut {NoStop}%
\bibitem [{\citenamefont {Whitfield}\ \emph {et~al.}(2011)\citenamefont
  {Whitfield}, \citenamefont {Biamonte},\ and\ \citenamefont
  {Aspuru-Guzik}}]{Whitfield_2011}%
  \BibitemOpen
  \bibfield  {author} {\bibinfo {author} {\bibfnamefont {J.~D.}\ \bibnamefont
  {Whitfield}}, \bibinfo {author} {\bibfnamefont {J.}~\bibnamefont
  {Biamonte}},\ and\ \bibinfo {author} {\bibfnamefont {A.}~\bibnamefont
  {Aspuru-Guzik}},\ }\bibfield  {title} {\bibinfo {title} {Simulation of
  electronic structure hamiltonians using quantum computers},\ }\href
  {https://doi.org/10.1080/00268976.2011.552441} {\bibfield  {journal}
  {\bibinfo  {journal} {Molecular Physics}\ }\textbf {\bibinfo {volume}
  {109}},\ \bibinfo {pages} {735} (\bibinfo {year} {2011})}\BibitemShut
  {NoStop}%
\bibitem [{\citenamefont {Ball}(2005)}]{Ball_2005}%
  \BibitemOpen
  \bibfield  {author} {\bibinfo {author} {\bibfnamefont {R.~C.}\ \bibnamefont
  {Ball}},\ }\bibfield  {title} {\bibinfo {title} {Fermions without fermion
  fields},\ }\href {https://doi.org/10.1103/PhysRevLett.95.176407} {\bibfield
  {journal} {\bibinfo  {journal} {Phys. Rev. Lett.}\ }\textbf {\bibinfo
  {volume} {95}},\ \bibinfo {pages} {176407} (\bibinfo {year}
  {2005})}\BibitemShut {NoStop}%
\bibitem [{\citenamefont {Verstraete}\ and\ \citenamefont
  {Cirac}(2005)}]{Verstraete_2005}%
  \BibitemOpen
  \bibfield  {author} {\bibinfo {author} {\bibfnamefont {F.}~\bibnamefont
  {Verstraete}}\ and\ \bibinfo {author} {\bibfnamefont {J.~I.}\ \bibnamefont
  {Cirac}},\ }\bibfield  {title} {\bibinfo {title} {Mapping local hamiltonians
  of fermions to local hamiltonians of spins},\ }\href
  {https://doi.org/10.1088/1742-5468/2005/09/p09012} {\bibfield  {journal}
  {\bibinfo  {journal} {Journal of Statistical Mechanics: Theory and
  Experiment}\ }\textbf {\bibinfo {volume} {2005}},\ \bibinfo {pages} {P09012}
  (\bibinfo {year} {2005})}\BibitemShut {NoStop}%
\bibitem [{\citenamefont {Whitfield}\ \emph {et~al.}(2016)\citenamefont
  {Whitfield}, \citenamefont {Havl\'{\i}\ifmmode~\check{c}\else \v{c}\fi{}ek},\
  and\ \citenamefont {Troyer}}]{Whitfield_2016}%
  \BibitemOpen
  \bibfield  {author} {\bibinfo {author} {\bibfnamefont {J.~D.}\ \bibnamefont
  {Whitfield}}, \bibinfo {author} {\bibfnamefont {V.~c.~v.}\ \bibnamefont
  {Havl\'{\i}\ifmmode~\check{c}\else \v{c}\fi{}ek}},\ and\ \bibinfo {author}
  {\bibfnamefont {M.}~\bibnamefont {Troyer}},\ }\bibfield  {title} {\bibinfo
  {title} {Local spin operators for fermion simulations},\ }\href
  {https://doi.org/10.1103/PhysRevA.94.030301} {\bibfield  {journal} {\bibinfo
  {journal} {Phys. Rev. A}\ }\textbf {\bibinfo {volume} {94}},\ \bibinfo
  {pages} {030301} (\bibinfo {year} {2016})}\BibitemShut {NoStop}%
\bibitem [{\citenamefont {Peruzzo}\ \emph {et~al.}(2014)\citenamefont
  {Peruzzo}, \citenamefont {McClean}, \citenamefont {Shadbolt}, \citenamefont
  {Yung}, \citenamefont {Zhou}, \citenamefont {Love}, \citenamefont
  {Aspuru-Guzik},\ and\ \citenamefont {O'Brien}}]{Peruzzo2014}%
  \BibitemOpen
  \bibfield  {author} {\bibinfo {author} {\bibfnamefont {A.}~\bibnamefont
  {Peruzzo}}, \bibinfo {author} {\bibfnamefont {J.}~\bibnamefont {McClean}},
  \bibinfo {author} {\bibfnamefont {P.}~\bibnamefont {Shadbolt}}, \bibinfo
  {author} {\bibfnamefont {M.-H.}\ \bibnamefont {Yung}}, \bibinfo {author}
  {\bibfnamefont {X.-Q.}\ \bibnamefont {Zhou}}, \bibinfo {author}
  {\bibfnamefont {P.~J.}\ \bibnamefont {Love}}, \bibinfo {author}
  {\bibfnamefont {A.}~\bibnamefont {Aspuru-Guzik}},\ and\ \bibinfo {author}
  {\bibfnamefont {J.~L.}\ \bibnamefont {O'Brien}},\ }\bibfield  {title}
  {\bibinfo {title} {A variational eigenvalue solver on a photonic quantum
  processor},\ }\href {https://doi.org/10.1038/ncomms5213} {\bibfield
  {journal} {\bibinfo  {journal} {Nature Communications}\ }\textbf {\bibinfo
  {volume} {5}},\ \bibinfo {pages} {4213} (\bibinfo {year} {2014})}\BibitemShut
  {NoStop}%
\bibitem [{\citenamefont {O'Malley}\ \emph {et~al.}(2016)\citenamefont
  {O'Malley}, \citenamefont {Babbush}, \citenamefont {Kivlichan}, \citenamefont
  {Romero}, \citenamefont {McClean}, \citenamefont {Barends}, \citenamefont
  {Kelly}, \citenamefont {Roushan}, \citenamefont {Tranter}, \citenamefont
  {Ding}, \citenamefont {Campbell}, \citenamefont {Chen}, \citenamefont {Chen},
  \citenamefont {Chiaro}, \citenamefont {Dunsworth}, \citenamefont {Fowler},
  \citenamefont {Jeffrey}, \citenamefont {Lucero}, \citenamefont {Megrant},
  \citenamefont {Mutus}, \citenamefont {Neeley}, \citenamefont {Neill},
  \citenamefont {Quintana}, \citenamefont {Sank}, \citenamefont {Vainsencher},
  \citenamefont {Wenner}, \citenamefont {White}, \citenamefont {Coveney},
  \citenamefont {Love}, \citenamefont {Neven}, \citenamefont {Aspuru-Guzik},\
  and\ \citenamefont {Martinis}}]{OMalley_2016}%
  \BibitemOpen
  \bibfield  {author} {\bibinfo {author} {\bibfnamefont {P.~J.~J.}\
  \bibnamefont {O'Malley}}, \bibinfo {author} {\bibfnamefont {R.}~\bibnamefont
  {Babbush}}, \bibinfo {author} {\bibfnamefont {I.~D.}\ \bibnamefont
  {Kivlichan}}, \bibinfo {author} {\bibfnamefont {J.}~\bibnamefont {Romero}},
  \bibinfo {author} {\bibfnamefont {J.~R.}\ \bibnamefont {McClean}}, \bibinfo
  {author} {\bibfnamefont {R.}~\bibnamefont {Barends}}, \bibinfo {author}
  {\bibfnamefont {J.}~\bibnamefont {Kelly}}, \bibinfo {author} {\bibfnamefont
  {P.}~\bibnamefont {Roushan}}, \bibinfo {author} {\bibfnamefont
  {A.}~\bibnamefont {Tranter}}, \bibinfo {author} {\bibfnamefont
  {N.}~\bibnamefont {Ding}}, \bibinfo {author} {\bibfnamefont {B.}~\bibnamefont
  {Campbell}}, \bibinfo {author} {\bibfnamefont {Y.}~\bibnamefont {Chen}},
  \bibinfo {author} {\bibfnamefont {Z.}~\bibnamefont {Chen}}, \bibinfo {author}
  {\bibfnamefont {B.}~\bibnamefont {Chiaro}}, \bibinfo {author} {\bibfnamefont
  {A.}~\bibnamefont {Dunsworth}}, \bibinfo {author} {\bibfnamefont {A.~G.}\
  \bibnamefont {Fowler}}, \bibinfo {author} {\bibfnamefont {E.}~\bibnamefont
  {Jeffrey}}, \bibinfo {author} {\bibfnamefont {E.}~\bibnamefont {Lucero}},
  \bibinfo {author} {\bibfnamefont {A.}~\bibnamefont {Megrant}}, \bibinfo
  {author} {\bibfnamefont {J.~Y.}\ \bibnamefont {Mutus}}, \bibinfo {author}
  {\bibfnamefont {M.}~\bibnamefont {Neeley}}, \bibinfo {author} {\bibfnamefont
  {C.}~\bibnamefont {Neill}}, \bibinfo {author} {\bibfnamefont
  {C.}~\bibnamefont {Quintana}}, \bibinfo {author} {\bibfnamefont
  {D.}~\bibnamefont {Sank}}, \bibinfo {author} {\bibfnamefont {A.}~\bibnamefont
  {Vainsencher}}, \bibinfo {author} {\bibfnamefont {J.}~\bibnamefont {Wenner}},
  \bibinfo {author} {\bibfnamefont {T.~C.}\ \bibnamefont {White}}, \bibinfo
  {author} {\bibfnamefont {P.~V.}\ \bibnamefont {Coveney}}, \bibinfo {author}
  {\bibfnamefont {P.~J.}\ \bibnamefont {Love}}, \bibinfo {author}
  {\bibfnamefont {H.}~\bibnamefont {Neven}}, \bibinfo {author} {\bibfnamefont
  {A.}~\bibnamefont {Aspuru-Guzik}},\ and\ \bibinfo {author} {\bibfnamefont
  {J.~M.}\ \bibnamefont {Martinis}},\ }\bibfield  {title} {\bibinfo {title}
  {Scalable quantum simulation of molecular energies},\ }\href
  {https://doi.org/10.1103/PhysRevX.6.031007} {\bibfield  {journal} {\bibinfo
  {journal} {Phys. Rev. X}\ }\textbf {\bibinfo {volume} {6}},\ \bibinfo {pages}
  {031007} (\bibinfo {year} {2016})}\BibitemShut {NoStop}%
\bibitem [{\citenamefont {Kandala}\ \emph
  {et~al.}(2017{\natexlab{a}})\citenamefont {Kandala}, \citenamefont
  {Mezzacapo}, \citenamefont {Temme}, \citenamefont {Takita}, \citenamefont
  {Brink}, \citenamefont {Chow},\ and\ \citenamefont {Gambetta}}]{Kandala2017}%
  \BibitemOpen
  \bibfield  {author} {\bibinfo {author} {\bibfnamefont {A.}~\bibnamefont
  {Kandala}}, \bibinfo {author} {\bibfnamefont {A.}~\bibnamefont {Mezzacapo}},
  \bibinfo {author} {\bibfnamefont {K.}~\bibnamefont {Temme}}, \bibinfo
  {author} {\bibfnamefont {M.}~\bibnamefont {Takita}}, \bibinfo {author}
  {\bibfnamefont {M.}~\bibnamefont {Brink}}, \bibinfo {author} {\bibfnamefont
  {J.~M.}\ \bibnamefont {Chow}},\ and\ \bibinfo {author} {\bibfnamefont
  {J.~M.}\ \bibnamefont {Gambetta}},\ }\bibfield  {title} {\bibinfo {title}
  {Hardware-efficient variational quantum eigensolver for small molecules and
  quantum magnets},\ }\href {https://doi.org/10.1038/nature23879} {\bibfield
  {journal} {\bibinfo  {journal} {Nature}\ }\textbf {\bibinfo {volume} {549}},\
  \bibinfo {pages} {242} (\bibinfo {year} {2017}{\natexlab{a}})}\BibitemShut
  {NoStop}%
\bibitem [{\citenamefont {Hempel}\ \emph {et~al.}(2018)\citenamefont {Hempel},
  \citenamefont {Maier}, \citenamefont {Romero}, \citenamefont {McClean},
  \citenamefont {Monz}, \citenamefont {Shen}, \citenamefont {Jurcevic},
  \citenamefont {Lanyon}, \citenamefont {Love}, \citenamefont {Babbush} \emph
  {et~al.}}]{Hempel2018}%
  \BibitemOpen
  \bibfield  {author} {\bibinfo {author} {\bibfnamefont {C.}~\bibnamefont
  {Hempel}}, \bibinfo {author} {\bibfnamefont {C.}~\bibnamefont {Maier}},
  \bibinfo {author} {\bibfnamefont {J.}~\bibnamefont {Romero}}, \bibinfo
  {author} {\bibfnamefont {J.}~\bibnamefont {McClean}}, \bibinfo {author}
  {\bibfnamefont {T.}~\bibnamefont {Monz}}, \bibinfo {author} {\bibfnamefont
  {H.}~\bibnamefont {Shen}}, \bibinfo {author} {\bibfnamefont {P.}~\bibnamefont
  {Jurcevic}}, \bibinfo {author} {\bibfnamefont {B.~P.}\ \bibnamefont
  {Lanyon}}, \bibinfo {author} {\bibfnamefont {P.}~\bibnamefont {Love}},
  \bibinfo {author} {\bibfnamefont {R.}~\bibnamefont {Babbush}}, \emph
  {et~al.},\ }\bibfield  {title} {\bibinfo {title} {Quantum chemistry
  calculations on a trapped-ion quantum simulator},\ }\href
  {https://doi.org/10.1103/PhysRevX.8.031022} {\bibfield  {journal} {\bibinfo
  {journal} {Physical Review X}\ }\textbf {\bibinfo {volume} {8}},\ \bibinfo
  {pages} {031022} (\bibinfo {year} {2018})}\BibitemShut {NoStop}%
\bibitem [{\citenamefont {{Google AI QUANTUM}}\ \emph
  {et~al.}(2020)\citenamefont {{Google AI QUANTUM}}, \citenamefont {Arute},
  \citenamefont {Arya}, \citenamefont {Babbush}, \citenamefont {Bacon},
  \citenamefont {Bardin}, \citenamefont {Barends}, \citenamefont {Boixo},
  \citenamefont {Broughton} \emph {et~al.}}]{Arute2020}%
  \BibitemOpen
  \bibfield  {author} {\bibinfo {author} {\bibnamefont {{Google AI QUANTUM}}},
  \bibinfo {author} {\bibfnamefont {F.}~\bibnamefont {Arute}}, \bibinfo
  {author} {\bibfnamefont {K.}~\bibnamefont {Arya}}, \bibinfo {author}
  {\bibfnamefont {R.}~\bibnamefont {Babbush}}, \bibinfo {author} {\bibfnamefont
  {D.}~\bibnamefont {Bacon}}, \bibinfo {author} {\bibfnamefont {J.~C.}\
  \bibnamefont {Bardin}}, \bibinfo {author} {\bibfnamefont {R.}~\bibnamefont
  {Barends}}, \bibinfo {author} {\bibfnamefont {S.}~\bibnamefont {Boixo}},
  \bibinfo {author} {\bibfnamefont {M.}~\bibnamefont {Broughton}}, \emph
  {et~al.},\ }\bibfield  {title} {\bibinfo {title} {Hartree-fock on a
  superconducting qubit quantum computer},\ }\href
  {https://doi.org/10.1126/science.abb9811} {\bibfield  {journal} {\bibinfo
  {journal} {Science}\ }\textbf {\bibinfo {volume} {369}},\ \bibinfo {pages}
  {1084} (\bibinfo {year} {2020})}\BibitemShut {NoStop}%
\bibitem [{\citenamefont {Lanyon}\ \emph {et~al.}(2011)\citenamefont {Lanyon},
  \citenamefont {Hempel}, \citenamefont {Nigg}, \citenamefont {M{\"u}ller},
  \citenamefont {Gerritsma}, \citenamefont {Z{\"a}hringer}, \citenamefont
  {Schindler}, \citenamefont {Barreiro}, \citenamefont {Rambach}, \citenamefont
  {Kirchmair}, \citenamefont {Hennrich}, \citenamefont {Zoller}, \citenamefont
  {Blatt},\ and\ \citenamefont {Roos}}]{Lanyon_2011}%
  \BibitemOpen
  \bibfield  {author} {\bibinfo {author} {\bibfnamefont {B.~P.}\ \bibnamefont
  {Lanyon}}, \bibinfo {author} {\bibfnamefont {C.}~\bibnamefont {Hempel}},
  \bibinfo {author} {\bibfnamefont {D.}~\bibnamefont {Nigg}}, \bibinfo {author}
  {\bibfnamefont {M.}~\bibnamefont {M{\"u}ller}}, \bibinfo {author}
  {\bibfnamefont {R.}~\bibnamefont {Gerritsma}}, \bibinfo {author}
  {\bibfnamefont {F.}~\bibnamefont {Z{\"a}hringer}}, \bibinfo {author}
  {\bibfnamefont {P.}~\bibnamefont {Schindler}}, \bibinfo {author}
  {\bibfnamefont {J.~T.}\ \bibnamefont {Barreiro}}, \bibinfo {author}
  {\bibfnamefont {M.}~\bibnamefont {Rambach}}, \bibinfo {author} {\bibfnamefont
  {G.}~\bibnamefont {Kirchmair}}, \bibinfo {author} {\bibfnamefont
  {M.}~\bibnamefont {Hennrich}}, \bibinfo {author} {\bibfnamefont
  {P.}~\bibnamefont {Zoller}}, \bibinfo {author} {\bibfnamefont
  {R.}~\bibnamefont {Blatt}},\ and\ \bibinfo {author} {\bibfnamefont {C.~F.}\
  \bibnamefont {Roos}},\ }\bibfield  {title} {\bibinfo {title} {Universal
  digital quantum simulation with trapped ions},\ }\href
  {https://doi.org/10.1126/science.1208001} {\bibfield  {journal} {\bibinfo
  {journal} {Science}\ }\textbf {\bibinfo {volume} {334}},\ \bibinfo {pages}
  {57} (\bibinfo {year} {2011})}\BibitemShut {NoStop}%
\bibitem [{\citenamefont {Salath\'e}\ \emph {et~al.}(2015)\citenamefont
  {Salath\'e}, \citenamefont {Mondal}, \citenamefont {Oppliger}, \citenamefont
  {Heinsoo}, \citenamefont {Kurpiers}, \citenamefont
  {Poto\ifmmode~\check{c}\else \v{c}\fi{}nik}, \citenamefont {Mezzacapo},
  \citenamefont {Las~Heras}, \citenamefont {Lamata}, \citenamefont {Solano},
  \citenamefont {Filipp},\ and\ \citenamefont {Wallraff}}]{Salathe_2015}%
  \BibitemOpen
  \bibfield  {author} {\bibinfo {author} {\bibfnamefont {Y.}~\bibnamefont
  {Salath\'e}}, \bibinfo {author} {\bibfnamefont {M.}~\bibnamefont {Mondal}},
  \bibinfo {author} {\bibfnamefont {M.}~\bibnamefont {Oppliger}}, \bibinfo
  {author} {\bibfnamefont {J.}~\bibnamefont {Heinsoo}}, \bibinfo {author}
  {\bibfnamefont {P.}~\bibnamefont {Kurpiers}}, \bibinfo {author}
  {\bibfnamefont {A.}~\bibnamefont {Poto\ifmmode~\check{c}\else
  \v{c}\fi{}nik}}, \bibinfo {author} {\bibfnamefont {A.}~\bibnamefont
  {Mezzacapo}}, \bibinfo {author} {\bibfnamefont {U.}~\bibnamefont
  {Las~Heras}}, \bibinfo {author} {\bibfnamefont {L.}~\bibnamefont {Lamata}},
  \bibinfo {author} {\bibfnamefont {E.}~\bibnamefont {Solano}}, \bibinfo
  {author} {\bibfnamefont {S.}~\bibnamefont {Filipp}},\ and\ \bibinfo {author}
  {\bibfnamefont {A.}~\bibnamefont {Wallraff}},\ }\bibfield  {title} {\bibinfo
  {title} {Digital quantum simulation of spin models with circuit quantum
  electrodynamics},\ }\href {https://doi.org/10.1103/PhysRevX.5.021027}
  {\bibfield  {journal} {\bibinfo  {journal} {Phys. Rev. X}\ }\textbf {\bibinfo
  {volume} {5}},\ \bibinfo {pages} {021027} (\bibinfo {year}
  {2015})}\BibitemShut {NoStop}%
\bibitem [{\citenamefont {Barends}\ \emph {et~al.}(2015)\citenamefont
  {Barends}, \citenamefont {Lamata}, \citenamefont {Kelly}, \citenamefont
  {Garc{\'\i}a-{\'A}lvarez}, \citenamefont {Fowler}, \citenamefont {Megrant},
  \citenamefont {Jeffrey}, \citenamefont {White}, \citenamefont {Sank},
  \citenamefont {Mutus}, \citenamefont {Campbell}, \citenamefont {Chen},
  \citenamefont {Chen}, \citenamefont {Chiaro}, \citenamefont {Dunsworth},
  \citenamefont {Hoi}, \citenamefont {Neill}, \citenamefont {O'Malley},
  \citenamefont {Quintana}, \citenamefont {Roushan}, \citenamefont
  {Vainsencher}, \citenamefont {Wenner}, \citenamefont {Solano},\ and\
  \citenamefont {Martinis}}]{Barends_2015}%
  \BibitemOpen
  \bibfield  {author} {\bibinfo {author} {\bibfnamefont {R.}~\bibnamefont
  {Barends}}, \bibinfo {author} {\bibfnamefont {L.}~\bibnamefont {Lamata}},
  \bibinfo {author} {\bibfnamefont {J.}~\bibnamefont {Kelly}}, \bibinfo
  {author} {\bibfnamefont {L.}~\bibnamefont {Garc{\'\i}a-{\'A}lvarez}},
  \bibinfo {author} {\bibfnamefont {A.~G.}\ \bibnamefont {Fowler}}, \bibinfo
  {author} {\bibfnamefont {A.}~\bibnamefont {Megrant}}, \bibinfo {author}
  {\bibfnamefont {E.}~\bibnamefont {Jeffrey}}, \bibinfo {author} {\bibfnamefont
  {T.~C.}\ \bibnamefont {White}}, \bibinfo {author} {\bibfnamefont
  {D.}~\bibnamefont {Sank}}, \bibinfo {author} {\bibfnamefont {J.~Y.}\
  \bibnamefont {Mutus}}, \bibinfo {author} {\bibfnamefont {B.}~\bibnamefont
  {Campbell}}, \bibinfo {author} {\bibfnamefont {Y.}~\bibnamefont {Chen}},
  \bibinfo {author} {\bibfnamefont {Z.}~\bibnamefont {Chen}}, \bibinfo {author}
  {\bibfnamefont {B.}~\bibnamefont {Chiaro}}, \bibinfo {author} {\bibfnamefont
  {A.}~\bibnamefont {Dunsworth}}, \bibinfo {author} {\bibfnamefont {I.~C.}\
  \bibnamefont {Hoi}}, \bibinfo {author} {\bibfnamefont {C.}~\bibnamefont
  {Neill}}, \bibinfo {author} {\bibfnamefont {P.~J.~J.}\ \bibnamefont
  {O'Malley}}, \bibinfo {author} {\bibfnamefont {C.}~\bibnamefont {Quintana}},
  \bibinfo {author} {\bibfnamefont {P.}~\bibnamefont {Roushan}}, \bibinfo
  {author} {\bibfnamefont {A.}~\bibnamefont {Vainsencher}}, \bibinfo {author}
  {\bibfnamefont {J.}~\bibnamefont {Wenner}}, \bibinfo {author} {\bibfnamefont
  {E.}~\bibnamefont {Solano}},\ and\ \bibinfo {author} {\bibfnamefont {J.~M.}\
  \bibnamefont {Martinis}},\ }\bibfield  {title} {\bibinfo {title} {Digital
  quantum simulation of fermionic models with a superconducting circuit},\
  }\href {https://doi.org/10.1038/ncomms8654} {\bibfield  {journal} {\bibinfo
  {journal} {Nature Communications}\ }\textbf {\bibinfo {volume} {6}},\
  \bibinfo {pages} {7654} (\bibinfo {year} {2015})}\BibitemShut {NoStop}%
\bibitem [{\citenamefont {Choi}\ \emph {et~al.}(2017)\citenamefont {Choi},
  \citenamefont {Choi}, \citenamefont {Landig}, \citenamefont {Kucsko},
  \citenamefont {Zhou}, \citenamefont {Isoya}, \citenamefont {Jelezko},
  \citenamefont {Onoda}, \citenamefont {Sumiya}, \citenamefont {Khemani},
  \citenamefont {von Keyserlingk}, \citenamefont {Yao}, \citenamefont
  {Demler},\ and\ \citenamefont {Lukin}}]{Choi_2017}%
  \BibitemOpen
  \bibfield  {author} {\bibinfo {author} {\bibfnamefont {S.}~\bibnamefont
  {Choi}}, \bibinfo {author} {\bibfnamefont {J.}~\bibnamefont {Choi}}, \bibinfo
  {author} {\bibfnamefont {R.}~\bibnamefont {Landig}}, \bibinfo {author}
  {\bibfnamefont {G.}~\bibnamefont {Kucsko}}, \bibinfo {author} {\bibfnamefont
  {H.}~\bibnamefont {Zhou}}, \bibinfo {author} {\bibfnamefont {J.}~\bibnamefont
  {Isoya}}, \bibinfo {author} {\bibfnamefont {F.}~\bibnamefont {Jelezko}},
  \bibinfo {author} {\bibfnamefont {S.}~\bibnamefont {Onoda}}, \bibinfo
  {author} {\bibfnamefont {H.}~\bibnamefont {Sumiya}}, \bibinfo {author}
  {\bibfnamefont {V.}~\bibnamefont {Khemani}}, \bibinfo {author} {\bibfnamefont
  {C.}~\bibnamefont {von Keyserlingk}}, \bibinfo {author} {\bibfnamefont
  {N.~Y.}\ \bibnamefont {Yao}}, \bibinfo {author} {\bibfnamefont
  {E.}~\bibnamefont {Demler}},\ and\ \bibinfo {author} {\bibfnamefont {M.~D.}\
  \bibnamefont {Lukin}},\ }\bibfield  {title} {\bibinfo {title} {Observation of
  discrete time-crystalline order in a disordered dipolar many-body system},\
  }\href {https://doi.org/10.1038/nature21426} {\bibfield  {journal} {\bibinfo
  {journal} {Nature}\ }\textbf {\bibinfo {volume} {543}},\ \bibinfo {pages}
  {221} (\bibinfo {year} {2017})}\BibitemShut {NoStop}%
\bibitem [{\citenamefont {Zhang}\ \emph {et~al.}(2017)\citenamefont {Zhang},
  \citenamefont {Hess}, \citenamefont {Kyprianidis}, \citenamefont {Becker},
  \citenamefont {Lee}, \citenamefont {Smith}, \citenamefont {Pagano},
  \citenamefont {Potirniche}, \citenamefont {Potter}, \citenamefont
  {Vishwanath}, \citenamefont {Yao},\ and\ \citenamefont
  {Monroe}}]{Zhang_2017}%
  \BibitemOpen
  \bibfield  {author} {\bibinfo {author} {\bibfnamefont {J.}~\bibnamefont
  {Zhang}}, \bibinfo {author} {\bibfnamefont {P.~W.}\ \bibnamefont {Hess}},
  \bibinfo {author} {\bibfnamefont {A.}~\bibnamefont {Kyprianidis}}, \bibinfo
  {author} {\bibfnamefont {P.}~\bibnamefont {Becker}}, \bibinfo {author}
  {\bibfnamefont {A.}~\bibnamefont {Lee}}, \bibinfo {author} {\bibfnamefont
  {J.}~\bibnamefont {Smith}}, \bibinfo {author} {\bibfnamefont
  {G.}~\bibnamefont {Pagano}}, \bibinfo {author} {\bibfnamefont {I.~D.}\
  \bibnamefont {Potirniche}}, \bibinfo {author} {\bibfnamefont {A.~C.}\
  \bibnamefont {Potter}}, \bibinfo {author} {\bibfnamefont {A.}~\bibnamefont
  {Vishwanath}}, \bibinfo {author} {\bibfnamefont {N.~Y.}\ \bibnamefont
  {Yao}},\ and\ \bibinfo {author} {\bibfnamefont {C.}~\bibnamefont {Monroe}},\
  }\bibfield  {title} {\bibinfo {title} {Observation of a discrete time
  crystal},\ }\href {https://doi.org/10.1038/nature21413} {\bibfield  {journal}
  {\bibinfo  {journal} {Nature}\ }\textbf {\bibinfo {volume} {543}},\ \bibinfo
  {pages} {217} (\bibinfo {year} {2017})}\BibitemShut {NoStop}%
\bibitem [{\citenamefont {Satzinger}\ \emph {et~al.}(2021)\citenamefont
  {Satzinger}, \citenamefont {Liu}, \citenamefont {Smith}, \citenamefont
  {Knapp}, \citenamefont {Newman}, \citenamefont {Jones}, \citenamefont {Chen},
  \citenamefont {Quintana}, \citenamefont {Mi}, \citenamefont {Dunsworth},
  \citenamefont {Gidney}, \citenamefont {Aleiner}, \citenamefont {Arute},
  \citenamefont {Arya}, \citenamefont {Atalaya}, \citenamefont {Babbush},
  \citenamefont {Bardin}, \citenamefont {Barends}, \citenamefont {Basso},
  \citenamefont {Bengtsson}, \citenamefont {Bilmes}, \citenamefont {Broughton},
  \citenamefont {Buckley}, \citenamefont {Buell}, \citenamefont {Burkett},
  \citenamefont {Bushnell}, \citenamefont {Chiaro}, \citenamefont {Collins},
  \citenamefont {Courtney}, \citenamefont {Demura}, \citenamefont {Derk},
  \citenamefont {Eppens}, \citenamefont {Erickson}, \citenamefont {Faoro},
  \citenamefont {Farhi}, \citenamefont {Fowler}, \citenamefont {Foxen},
  \citenamefont {Giustina}, \citenamefont {Greene}, \citenamefont {Gross},
  \citenamefont {Harrigan}, \citenamefont {Harrington}, \citenamefont {Hilton},
  \citenamefont {Hong}, \citenamefont {Huang}, \citenamefont {Huggins},
  \citenamefont {Ioffe}, \citenamefont {Isakov}, \citenamefont {Jeffrey},
  \citenamefont {Jiang}, \citenamefont {Kafri}, \citenamefont {Kechedzhi},
  \citenamefont {Khattar}, \citenamefont {Kim}, \citenamefont {Klimov},
  \citenamefont {Korotkov}, \citenamefont {Kostritsa}, \citenamefont
  {Landhuis}, \citenamefont {Laptev}, \citenamefont {Locharla}, \citenamefont
  {Lucero}, \citenamefont {Martin}, \citenamefont {McClean}, \citenamefont
  {McEwen}, \citenamefont {Miao}, \citenamefont {Mohseni}, \citenamefont
  {Montazeri}, \citenamefont {Mruczkiewicz}, \citenamefont {Mutus},
  \citenamefont {Naaman}, \citenamefont {Neeley}, \citenamefont {Neill},
  \citenamefont {Niu}, \citenamefont {O'Brien}, \citenamefont {Opremcak},
  \citenamefont {Pat{\'o}}, \citenamefont {Petukhov}, \citenamefont {Rubin},
  \citenamefont {Sank}, \citenamefont {Shvarts}, \citenamefont {Strain},
  \citenamefont {Szalay}, \citenamefont {Villalonga}, \citenamefont {White},
  \citenamefont {Yao}, \citenamefont {Yeh}, \citenamefont {Yoo}, \citenamefont
  {Zalcman}, \citenamefont {Neven}, \citenamefont {Boixo}, \citenamefont
  {Megrant}, \citenamefont {Chen}, \citenamefont {Kelly}, \citenamefont
  {Smelyanskiy}, \citenamefont {Kitaev}, \citenamefont {Knap}, \citenamefont
  {Pollmann},\ and\ \citenamefont {Roushan}}]{Satzinger_2021}%
  \BibitemOpen
  \bibfield  {author} {\bibinfo {author} {\bibfnamefont {K.~J.}\ \bibnamefont
  {Satzinger}}, \bibinfo {author} {\bibfnamefont {Y.-J.}\ \bibnamefont {Liu}},
  \bibinfo {author} {\bibfnamefont {A.}~\bibnamefont {Smith}}, \bibinfo
  {author} {\bibfnamefont {C.}~\bibnamefont {Knapp}}, \bibinfo {author}
  {\bibfnamefont {M.}~\bibnamefont {Newman}}, \bibinfo {author} {\bibfnamefont
  {C.}~\bibnamefont {Jones}}, \bibinfo {author} {\bibfnamefont
  {Z.}~\bibnamefont {Chen}}, \bibinfo {author} {\bibfnamefont {C.}~\bibnamefont
  {Quintana}}, \bibinfo {author} {\bibfnamefont {X.}~\bibnamefont {Mi}},
  \bibinfo {author} {\bibfnamefont {A.}~\bibnamefont {Dunsworth}}, \bibinfo
  {author} {\bibfnamefont {C.}~\bibnamefont {Gidney}}, \bibinfo {author}
  {\bibfnamefont {I.}~\bibnamefont {Aleiner}}, \bibinfo {author} {\bibfnamefont
  {F.}~\bibnamefont {Arute}}, \bibinfo {author} {\bibfnamefont
  {K.}~\bibnamefont {Arya}}, \bibinfo {author} {\bibfnamefont {J.}~\bibnamefont
  {Atalaya}}, \bibinfo {author} {\bibfnamefont {R.}~\bibnamefont {Babbush}},
  \bibinfo {author} {\bibfnamefont {J.~C.}\ \bibnamefont {Bardin}}, \bibinfo
  {author} {\bibfnamefont {R.}~\bibnamefont {Barends}}, \bibinfo {author}
  {\bibfnamefont {J.}~\bibnamefont {Basso}}, \bibinfo {author} {\bibfnamefont
  {A.}~\bibnamefont {Bengtsson}}, \bibinfo {author} {\bibfnamefont
  {A.}~\bibnamefont {Bilmes}}, \bibinfo {author} {\bibfnamefont
  {M.}~\bibnamefont {Broughton}}, \bibinfo {author} {\bibfnamefont {B.~B.}\
  \bibnamefont {Buckley}}, \bibinfo {author} {\bibfnamefont {D.~A.}\
  \bibnamefont {Buell}}, \bibinfo {author} {\bibfnamefont {B.}~\bibnamefont
  {Burkett}}, \bibinfo {author} {\bibfnamefont {N.}~\bibnamefont {Bushnell}},
  \bibinfo {author} {\bibfnamefont {B.}~\bibnamefont {Chiaro}}, \bibinfo
  {author} {\bibfnamefont {R.}~\bibnamefont {Collins}}, \bibinfo {author}
  {\bibfnamefont {W.}~\bibnamefont {Courtney}}, \bibinfo {author}
  {\bibfnamefont {S.}~\bibnamefont {Demura}}, \bibinfo {author} {\bibfnamefont
  {A.~R.}\ \bibnamefont {Derk}}, \bibinfo {author} {\bibfnamefont
  {D.}~\bibnamefont {Eppens}}, \bibinfo {author} {\bibfnamefont
  {C.}~\bibnamefont {Erickson}}, \bibinfo {author} {\bibfnamefont
  {L.}~\bibnamefont {Faoro}}, \bibinfo {author} {\bibfnamefont
  {E.}~\bibnamefont {Farhi}}, \bibinfo {author} {\bibfnamefont {A.~G.}\
  \bibnamefont {Fowler}}, \bibinfo {author} {\bibfnamefont {B.}~\bibnamefont
  {Foxen}}, \bibinfo {author} {\bibfnamefont {M.}~\bibnamefont {Giustina}},
  \bibinfo {author} {\bibfnamefont {A.}~\bibnamefont {Greene}}, \bibinfo
  {author} {\bibfnamefont {J.~A.}\ \bibnamefont {Gross}}, \bibinfo {author}
  {\bibfnamefont {M.~P.}\ \bibnamefont {Harrigan}}, \bibinfo {author}
  {\bibfnamefont {S.~D.}\ \bibnamefont {Harrington}}, \bibinfo {author}
  {\bibfnamefont {J.}~\bibnamefont {Hilton}}, \bibinfo {author} {\bibfnamefont
  {S.}~\bibnamefont {Hong}}, \bibinfo {author} {\bibfnamefont {T.}~\bibnamefont
  {Huang}}, \bibinfo {author} {\bibfnamefont {W.~J.}\ \bibnamefont {Huggins}},
  \bibinfo {author} {\bibfnamefont {L.~B.}\ \bibnamefont {Ioffe}}, \bibinfo
  {author} {\bibfnamefont {S.~V.}\ \bibnamefont {Isakov}}, \bibinfo {author}
  {\bibfnamefont {E.}~\bibnamefont {Jeffrey}}, \bibinfo {author} {\bibfnamefont
  {Z.}~\bibnamefont {Jiang}}, \bibinfo {author} {\bibfnamefont
  {D.}~\bibnamefont {Kafri}}, \bibinfo {author} {\bibfnamefont
  {K.}~\bibnamefont {Kechedzhi}}, \bibinfo {author} {\bibfnamefont
  {T.}~\bibnamefont {Khattar}}, \bibinfo {author} {\bibfnamefont
  {S.}~\bibnamefont {Kim}}, \bibinfo {author} {\bibfnamefont {P.~V.}\
  \bibnamefont {Klimov}}, \bibinfo {author} {\bibfnamefont {A.~N.}\
  \bibnamefont {Korotkov}}, \bibinfo {author} {\bibfnamefont {F.}~\bibnamefont
  {Kostritsa}}, \bibinfo {author} {\bibfnamefont {D.}~\bibnamefont {Landhuis}},
  \bibinfo {author} {\bibfnamefont {P.}~\bibnamefont {Laptev}}, \bibinfo
  {author} {\bibfnamefont {A.}~\bibnamefont {Locharla}}, \bibinfo {author}
  {\bibfnamefont {E.}~\bibnamefont {Lucero}}, \bibinfo {author} {\bibfnamefont
  {O.}~\bibnamefont {Martin}}, \bibinfo {author} {\bibfnamefont {J.~R.}\
  \bibnamefont {McClean}}, \bibinfo {author} {\bibfnamefont {M.}~\bibnamefont
  {McEwen}}, \bibinfo {author} {\bibfnamefont {K.~C.}\ \bibnamefont {Miao}},
  \bibinfo {author} {\bibfnamefont {M.}~\bibnamefont {Mohseni}}, \bibinfo
  {author} {\bibfnamefont {S.}~\bibnamefont {Montazeri}}, \bibinfo {author}
  {\bibfnamefont {W.}~\bibnamefont {Mruczkiewicz}}, \bibinfo {author}
  {\bibfnamefont {J.}~\bibnamefont {Mutus}}, \bibinfo {author} {\bibfnamefont
  {O.}~\bibnamefont {Naaman}}, \bibinfo {author} {\bibfnamefont
  {M.}~\bibnamefont {Neeley}}, \bibinfo {author} {\bibfnamefont
  {C.}~\bibnamefont {Neill}}, \bibinfo {author} {\bibfnamefont {M.~Y.}\
  \bibnamefont {Niu}}, \bibinfo {author} {\bibfnamefont {T.~E.}\ \bibnamefont
  {O'Brien}}, \bibinfo {author} {\bibfnamefont {A.}~\bibnamefont {Opremcak}},
  \bibinfo {author} {\bibfnamefont {B.}~\bibnamefont {Pat{\'o}}}, \bibinfo
  {author} {\bibfnamefont {A.}~\bibnamefont {Petukhov}}, \bibinfo {author}
  {\bibfnamefont {N.~C.}\ \bibnamefont {Rubin}}, \bibinfo {author}
  {\bibfnamefont {D.}~\bibnamefont {Sank}}, \bibinfo {author} {\bibfnamefont
  {V.}~\bibnamefont {Shvarts}}, \bibinfo {author} {\bibfnamefont
  {D.}~\bibnamefont {Strain}}, \bibinfo {author} {\bibfnamefont
  {M.}~\bibnamefont {Szalay}}, \bibinfo {author} {\bibfnamefont
  {B.}~\bibnamefont {Villalonga}}, \bibinfo {author} {\bibfnamefont {T.~C.}\
  \bibnamefont {White}}, \bibinfo {author} {\bibfnamefont {Z.}~\bibnamefont
  {Yao}}, \bibinfo {author} {\bibfnamefont {P.}~\bibnamefont {Yeh}}, \bibinfo
  {author} {\bibfnamefont {J.}~\bibnamefont {Yoo}}, \bibinfo {author}
  {\bibfnamefont {A.}~\bibnamefont {Zalcman}}, \bibinfo {author} {\bibfnamefont
  {H.}~\bibnamefont {Neven}}, \bibinfo {author} {\bibfnamefont
  {S.}~\bibnamefont {Boixo}}, \bibinfo {author} {\bibfnamefont
  {A.}~\bibnamefont {Megrant}}, \bibinfo {author} {\bibfnamefont
  {Y.}~\bibnamefont {Chen}}, \bibinfo {author} {\bibfnamefont {J.}~\bibnamefont
  {Kelly}}, \bibinfo {author} {\bibfnamefont {V.}~\bibnamefont {Smelyanskiy}},
  \bibinfo {author} {\bibfnamefont {A.}~\bibnamefont {Kitaev}}, \bibinfo
  {author} {\bibfnamefont {M.}~\bibnamefont {Knap}}, \bibinfo {author}
  {\bibfnamefont {F.}~\bibnamefont {Pollmann}},\ and\ \bibinfo {author}
  {\bibfnamefont {P.}~\bibnamefont {Roushan}},\ }\bibfield  {title} {\bibinfo
  {title} {Realizing topologically ordered states on a quantum processor},\
  }\href {https://doi.org/10.1126/science.abi8378} {\bibfield  {journal}
  {\bibinfo  {journal} {Science}\ }\textbf {\bibinfo {volume} {374}},\ \bibinfo
  {pages} {1237} (\bibinfo {year} {2021})}\BibitemShut {NoStop}%
\bibitem [{\citenamefont {Kyprianidis}\ \emph {et~al.}(2021)\citenamefont
  {Kyprianidis}, \citenamefont {Machado}, \citenamefont {Morong}, \citenamefont
  {Becker}, \citenamefont {Collins}, \citenamefont {Else}, \citenamefont
  {Feng}, \citenamefont {Hess}, \citenamefont {Nayak}, \citenamefont {Pagano},
  \citenamefont {Yao},\ and\ \citenamefont {Monroe}}]{Kyprianidis_2021}%
  \BibitemOpen
  \bibfield  {author} {\bibinfo {author} {\bibfnamefont {A.}~\bibnamefont
  {Kyprianidis}}, \bibinfo {author} {\bibfnamefont {F.}~\bibnamefont
  {Machado}}, \bibinfo {author} {\bibfnamefont {W.}~\bibnamefont {Morong}},
  \bibinfo {author} {\bibfnamefont {P.}~\bibnamefont {Becker}}, \bibinfo
  {author} {\bibfnamefont {K.~S.}\ \bibnamefont {Collins}}, \bibinfo {author}
  {\bibfnamefont {D.~V.}\ \bibnamefont {Else}}, \bibinfo {author}
  {\bibfnamefont {L.}~\bibnamefont {Feng}}, \bibinfo {author} {\bibfnamefont
  {P.~W.}\ \bibnamefont {Hess}}, \bibinfo {author} {\bibfnamefont
  {C.}~\bibnamefont {Nayak}}, \bibinfo {author} {\bibfnamefont
  {G.}~\bibnamefont {Pagano}}, \bibinfo {author} {\bibfnamefont {N.~Y.}\
  \bibnamefont {Yao}},\ and\ \bibinfo {author} {\bibfnamefont {C.}~\bibnamefont
  {Monroe}},\ }\bibfield  {title} {\bibinfo {title} {Observation of a
  prethermal discrete time crystal},\ }\href
  {https://doi.org/10.1126/science.abg8102} {\bibfield  {journal} {\bibinfo
  {journal} {Science}\ }\textbf {\bibinfo {volume} {372}},\ \bibinfo {pages}
  {1192} (\bibinfo {year} {2021})}\BibitemShut {NoStop}%
\bibitem [{\citenamefont {Randall}\ \emph {et~al.}(2021)\citenamefont
  {Randall}, \citenamefont {Bradley}, \citenamefont {van~der Gronden},
  \citenamefont {Galicia}, \citenamefont {Abobeih}, \citenamefont {Markham},
  \citenamefont {Twitchen}, \citenamefont {Machado}, \citenamefont {Yao},\ and\
  \citenamefont {Taminiau}}]{Randall_2021}%
  \BibitemOpen
  \bibfield  {author} {\bibinfo {author} {\bibfnamefont {J.}~\bibnamefont
  {Randall}}, \bibinfo {author} {\bibfnamefont {C.~E.}\ \bibnamefont
  {Bradley}}, \bibinfo {author} {\bibfnamefont {F.~V.}\ \bibnamefont {van~der
  Gronden}}, \bibinfo {author} {\bibfnamefont {A.}~\bibnamefont {Galicia}},
  \bibinfo {author} {\bibfnamefont {M.~H.}\ \bibnamefont {Abobeih}}, \bibinfo
  {author} {\bibfnamefont {M.}~\bibnamefont {Markham}}, \bibinfo {author}
  {\bibfnamefont {D.~J.}\ \bibnamefont {Twitchen}}, \bibinfo {author}
  {\bibfnamefont {F.}~\bibnamefont {Machado}}, \bibinfo {author} {\bibfnamefont
  {N.~Y.}\ \bibnamefont {Yao}},\ and\ \bibinfo {author} {\bibfnamefont {T.~H.}\
  \bibnamefont {Taminiau}},\ }\bibfield  {title} {\bibinfo {title}
  {Many-body\&\#x2013;localized discrete time crystal with a programmable
  spin-based quantum simulator},\ }\href
  {https://doi.org/10.1126/science.abk0603} {\bibfield  {journal} {\bibinfo
  {journal} {Science}\ }\textbf {\bibinfo {volume} {374}},\ \bibinfo {pages}
  {1474} (\bibinfo {year} {2021})}\BibitemShut {NoStop}%
\bibitem [{\citenamefont {Mi}\ \emph {et~al.}(2022)\citenamefont {Mi},
  \citenamefont {Ippoliti}, \citenamefont {Quintana}, \citenamefont {Greene},
  \citenamefont {Chen}, \citenamefont {Gross}, \citenamefont {Arute},
  \citenamefont {Arya}, \citenamefont {Atalaya}, \citenamefont {Babbush},
  \citenamefont {Bardin}, \citenamefont {Basso}, \citenamefont {Bengtsson},
  \citenamefont {Bilmes}, \citenamefont {Bourassa}, \citenamefont {Brill},
  \citenamefont {Broughton}, \citenamefont {Buckley}, \citenamefont {Buell},
  \citenamefont {Burkett}, \citenamefont {Bushnell}, \citenamefont {Chiaro},
  \citenamefont {Collins}, \citenamefont {Courtney}, \citenamefont {Debroy},
  \citenamefont {Demura}, \citenamefont {Derk}, \citenamefont {Dunsworth},
  \citenamefont {Eppens}, \citenamefont {Erickson}, \citenamefont {Farhi},
  \citenamefont {Fowler}, \citenamefont {Foxen}, \citenamefont {Gidney},
  \citenamefont {Giustina}, \citenamefont {Harrigan}, \citenamefont
  {Harrington}, \citenamefont {Hilton}, \citenamefont {Ho}, \citenamefont
  {Hong}, \citenamefont {Huang}, \citenamefont {Huff}, \citenamefont {Huggins},
  \citenamefont {Ioffe}, \citenamefont {Isakov}, \citenamefont {Iveland},
  \citenamefont {Jeffrey}, \citenamefont {Jiang}, \citenamefont {Jones},
  \citenamefont {Kafri}, \citenamefont {Khattar}, \citenamefont {Kim},
  \citenamefont {Kitaev}, \citenamefont {Klimov}, \citenamefont {Korotkov},
  \citenamefont {Kostritsa}, \citenamefont {Landhuis}, \citenamefont {Laptev},
  \citenamefont {Lee}, \citenamefont {Lee}, \citenamefont {Locharla},
  \citenamefont {Lucero}, \citenamefont {Martin}, \citenamefont {McClean},
  \citenamefont {McCourt}, \citenamefont {McEwen}, \citenamefont {Miao},
  \citenamefont {Mohseni}, \citenamefont {Montazeri}, \citenamefont
  {Mruczkiewicz}, \citenamefont {Naaman}, \citenamefont {Neeley}, \citenamefont
  {Neill}, \citenamefont {Newman}, \citenamefont {Niu}, \citenamefont
  {O'Brien}, \citenamefont {Opremcak}, \citenamefont {Ostby}, \citenamefont
  {Pato}, \citenamefont {Petukhov}, \citenamefont {Rubin}, \citenamefont
  {Sank}, \citenamefont {Satzinger}, \citenamefont {Shvarts}, \citenamefont
  {Su}, \citenamefont {Strain}, \citenamefont {Szalay}, \citenamefont
  {Trevithick}, \citenamefont {Villalonga}, \citenamefont {White},
  \citenamefont {Yao}, \citenamefont {Yeh}, \citenamefont {Yoo}, \citenamefont
  {Zalcman}, \citenamefont {Neven}, \citenamefont {Boixo}, \citenamefont
  {Smelyanskiy}, \citenamefont {Megrant}, \citenamefont {Kelly}, \citenamefont
  {Chen}, \citenamefont {Sondhi}, \citenamefont {Moessner}, \citenamefont
  {Kechedzhi}, \citenamefont {Khemani},\ and\ \citenamefont
  {Roushan}}]{Mi_2022}%
  \BibitemOpen
  \bibfield  {author} {\bibinfo {author} {\bibfnamefont {X.}~\bibnamefont
  {Mi}}, \bibinfo {author} {\bibfnamefont {M.}~\bibnamefont {Ippoliti}},
  \bibinfo {author} {\bibfnamefont {C.}~\bibnamefont {Quintana}}, \bibinfo
  {author} {\bibfnamefont {A.}~\bibnamefont {Greene}}, \bibinfo {author}
  {\bibfnamefont {Z.}~\bibnamefont {Chen}}, \bibinfo {author} {\bibfnamefont
  {J.}~\bibnamefont {Gross}}, \bibinfo {author} {\bibfnamefont
  {F.}~\bibnamefont {Arute}}, \bibinfo {author} {\bibfnamefont
  {K.}~\bibnamefont {Arya}}, \bibinfo {author} {\bibfnamefont {J.}~\bibnamefont
  {Atalaya}}, \bibinfo {author} {\bibfnamefont {R.}~\bibnamefont {Babbush}},
  \bibinfo {author} {\bibfnamefont {J.~C.}\ \bibnamefont {Bardin}}, \bibinfo
  {author} {\bibfnamefont {J.}~\bibnamefont {Basso}}, \bibinfo {author}
  {\bibfnamefont {A.}~\bibnamefont {Bengtsson}}, \bibinfo {author}
  {\bibfnamefont {A.}~\bibnamefont {Bilmes}}, \bibinfo {author} {\bibfnamefont
  {A.}~\bibnamefont {Bourassa}}, \bibinfo {author} {\bibfnamefont
  {L.}~\bibnamefont {Brill}}, \bibinfo {author} {\bibfnamefont
  {M.}~\bibnamefont {Broughton}}, \bibinfo {author} {\bibfnamefont {B.~B.}\
  \bibnamefont {Buckley}}, \bibinfo {author} {\bibfnamefont {D.~A.}\
  \bibnamefont {Buell}}, \bibinfo {author} {\bibfnamefont {B.}~\bibnamefont
  {Burkett}}, \bibinfo {author} {\bibfnamefont {N.}~\bibnamefont {Bushnell}},
  \bibinfo {author} {\bibfnamefont {B.}~\bibnamefont {Chiaro}}, \bibinfo
  {author} {\bibfnamefont {R.}~\bibnamefont {Collins}}, \bibinfo {author}
  {\bibfnamefont {W.}~\bibnamefont {Courtney}}, \bibinfo {author}
  {\bibfnamefont {D.}~\bibnamefont {Debroy}}, \bibinfo {author} {\bibfnamefont
  {S.}~\bibnamefont {Demura}}, \bibinfo {author} {\bibfnamefont {A.~R.}\
  \bibnamefont {Derk}}, \bibinfo {author} {\bibfnamefont {A.}~\bibnamefont
  {Dunsworth}}, \bibinfo {author} {\bibfnamefont {D.}~\bibnamefont {Eppens}},
  \bibinfo {author} {\bibfnamefont {C.}~\bibnamefont {Erickson}}, \bibinfo
  {author} {\bibfnamefont {E.}~\bibnamefont {Farhi}}, \bibinfo {author}
  {\bibfnamefont {A.~G.}\ \bibnamefont {Fowler}}, \bibinfo {author}
  {\bibfnamefont {B.}~\bibnamefont {Foxen}}, \bibinfo {author} {\bibfnamefont
  {C.}~\bibnamefont {Gidney}}, \bibinfo {author} {\bibfnamefont
  {M.}~\bibnamefont {Giustina}}, \bibinfo {author} {\bibfnamefont {M.~P.}\
  \bibnamefont {Harrigan}}, \bibinfo {author} {\bibfnamefont {S.~D.}\
  \bibnamefont {Harrington}}, \bibinfo {author} {\bibfnamefont
  {J.}~\bibnamefont {Hilton}}, \bibinfo {author} {\bibfnamefont
  {A.}~\bibnamefont {Ho}}, \bibinfo {author} {\bibfnamefont {S.}~\bibnamefont
  {Hong}}, \bibinfo {author} {\bibfnamefont {T.}~\bibnamefont {Huang}},
  \bibinfo {author} {\bibfnamefont {A.}~\bibnamefont {Huff}}, \bibinfo {author}
  {\bibfnamefont {W.~J.}\ \bibnamefont {Huggins}}, \bibinfo {author}
  {\bibfnamefont {L.~B.}\ \bibnamefont {Ioffe}}, \bibinfo {author}
  {\bibfnamefont {S.~V.}\ \bibnamefont {Isakov}}, \bibinfo {author}
  {\bibfnamefont {J.}~\bibnamefont {Iveland}}, \bibinfo {author} {\bibfnamefont
  {E.}~\bibnamefont {Jeffrey}}, \bibinfo {author} {\bibfnamefont
  {Z.}~\bibnamefont {Jiang}}, \bibinfo {author} {\bibfnamefont
  {C.}~\bibnamefont {Jones}}, \bibinfo {author} {\bibfnamefont
  {D.}~\bibnamefont {Kafri}}, \bibinfo {author} {\bibfnamefont
  {T.}~\bibnamefont {Khattar}}, \bibinfo {author} {\bibfnamefont
  {S.}~\bibnamefont {Kim}}, \bibinfo {author} {\bibfnamefont {A.}~\bibnamefont
  {Kitaev}}, \bibinfo {author} {\bibfnamefont {P.~V.}\ \bibnamefont {Klimov}},
  \bibinfo {author} {\bibfnamefont {A.~N.}\ \bibnamefont {Korotkov}}, \bibinfo
  {author} {\bibfnamefont {F.}~\bibnamefont {Kostritsa}}, \bibinfo {author}
  {\bibfnamefont {D.}~\bibnamefont {Landhuis}}, \bibinfo {author}
  {\bibfnamefont {P.}~\bibnamefont {Laptev}}, \bibinfo {author} {\bibfnamefont
  {J.}~\bibnamefont {Lee}}, \bibinfo {author} {\bibfnamefont {K.}~\bibnamefont
  {Lee}}, \bibinfo {author} {\bibfnamefont {A.}~\bibnamefont {Locharla}},
  \bibinfo {author} {\bibfnamefont {E.}~\bibnamefont {Lucero}}, \bibinfo
  {author} {\bibfnamefont {O.}~\bibnamefont {Martin}}, \bibinfo {author}
  {\bibfnamefont {J.~R.}\ \bibnamefont {McClean}}, \bibinfo {author}
  {\bibfnamefont {T.}~\bibnamefont {McCourt}}, \bibinfo {author} {\bibfnamefont
  {M.}~\bibnamefont {McEwen}}, \bibinfo {author} {\bibfnamefont {K.~C.}\
  \bibnamefont {Miao}}, \bibinfo {author} {\bibfnamefont {M.}~\bibnamefont
  {Mohseni}}, \bibinfo {author} {\bibfnamefont {S.}~\bibnamefont {Montazeri}},
  \bibinfo {author} {\bibfnamefont {W.}~\bibnamefont {Mruczkiewicz}}, \bibinfo
  {author} {\bibfnamefont {O.}~\bibnamefont {Naaman}}, \bibinfo {author}
  {\bibfnamefont {M.}~\bibnamefont {Neeley}}, \bibinfo {author} {\bibfnamefont
  {C.}~\bibnamefont {Neill}}, \bibinfo {author} {\bibfnamefont
  {M.}~\bibnamefont {Newman}}, \bibinfo {author} {\bibfnamefont {M.~Y.}\
  \bibnamefont {Niu}}, \bibinfo {author} {\bibfnamefont {T.~E.}\ \bibnamefont
  {O'Brien}}, \bibinfo {author} {\bibfnamefont {A.}~\bibnamefont {Opremcak}},
  \bibinfo {author} {\bibfnamefont {E.}~\bibnamefont {Ostby}}, \bibinfo
  {author} {\bibfnamefont {B.}~\bibnamefont {Pato}}, \bibinfo {author}
  {\bibfnamefont {A.}~\bibnamefont {Petukhov}}, \bibinfo {author}
  {\bibfnamefont {N.~C.}\ \bibnamefont {Rubin}}, \bibinfo {author}
  {\bibfnamefont {D.}~\bibnamefont {Sank}}, \bibinfo {author} {\bibfnamefont
  {K.~J.}\ \bibnamefont {Satzinger}}, \bibinfo {author} {\bibfnamefont
  {V.}~\bibnamefont {Shvarts}}, \bibinfo {author} {\bibfnamefont
  {Y.}~\bibnamefont {Su}}, \bibinfo {author} {\bibfnamefont {D.}~\bibnamefont
  {Strain}}, \bibinfo {author} {\bibfnamefont {M.}~\bibnamefont {Szalay}},
  \bibinfo {author} {\bibfnamefont {M.~D.}\ \bibnamefont {Trevithick}},
  \bibinfo {author} {\bibfnamefont {B.}~\bibnamefont {Villalonga}}, \bibinfo
  {author} {\bibfnamefont {T.}~\bibnamefont {White}}, \bibinfo {author}
  {\bibfnamefont {Z.~J.}\ \bibnamefont {Yao}}, \bibinfo {author} {\bibfnamefont
  {P.}~\bibnamefont {Yeh}}, \bibinfo {author} {\bibfnamefont {J.}~\bibnamefont
  {Yoo}}, \bibinfo {author} {\bibfnamefont {A.}~\bibnamefont {Zalcman}},
  \bibinfo {author} {\bibfnamefont {H.}~\bibnamefont {Neven}}, \bibinfo
  {author} {\bibfnamefont {S.}~\bibnamefont {Boixo}}, \bibinfo {author}
  {\bibfnamefont {V.}~\bibnamefont {Smelyanskiy}}, \bibinfo {author}
  {\bibfnamefont {A.}~\bibnamefont {Megrant}}, \bibinfo {author} {\bibfnamefont
  {J.}~\bibnamefont {Kelly}}, \bibinfo {author} {\bibfnamefont
  {Y.}~\bibnamefont {Chen}}, \bibinfo {author} {\bibfnamefont {S.~L.}\
  \bibnamefont {Sondhi}}, \bibinfo {author} {\bibfnamefont {R.}~\bibnamefont
  {Moessner}}, \bibinfo {author} {\bibfnamefont {K.}~\bibnamefont {Kechedzhi}},
  \bibinfo {author} {\bibfnamefont {V.}~\bibnamefont {Khemani}},\ and\ \bibinfo
  {author} {\bibfnamefont {P.}~\bibnamefont {Roushan}},\ }\bibfield  {title}
  {\bibinfo {title} {Time-crystalline eigenstate order on a quantum
  processor},\ }\href {https://doi.org/10.1038/s41586-021-04257-w} {\bibfield
  {journal} {\bibinfo  {journal} {Nature}\ }\textbf {\bibinfo {volume} {601}},\
  \bibinfo {pages} {531} (\bibinfo {year} {2022})}\BibitemShut {NoStop}%
\bibitem [{\citenamefont {Zhang}\ \emph
  {et~al.}(2022{\natexlab{a}})\citenamefont {Zhang}, \citenamefont {Jiang},
  \citenamefont {Deng}, \citenamefont {Wang}, \citenamefont {Chen},
  \citenamefont {Zhang}, \citenamefont {Ren}, \citenamefont {Dong},
  \citenamefont {Xu}, \citenamefont {Gao}, \citenamefont {Jin}, \citenamefont
  {Zhu}, \citenamefont {Guo}, \citenamefont {Li}, \citenamefont {Song},
  \citenamefont {Gorshkov}, \citenamefont {Iadecola}, \citenamefont {Liu},
  \citenamefont {Gong}, \citenamefont {Wang}, \citenamefont {Deng},\ and\
  \citenamefont {Wang}}]{Zhang_2022_2}%
  \BibitemOpen
  \bibfield  {author} {\bibinfo {author} {\bibfnamefont {X.}~\bibnamefont
  {Zhang}}, \bibinfo {author} {\bibfnamefont {W.}~\bibnamefont {Jiang}},
  \bibinfo {author} {\bibfnamefont {J.}~\bibnamefont {Deng}}, \bibinfo {author}
  {\bibfnamefont {K.}~\bibnamefont {Wang}}, \bibinfo {author} {\bibfnamefont
  {J.}~\bibnamefont {Chen}}, \bibinfo {author} {\bibfnamefont {P.}~\bibnamefont
  {Zhang}}, \bibinfo {author} {\bibfnamefont {W.}~\bibnamefont {Ren}}, \bibinfo
  {author} {\bibfnamefont {H.}~\bibnamefont {Dong}}, \bibinfo {author}
  {\bibfnamefont {S.}~\bibnamefont {Xu}}, \bibinfo {author} {\bibfnamefont
  {Y.}~\bibnamefont {Gao}}, \bibinfo {author} {\bibfnamefont {F.}~\bibnamefont
  {Jin}}, \bibinfo {author} {\bibfnamefont {X.}~\bibnamefont {Zhu}}, \bibinfo
  {author} {\bibfnamefont {Q.}~\bibnamefont {Guo}}, \bibinfo {author}
  {\bibfnamefont {H.}~\bibnamefont {Li}}, \bibinfo {author} {\bibfnamefont
  {C.}~\bibnamefont {Song}}, \bibinfo {author} {\bibfnamefont {A.~V.}\
  \bibnamefont {Gorshkov}}, \bibinfo {author} {\bibfnamefont {T.}~\bibnamefont
  {Iadecola}}, \bibinfo {author} {\bibfnamefont {F.}~\bibnamefont {Liu}},
  \bibinfo {author} {\bibfnamefont {Z.-X.}\ \bibnamefont {Gong}}, \bibinfo
  {author} {\bibfnamefont {Z.}~\bibnamefont {Wang}}, \bibinfo {author}
  {\bibfnamefont {D.-L.}\ \bibnamefont {Deng}},\ and\ \bibinfo {author}
  {\bibfnamefont {H.}~\bibnamefont {Wang}},\ }\bibfield  {title} {\bibinfo
  {title} {Digital quantum simulation of floquet symmetry-protected topological
  phases},\ }\href {https://doi.org/10.1038/s41586-022-04854-3} {\bibfield
  {journal} {\bibinfo  {journal} {Nature}\ }\textbf {\bibinfo {volume} {607}},\
  \bibinfo {pages} {468} (\bibinfo {year} {2022}{\natexlab{a}})}\BibitemShut
  {NoStop}%
\bibitem [{\citenamefont {Martinez}\ \emph {et~al.}(2016)\citenamefont
  {Martinez}, \citenamefont {Muschik}, \citenamefont {Schindler}, \citenamefont
  {Nigg}, \citenamefont {Erhard}, \citenamefont {Heyl}, \citenamefont {Hauke},
  \citenamefont {Dalmonte}, \citenamefont {Monz}, \citenamefont {Zoller},\ and\
  \citenamefont {Blatt}}]{Martinez_2016}%
  \BibitemOpen
  \bibfield  {author} {\bibinfo {author} {\bibfnamefont {E.~A.}\ \bibnamefont
  {Martinez}}, \bibinfo {author} {\bibfnamefont {C.~A.}\ \bibnamefont
  {Muschik}}, \bibinfo {author} {\bibfnamefont {P.}~\bibnamefont {Schindler}},
  \bibinfo {author} {\bibfnamefont {D.}~\bibnamefont {Nigg}}, \bibinfo {author}
  {\bibfnamefont {A.}~\bibnamefont {Erhard}}, \bibinfo {author} {\bibfnamefont
  {M.}~\bibnamefont {Heyl}}, \bibinfo {author} {\bibfnamefont {P.}~\bibnamefont
  {Hauke}}, \bibinfo {author} {\bibfnamefont {M.}~\bibnamefont {Dalmonte}},
  \bibinfo {author} {\bibfnamefont {T.}~\bibnamefont {Monz}}, \bibinfo {author}
  {\bibfnamefont {P.}~\bibnamefont {Zoller}},\ and\ \bibinfo {author}
  {\bibfnamefont {R.}~\bibnamefont {Blatt}},\ }\bibfield  {title} {\bibinfo
  {title} {Real-time dynamics of lattice gauge theories with a few-qubit
  quantum computer},\ }\href {https://doi.org/10.1038/nature18318} {\bibfield
  {journal} {\bibinfo  {journal} {Nature}\ }\textbf {\bibinfo {volume} {534}},\
  \bibinfo {pages} {516} (\bibinfo {year} {2016})}\BibitemShut {NoStop}%
\bibitem [{\citenamefont {Klco}\ \emph {et~al.}(2018)\citenamefont {Klco},
  \citenamefont {Dumitrescu}, \citenamefont {McCaskey}, \citenamefont {Morris},
  \citenamefont {Pooser}, \citenamefont {Sanz}, \citenamefont {Solano},
  \citenamefont {Lougovski},\ and\ \citenamefont {Savage}}]{Klco_2018}%
  \BibitemOpen
  \bibfield  {author} {\bibinfo {author} {\bibfnamefont {N.}~\bibnamefont
  {Klco}}, \bibinfo {author} {\bibfnamefont {E.~F.}\ \bibnamefont
  {Dumitrescu}}, \bibinfo {author} {\bibfnamefont {A.~J.}\ \bibnamefont
  {McCaskey}}, \bibinfo {author} {\bibfnamefont {T.~D.}\ \bibnamefont
  {Morris}}, \bibinfo {author} {\bibfnamefont {R.~C.}\ \bibnamefont {Pooser}},
  \bibinfo {author} {\bibfnamefont {M.}~\bibnamefont {Sanz}}, \bibinfo {author}
  {\bibfnamefont {E.}~\bibnamefont {Solano}}, \bibinfo {author} {\bibfnamefont
  {P.}~\bibnamefont {Lougovski}},\ and\ \bibinfo {author} {\bibfnamefont
  {M.~J.}\ \bibnamefont {Savage}},\ }\bibfield  {title} {\bibinfo {title}
  {Quantum-classical computation of schwinger model dynamics using quantum
  computers},\ }\href {https://doi.org/10.1103/PhysRevA.98.032331} {\bibfield
  {journal} {\bibinfo  {journal} {Phys. Rev. A}\ }\textbf {\bibinfo {volume}
  {98}},\ \bibinfo {pages} {032331} (\bibinfo {year} {2018})}\BibitemShut
  {NoStop}%
\bibitem [{\citenamefont {Kokail}\ \emph {et~al.}(2019)\citenamefont {Kokail},
  \citenamefont {Maier}, \citenamefont {van Bijnen}, \citenamefont {Brydges},
  \citenamefont {Joshi}, \citenamefont {Jurcevic}, \citenamefont {Muschik},
  \citenamefont {Silvi}, \citenamefont {Blatt}, \citenamefont {Roos},\ and\
  \citenamefont {Zoller}}]{Kokail_2019}%
  \BibitemOpen
  \bibfield  {author} {\bibinfo {author} {\bibfnamefont {C.}~\bibnamefont
  {Kokail}}, \bibinfo {author} {\bibfnamefont {C.}~\bibnamefont {Maier}},
  \bibinfo {author} {\bibfnamefont {R.}~\bibnamefont {van Bijnen}}, \bibinfo
  {author} {\bibfnamefont {T.}~\bibnamefont {Brydges}}, \bibinfo {author}
  {\bibfnamefont {M.~K.}\ \bibnamefont {Joshi}}, \bibinfo {author}
  {\bibfnamefont {P.}~\bibnamefont {Jurcevic}}, \bibinfo {author}
  {\bibfnamefont {C.~A.}\ \bibnamefont {Muschik}}, \bibinfo {author}
  {\bibfnamefont {P.}~\bibnamefont {Silvi}}, \bibinfo {author} {\bibfnamefont
  {R.}~\bibnamefont {Blatt}}, \bibinfo {author} {\bibfnamefont {C.~F.}\
  \bibnamefont {Roos}},\ and\ \bibinfo {author} {\bibfnamefont
  {P.}~\bibnamefont {Zoller}},\ }\bibfield  {title} {\bibinfo {title}
  {Self-verifying variational quantum simulation of lattice models},\ }\href
  {https://doi.org/10.1038/s41586-019-1177-4} {\bibfield  {journal} {\bibinfo
  {journal} {Nature}\ }\textbf {\bibinfo {volume} {569}},\ \bibinfo {pages}
  {355} (\bibinfo {year} {2019})}\BibitemShut {NoStop}%
\bibitem [{\citenamefont {Jaksch}\ and\ \citenamefont
  {Zoller}(2005)}]{Jaksch_2005}%
  \BibitemOpen
  \bibfield  {author} {\bibinfo {author} {\bibfnamefont {D.}~\bibnamefont
  {Jaksch}}\ and\ \bibinfo {author} {\bibfnamefont {P.}~\bibnamefont
  {Zoller}},\ }\bibfield  {title} {\bibinfo {title} {The cold atom hubbard
  toolbox},\ }\href {https://doi.org/https://doi.org/10.1016/j.aop.2004.09.010}
  {\bibfield  {journal} {\bibinfo  {journal} {Annals of Physics}\ }\textbf
  {\bibinfo {volume} {315}},\ \bibinfo {pages} {52} (\bibinfo {year} {2005})},\
  \bibinfo {note} {special Issue}\BibitemShut {NoStop}%
\bibitem [{\citenamefont {Lewenstein}\ \emph {et~al.}(2007)\citenamefont
  {Lewenstein}, \citenamefont {Sanpera}, \citenamefont {Ahufinger},
  \citenamefont {Damski}, \citenamefont {Sen(De)},\ and\ \citenamefont
  {Sen}}]{Lewenstein_2007}%
  \BibitemOpen
  \bibfield  {author} {\bibinfo {author} {\bibfnamefont {M.}~\bibnamefont
  {Lewenstein}}, \bibinfo {author} {\bibfnamefont {A.}~\bibnamefont {Sanpera}},
  \bibinfo {author} {\bibfnamefont {V.}~\bibnamefont {Ahufinger}}, \bibinfo
  {author} {\bibfnamefont {B.}~\bibnamefont {Damski}}, \bibinfo {author}
  {\bibfnamefont {A.}~\bibnamefont {Sen(De)}},\ and\ \bibinfo {author}
  {\bibfnamefont {U.}~\bibnamefont {Sen}},\ }\bibfield  {title} {\bibinfo
  {title} {Ultracold atomic gases in optical lattices: mimicking condensed
  matter physics and beyond},\ }\href
  {https://doi.org/10.1080/00018730701223200} {\bibfield  {journal} {\bibinfo
  {journal} {Advances in Physics}\ }\textbf {\bibinfo {volume} {56}},\ \bibinfo
  {pages} {243} (\bibinfo {year} {2007})}\BibitemShut {NoStop}%
\bibitem [{\citenamefont {Gross}\ and\ \citenamefont
  {Bloch}(2017)}]{Gross2017}%
  \BibitemOpen
  \bibfield  {author} {\bibinfo {author} {\bibfnamefont {C.}~\bibnamefont
  {Gross}}\ and\ \bibinfo {author} {\bibfnamefont {I.}~\bibnamefont {Bloch}},\
  }\bibfield  {title} {\bibinfo {title} {Quantum simulations with ultracold
  atoms in optical lattices},\ }\href {https://doi.org/10.1126/science.aal3837}
  {\bibfield  {journal} {\bibinfo  {journal} {Science}\ }\textbf {\bibinfo
  {volume} {357}},\ \bibinfo {pages} {995} (\bibinfo {year}
  {2017})}\BibitemShut {NoStop}%
\bibitem [{\citenamefont {Wilk}\ \emph {et~al.}(2010)\citenamefont {Wilk},
  \citenamefont {Ga\"etan}, \citenamefont {Evellin}, \citenamefont {Wolters},
  \citenamefont {Miroshnychenko}, \citenamefont {Grangier},\ and\ \citenamefont
  {Browaeys}}]{Wilk_2010}%
  \BibitemOpen
  \bibfield  {author} {\bibinfo {author} {\bibfnamefont {T.}~\bibnamefont
  {Wilk}}, \bibinfo {author} {\bibfnamefont {A.}~\bibnamefont {Ga\"etan}},
  \bibinfo {author} {\bibfnamefont {C.}~\bibnamefont {Evellin}}, \bibinfo
  {author} {\bibfnamefont {J.}~\bibnamefont {Wolters}}, \bibinfo {author}
  {\bibfnamefont {Y.}~\bibnamefont {Miroshnychenko}}, \bibinfo {author}
  {\bibfnamefont {P.}~\bibnamefont {Grangier}},\ and\ \bibinfo {author}
  {\bibfnamefont {A.}~\bibnamefont {Browaeys}},\ }\bibfield  {title} {\bibinfo
  {title} {Entanglement of two individual neutral atoms using rydberg
  blockade},\ }\href {https://doi.org/10.1103/PhysRevLett.104.010502}
  {\bibfield  {journal} {\bibinfo  {journal} {Phys. Rev. Lett.}\ }\textbf
  {\bibinfo {volume} {104}},\ \bibinfo {pages} {010502} (\bibinfo {year}
  {2010})}\BibitemShut {NoStop}%
\bibitem [{\citenamefont {Saffman}\ \emph {et~al.}(2010)\citenamefont
  {Saffman}, \citenamefont {Walker},\ and\ \citenamefont
  {M\o{}lmer}}]{Saffman_2010}%
  \BibitemOpen
  \bibfield  {author} {\bibinfo {author} {\bibfnamefont {M.}~\bibnamefont
  {Saffman}}, \bibinfo {author} {\bibfnamefont {T.~G.}\ \bibnamefont
  {Walker}},\ and\ \bibinfo {author} {\bibfnamefont {K.}~\bibnamefont
  {M\o{}lmer}},\ }\bibfield  {title} {\bibinfo {title} {Quantum information
  with rydberg atoms},\ }\href {https://doi.org/10.1103/RevModPhys.82.2313}
  {\bibfield  {journal} {\bibinfo  {journal} {Rev. Mod. Phys.}\ }\textbf
  {\bibinfo {volume} {82}},\ \bibinfo {pages} {2313} (\bibinfo {year}
  {2010})}\BibitemShut {NoStop}%
\bibitem [{\citenamefont {Saffman}(2016)}]{Saffman_2016}%
  \BibitemOpen
  \bibfield  {author} {\bibinfo {author} {\bibfnamefont {M.}~\bibnamefont
  {Saffman}},\ }\bibfield  {title} {\bibinfo {title} {Quantum computing with
  atomic qubits and rydberg interactions: progress and challenges},\ }\href
  {https://doi.org/10.1088/0953-4075/49/20/202001} {\bibfield  {journal}
  {\bibinfo  {journal} {Journal of Physics B: Atomic, Molecular and Optical
  Physics}\ }\textbf {\bibinfo {volume} {49}},\ \bibinfo {pages} {202001}
  (\bibinfo {year} {2016})}\BibitemShut {NoStop}%
\bibitem [{\citenamefont {Levine}\ \emph {et~al.}(2019)\citenamefont {Levine},
  \citenamefont {Keesling}, \citenamefont {Semeghini}, \citenamefont {Omran},
  \citenamefont {Wang}, \citenamefont {Ebadi}, \citenamefont {Bernien},
  \citenamefont {Greiner}, \citenamefont {Vuleti\ifmmode~\acute{c}\else
  \'{c}\fi{}}, \citenamefont {Pichler},\ and\ \citenamefont
  {Lukin}}]{Levine_2019}%
  \BibitemOpen
  \bibfield  {author} {\bibinfo {author} {\bibfnamefont {H.}~\bibnamefont
  {Levine}}, \bibinfo {author} {\bibfnamefont {A.}~\bibnamefont {Keesling}},
  \bibinfo {author} {\bibfnamefont {G.}~\bibnamefont {Semeghini}}, \bibinfo
  {author} {\bibfnamefont {A.}~\bibnamefont {Omran}}, \bibinfo {author}
  {\bibfnamefont {T.~T.}\ \bibnamefont {Wang}}, \bibinfo {author}
  {\bibfnamefont {S.}~\bibnamefont {Ebadi}}, \bibinfo {author} {\bibfnamefont
  {H.}~\bibnamefont {Bernien}}, \bibinfo {author} {\bibfnamefont
  {M.}~\bibnamefont {Greiner}}, \bibinfo {author} {\bibfnamefont
  {V.}~\bibnamefont {Vuleti\ifmmode~\acute{c}\else \'{c}\fi{}}}, \bibinfo
  {author} {\bibfnamefont {H.}~\bibnamefont {Pichler}},\ and\ \bibinfo {author}
  {\bibfnamefont {M.~D.}\ \bibnamefont {Lukin}},\ }\bibfield  {title} {\bibinfo
  {title} {Parallel implementation of high-fidelity multiqubit gates with
  neutral atoms},\ }\href {https://doi.org/10.1103/PhysRevLett.123.170503}
  {\bibfield  {journal} {\bibinfo  {journal} {Phys. Rev. Lett.}\ }\textbf
  {\bibinfo {volume} {123}},\ \bibinfo {pages} {170503} (\bibinfo {year}
  {2019})}\BibitemShut {NoStop}%
\bibitem [{\citenamefont {Graham}\ \emph {et~al.}(2019)\citenamefont {Graham},
  \citenamefont {Kwon}, \citenamefont {Grinkemeyer}, \citenamefont {Marra},
  \citenamefont {Jiang}, \citenamefont {Lichtman}, \citenamefont {Sun},
  \citenamefont {Ebert},\ and\ \citenamefont {Saffman}}]{Graham_2019}%
  \BibitemOpen
  \bibfield  {author} {\bibinfo {author} {\bibfnamefont {T.~M.}\ \bibnamefont
  {Graham}}, \bibinfo {author} {\bibfnamefont {M.}~\bibnamefont {Kwon}},
  \bibinfo {author} {\bibfnamefont {B.}~\bibnamefont {Grinkemeyer}}, \bibinfo
  {author} {\bibfnamefont {Z.}~\bibnamefont {Marra}}, \bibinfo {author}
  {\bibfnamefont {X.}~\bibnamefont {Jiang}}, \bibinfo {author} {\bibfnamefont
  {M.~T.}\ \bibnamefont {Lichtman}}, \bibinfo {author} {\bibfnamefont
  {Y.}~\bibnamefont {Sun}}, \bibinfo {author} {\bibfnamefont {M.}~\bibnamefont
  {Ebert}},\ and\ \bibinfo {author} {\bibfnamefont {M.}~\bibnamefont
  {Saffman}},\ }\bibfield  {title} {\bibinfo {title} {Rydberg-mediated
  entanglement in a two-dimensional neutral atom qubit array},\ }\href
  {https://doi.org/10.1103/PhysRevLett.123.230501} {\bibfield  {journal}
  {\bibinfo  {journal} {Phys. Rev. Lett.}\ }\textbf {\bibinfo {volume} {123}},\
  \bibinfo {pages} {230501} (\bibinfo {year} {2019})}\BibitemShut {NoStop}%
\bibitem [{\citenamefont {Madjarov}\ \emph {et~al.}(2020)\citenamefont
  {Madjarov}, \citenamefont {Covey}, \citenamefont {Shaw}, \citenamefont
  {Choi}, \citenamefont {Kale}, \citenamefont {Cooper}, \citenamefont
  {Pichler}, \citenamefont {Schkolnik}, \citenamefont {Williams},\ and\
  \citenamefont {Endres}}]{Madjarov_2020}%
  \BibitemOpen
  \bibfield  {author} {\bibinfo {author} {\bibfnamefont {I.~S.}\ \bibnamefont
  {Madjarov}}, \bibinfo {author} {\bibfnamefont {J.~P.}\ \bibnamefont {Covey}},
  \bibinfo {author} {\bibfnamefont {A.~L.}\ \bibnamefont {Shaw}}, \bibinfo
  {author} {\bibfnamefont {J.}~\bibnamefont {Choi}}, \bibinfo {author}
  {\bibfnamefont {A.}~\bibnamefont {Kale}}, \bibinfo {author} {\bibfnamefont
  {A.}~\bibnamefont {Cooper}}, \bibinfo {author} {\bibfnamefont
  {H.}~\bibnamefont {Pichler}}, \bibinfo {author} {\bibfnamefont
  {V.}~\bibnamefont {Schkolnik}}, \bibinfo {author} {\bibfnamefont {J.~R.}\
  \bibnamefont {Williams}},\ and\ \bibinfo {author} {\bibfnamefont
  {M.}~\bibnamefont {Endres}},\ }\bibfield  {title} {\bibinfo {title}
  {High-fidelity entanglement and detection of alkaline-earth rydberg atoms},\
  }\href {https://doi.org/10.1038/s41567-020-0903-z} {\bibfield  {journal}
  {\bibinfo  {journal} {Nature Physics}\ }\textbf {\bibinfo {volume} {16}},\
  \bibinfo {pages} {857} (\bibinfo {year} {2020})}\BibitemShut {NoStop}%
\bibitem [{\citenamefont {Henriet}\ \emph {et~al.}(2020)\citenamefont
  {Henriet}, \citenamefont {Beguin}, \citenamefont {Signoles}, \citenamefont
  {Lahaye}, \citenamefont {Browaeys}, \citenamefont {Reymond},\ and\
  \citenamefont {Jurczak}}]{Henriet_2020}%
  \BibitemOpen
  \bibfield  {author} {\bibinfo {author} {\bibfnamefont {L.}~\bibnamefont
  {Henriet}}, \bibinfo {author} {\bibfnamefont {L.}~\bibnamefont {Beguin}},
  \bibinfo {author} {\bibfnamefont {A.}~\bibnamefont {Signoles}}, \bibinfo
  {author} {\bibfnamefont {T.}~\bibnamefont {Lahaye}}, \bibinfo {author}
  {\bibfnamefont {A.}~\bibnamefont {Browaeys}}, \bibinfo {author}
  {\bibfnamefont {G.-O.}\ \bibnamefont {Reymond}},\ and\ \bibinfo {author}
  {\bibfnamefont {C.}~\bibnamefont {Jurczak}},\ }\bibfield  {title} {\bibinfo
  {title} {Quantum computing with neutral atoms},\ }\href
  {https://doi.org/10.22331/q-2020-09-21-327} {\bibfield  {journal} {\bibinfo
  {journal} {{Quantum}}\ }\textbf {\bibinfo {volume} {4}},\ \bibinfo {pages}
  {327} (\bibinfo {year} {2020})}\BibitemShut {NoStop}%
\bibitem [{\citenamefont {Bluvstein}\ \emph {et~al.}(2022)\citenamefont
  {Bluvstein}, \citenamefont {Levine}, \citenamefont {Semeghini}, \citenamefont
  {Wang}, \citenamefont {Ebadi}, \citenamefont {Kalinowski}, \citenamefont
  {Keesling}, \citenamefont {Maskara}, \citenamefont {Pichler}, \citenamefont
  {Greiner}, \citenamefont {Vuleti{\'c}},\ and\ \citenamefont
  {Lukin}}]{Bluvstein_2021}%
  \BibitemOpen
  \bibfield  {author} {\bibinfo {author} {\bibfnamefont {D.}~\bibnamefont
  {Bluvstein}}, \bibinfo {author} {\bibfnamefont {H.}~\bibnamefont {Levine}},
  \bibinfo {author} {\bibfnamefont {G.}~\bibnamefont {Semeghini}}, \bibinfo
  {author} {\bibfnamefont {T.~T.}\ \bibnamefont {Wang}}, \bibinfo {author}
  {\bibfnamefont {S.}~\bibnamefont {Ebadi}}, \bibinfo {author} {\bibfnamefont
  {M.}~\bibnamefont {Kalinowski}}, \bibinfo {author} {\bibfnamefont
  {A.}~\bibnamefont {Keesling}}, \bibinfo {author} {\bibfnamefont
  {N.}~\bibnamefont {Maskara}}, \bibinfo {author} {\bibfnamefont
  {H.}~\bibnamefont {Pichler}}, \bibinfo {author} {\bibfnamefont
  {M.}~\bibnamefont {Greiner}}, \bibinfo {author} {\bibfnamefont
  {V.}~\bibnamefont {Vuleti{\'c}}},\ and\ \bibinfo {author} {\bibfnamefont
  {M.~D.}\ \bibnamefont {Lukin}},\ }\bibfield  {title} {\bibinfo {title} {A
  quantum processor based on coherent transport of entangled atom arrays},\
  }\href {https://doi.org/10.1038/s41586-022-04592-6} {\bibfield  {journal}
  {\bibinfo  {journal} {Nature}\ }\textbf {\bibinfo {volume} {604}},\ \bibinfo
  {pages} {451} (\bibinfo {year} {2022})}\BibitemShut {NoStop}%
\bibitem [{\citenamefont {Morgado}\ and\ \citenamefont
  {Whitlock}(2021)}]{morgado2021quantum}%
  \BibitemOpen
  \bibfield  {author} {\bibinfo {author} {\bibfnamefont {M.}~\bibnamefont
  {Morgado}}\ and\ \bibinfo {author} {\bibfnamefont {S.}~\bibnamefont
  {Whitlock}},\ }\bibfield  {title} {\bibinfo {title} {Quantum simulation and
  computing with rydberg-interacting qubits},\ }\href@noop {} {\bibfield
  {journal} {\bibinfo  {journal} {AVS Quantum Science}\ }\textbf {\bibinfo
  {volume} {3}},\ \bibinfo {pages} {023501} (\bibinfo {year}
  {2021})}\BibitemShut {NoStop}%
\bibitem [{\citenamefont {Labuhn}\ \emph {et~al.}(2016)\citenamefont {Labuhn},
  \citenamefont {Barredo}, \citenamefont {Ravets}, \citenamefont
  {de~L{\'e}s{\'e}leuc}, \citenamefont {Macr{\`\i}}, \citenamefont {Lahaye},\
  and\ \citenamefont {Browaeys}}]{Labuhn_2016}%
  \BibitemOpen
  \bibfield  {author} {\bibinfo {author} {\bibfnamefont {H.}~\bibnamefont
  {Labuhn}}, \bibinfo {author} {\bibfnamefont {D.}~\bibnamefont {Barredo}},
  \bibinfo {author} {\bibfnamefont {S.}~\bibnamefont {Ravets}}, \bibinfo
  {author} {\bibfnamefont {S.}~\bibnamefont {de~L{\'e}s{\'e}leuc}}, \bibinfo
  {author} {\bibfnamefont {T.}~\bibnamefont {Macr{\`\i}}}, \bibinfo {author}
  {\bibfnamefont {T.}~\bibnamefont {Lahaye}},\ and\ \bibinfo {author}
  {\bibfnamefont {A.}~\bibnamefont {Browaeys}},\ }\bibfield  {title} {\bibinfo
  {title} {Tunable two-dimensional arrays of single rydberg atoms for realizing
  quantum ising models},\ }\href {https://doi.org/10.1038/nature18274}
  {\bibfield  {journal} {\bibinfo  {journal} {Nature}\ }\textbf {\bibinfo
  {volume} {534}},\ \bibinfo {pages} {667} (\bibinfo {year}
  {2016})}\BibitemShut {NoStop}%
\bibitem [{\citenamefont {Bernien}\ \emph {et~al.}(2017)\citenamefont
  {Bernien}, \citenamefont {Schwartz}, \citenamefont {Keesling}, \citenamefont
  {Levine}, \citenamefont {Omran}, \citenamefont {Pichler}, \citenamefont
  {Choi}, \citenamefont {Zibrov}, \citenamefont {Endres}, \citenamefont
  {Greiner}, \citenamefont {Vuleti{\'c}},\ and\ \citenamefont
  {Lukin}}]{Bernien_2017}%
  \BibitemOpen
  \bibfield  {author} {\bibinfo {author} {\bibfnamefont {H.}~\bibnamefont
  {Bernien}}, \bibinfo {author} {\bibfnamefont {S.}~\bibnamefont {Schwartz}},
  \bibinfo {author} {\bibfnamefont {A.}~\bibnamefont {Keesling}}, \bibinfo
  {author} {\bibfnamefont {H.}~\bibnamefont {Levine}}, \bibinfo {author}
  {\bibfnamefont {A.}~\bibnamefont {Omran}}, \bibinfo {author} {\bibfnamefont
  {H.}~\bibnamefont {Pichler}}, \bibinfo {author} {\bibfnamefont
  {S.}~\bibnamefont {Choi}}, \bibinfo {author} {\bibfnamefont {A.~S.}\
  \bibnamefont {Zibrov}}, \bibinfo {author} {\bibfnamefont {M.}~\bibnamefont
  {Endres}}, \bibinfo {author} {\bibfnamefont {M.}~\bibnamefont {Greiner}},
  \bibinfo {author} {\bibfnamefont {V.}~\bibnamefont {Vuleti{\'c}}},\ and\
  \bibinfo {author} {\bibfnamefont {M.~D.}\ \bibnamefont {Lukin}},\ }\bibfield
  {title} {\bibinfo {title} {Probing many-body dynamics on a 51-atom quantum
  simulator},\ }\href {https://doi.org/10.1038/nature24622} {\bibfield
  {journal} {\bibinfo  {journal} {Nature}\ }\textbf {\bibinfo {volume} {551}},\
  \bibinfo {pages} {579} (\bibinfo {year} {2017})}\BibitemShut {NoStop}%
\bibitem [{\citenamefont {Keesling}\ \emph {et~al.}(2019)\citenamefont
  {Keesling}, \citenamefont {Omran}, \citenamefont {Levine}, \citenamefont
  {Bernien}, \citenamefont {Pichler}, \citenamefont {Choi}, \citenamefont
  {Samajdar}, \citenamefont {Schwartz}, \citenamefont {Silvi}, \citenamefont
  {Sachdev}, \citenamefont {Zoller}, \citenamefont {Endres}, \citenamefont
  {Greiner}, \citenamefont {Vuleti{\'c}},\ and\ \citenamefont
  {Lukin}}]{Keesling_2019}%
  \BibitemOpen
  \bibfield  {author} {\bibinfo {author} {\bibfnamefont {A.}~\bibnamefont
  {Keesling}}, \bibinfo {author} {\bibfnamefont {A.}~\bibnamefont {Omran}},
  \bibinfo {author} {\bibfnamefont {H.}~\bibnamefont {Levine}}, \bibinfo
  {author} {\bibfnamefont {H.}~\bibnamefont {Bernien}}, \bibinfo {author}
  {\bibfnamefont {H.}~\bibnamefont {Pichler}}, \bibinfo {author} {\bibfnamefont
  {S.}~\bibnamefont {Choi}}, \bibinfo {author} {\bibfnamefont {R.}~\bibnamefont
  {Samajdar}}, \bibinfo {author} {\bibfnamefont {S.}~\bibnamefont {Schwartz}},
  \bibinfo {author} {\bibfnamefont {P.}~\bibnamefont {Silvi}}, \bibinfo
  {author} {\bibfnamefont {S.}~\bibnamefont {Sachdev}}, \bibinfo {author}
  {\bibfnamefont {P.}~\bibnamefont {Zoller}}, \bibinfo {author} {\bibfnamefont
  {M.}~\bibnamefont {Endres}}, \bibinfo {author} {\bibfnamefont
  {M.}~\bibnamefont {Greiner}}, \bibinfo {author} {\bibfnamefont
  {V.}~\bibnamefont {Vuleti{\'c}}},\ and\ \bibinfo {author} {\bibfnamefont
  {M.~D.}\ \bibnamefont {Lukin}},\ }\bibfield  {title} {\bibinfo {title}
  {Quantum kibble--zurek mechanism and critical dynamics on a programmable
  rydberg simulator},\ }\href {https://doi.org/10.1038/s41586-019-1070-1}
  {\bibfield  {journal} {\bibinfo  {journal} {Nature}\ }\textbf {\bibinfo
  {volume} {568}},\ \bibinfo {pages} {207} (\bibinfo {year}
  {2019})}\BibitemShut {NoStop}%
\bibitem [{\citenamefont {de~L{\'e}s{\'e}leuc}\ \emph
  {et~al.}(2019)\citenamefont {de~L{\'e}s{\'e}leuc}, \citenamefont {Lienhard},
  \citenamefont {Scholl}, \citenamefont {Barredo}, \citenamefont {Weber},
  \citenamefont {Lang}, \citenamefont {B{\"u}chler}, \citenamefont {Lahaye},\
  and\ \citenamefont {Browaeys}}]{Leseleuc_2019}%
  \BibitemOpen
  \bibfield  {author} {\bibinfo {author} {\bibfnamefont {S.}~\bibnamefont
  {de~L{\'e}s{\'e}leuc}}, \bibinfo {author} {\bibfnamefont {V.}~\bibnamefont
  {Lienhard}}, \bibinfo {author} {\bibfnamefont {P.}~\bibnamefont {Scholl}},
  \bibinfo {author} {\bibfnamefont {D.}~\bibnamefont {Barredo}}, \bibinfo
  {author} {\bibfnamefont {S.}~\bibnamefont {Weber}}, \bibinfo {author}
  {\bibfnamefont {N.}~\bibnamefont {Lang}}, \bibinfo {author} {\bibfnamefont
  {H.~P.}\ \bibnamefont {B{\"u}chler}}, \bibinfo {author} {\bibfnamefont
  {T.}~\bibnamefont {Lahaye}},\ and\ \bibinfo {author} {\bibfnamefont
  {A.}~\bibnamefont {Browaeys}},\ }\bibfield  {title} {\bibinfo {title}
  {Observation of a symmetry-protected topological phase of interacting bosons
  with rydberg atoms},\ }\href {https://doi.org/10.1126/science.aav9105}
  {\bibfield  {journal} {\bibinfo  {journal} {Science}\ }\textbf {\bibinfo
  {volume} {365}},\ \bibinfo {pages} {775} (\bibinfo {year}
  {2019})}\BibitemShut {NoStop}%
\bibitem [{\citenamefont {Ebadi}\ \emph {et~al.}(2021)\citenamefont {Ebadi},
  \citenamefont {Wang}, \citenamefont {Levine}, \citenamefont {Keesling},
  \citenamefont {Semeghini}, \citenamefont {Omran}, \citenamefont {Bluvstein},
  \citenamefont {Samajdar}, \citenamefont {Pichler}, \citenamefont {Ho},
  \citenamefont {Choi}, \citenamefont {Sachdev}, \citenamefont {Greiner},
  \citenamefont {Vuleti{\'c}},\ and\ \citenamefont {Lukin}}]{Ebadi_2021}%
  \BibitemOpen
  \bibfield  {author} {\bibinfo {author} {\bibfnamefont {S.}~\bibnamefont
  {Ebadi}}, \bibinfo {author} {\bibfnamefont {T.~T.}\ \bibnamefont {Wang}},
  \bibinfo {author} {\bibfnamefont {H.}~\bibnamefont {Levine}}, \bibinfo
  {author} {\bibfnamefont {A.}~\bibnamefont {Keesling}}, \bibinfo {author}
  {\bibfnamefont {G.}~\bibnamefont {Semeghini}}, \bibinfo {author}
  {\bibfnamefont {A.}~\bibnamefont {Omran}}, \bibinfo {author} {\bibfnamefont
  {D.}~\bibnamefont {Bluvstein}}, \bibinfo {author} {\bibfnamefont
  {R.}~\bibnamefont {Samajdar}}, \bibinfo {author} {\bibfnamefont
  {H.}~\bibnamefont {Pichler}}, \bibinfo {author} {\bibfnamefont {W.~W.}\
  \bibnamefont {Ho}}, \bibinfo {author} {\bibfnamefont {S.}~\bibnamefont
  {Choi}}, \bibinfo {author} {\bibfnamefont {S.}~\bibnamefont {Sachdev}},
  \bibinfo {author} {\bibfnamefont {M.}~\bibnamefont {Greiner}}, \bibinfo
  {author} {\bibfnamefont {V.}~\bibnamefont {Vuleti{\'c}}},\ and\ \bibinfo
  {author} {\bibfnamefont {M.~D.}\ \bibnamefont {Lukin}},\ }\bibfield  {title}
  {\bibinfo {title} {Quantum phases of matter on a 256-atom programmable
  quantum simulator},\ }\href {https://doi.org/10.1038/s41586-021-03582-4}
  {\bibfield  {journal} {\bibinfo  {journal} {Nature}\ }\textbf {\bibinfo
  {volume} {595}},\ \bibinfo {pages} {227} (\bibinfo {year}
  {2021})}\BibitemShut {NoStop}%
\bibitem [{\citenamefont {Scholl}\ \emph {et~al.}(2021)\citenamefont {Scholl},
  \citenamefont {Schuler}, \citenamefont {Williams}, \citenamefont
  {Eberharter}, \citenamefont {Barredo}, \citenamefont {Schymik}, \citenamefont
  {Lienhard}, \citenamefont {Henry}, \citenamefont {Lang}, \citenamefont
  {Lahaye}, \citenamefont {L{\"a}uchli},\ and\ \citenamefont
  {Browaeys}}]{Scholl_2021}%
  \BibitemOpen
  \bibfield  {author} {\bibinfo {author} {\bibfnamefont {P.}~\bibnamefont
  {Scholl}}, \bibinfo {author} {\bibfnamefont {M.}~\bibnamefont {Schuler}},
  \bibinfo {author} {\bibfnamefont {H.~J.}\ \bibnamefont {Williams}}, \bibinfo
  {author} {\bibfnamefont {A.~A.}\ \bibnamefont {Eberharter}}, \bibinfo
  {author} {\bibfnamefont {D.}~\bibnamefont {Barredo}}, \bibinfo {author}
  {\bibfnamefont {K.-N.}\ \bibnamefont {Schymik}}, \bibinfo {author}
  {\bibfnamefont {V.}~\bibnamefont {Lienhard}}, \bibinfo {author}
  {\bibfnamefont {L.-P.}\ \bibnamefont {Henry}}, \bibinfo {author}
  {\bibfnamefont {T.~C.}\ \bibnamefont {Lang}}, \bibinfo {author}
  {\bibfnamefont {T.}~\bibnamefont {Lahaye}}, \bibinfo {author} {\bibfnamefont
  {A.~M.}\ \bibnamefont {L{\"a}uchli}},\ and\ \bibinfo {author} {\bibfnamefont
  {A.}~\bibnamefont {Browaeys}},\ }\bibfield  {title} {\bibinfo {title}
  {Quantum simulation of 2d antiferromagnets with hundreds of rydberg atoms},\
  }\href {https://doi.org/10.1038/s41586-021-03585-1} {\bibfield  {journal}
  {\bibinfo  {journal} {Nature}\ }\textbf {\bibinfo {volume} {595}},\ \bibinfo
  {pages} {233} (\bibinfo {year} {2021})}\BibitemShut {NoStop}%
\bibitem [{\citenamefont {Semeghini}\ \emph {et~al.}(2021)\citenamefont
  {Semeghini}, \citenamefont {Levine}, \citenamefont {Keesling}, \citenamefont
  {Ebadi}, \citenamefont {Wang}, \citenamefont {Bluvstein}, \citenamefont
  {Verresen}, \citenamefont {Pichler}, \citenamefont {Kalinowski},
  \citenamefont {Samajdar}, \citenamefont {Omran}, \citenamefont {Sachdev},
  \citenamefont {Vishwanath}, \citenamefont {Greiner}, \citenamefont
  {Vuleti{\'c}},\ and\ \citenamefont {Lukin}}]{Semeghini_2021}%
  \BibitemOpen
  \bibfield  {author} {\bibinfo {author} {\bibfnamefont {G.}~\bibnamefont
  {Semeghini}}, \bibinfo {author} {\bibfnamefont {H.}~\bibnamefont {Levine}},
  \bibinfo {author} {\bibfnamefont {A.}~\bibnamefont {Keesling}}, \bibinfo
  {author} {\bibfnamefont {S.}~\bibnamefont {Ebadi}}, \bibinfo {author}
  {\bibfnamefont {T.~T.}\ \bibnamefont {Wang}}, \bibinfo {author}
  {\bibfnamefont {D.}~\bibnamefont {Bluvstein}}, \bibinfo {author}
  {\bibfnamefont {R.}~\bibnamefont {Verresen}}, \bibinfo {author}
  {\bibfnamefont {H.}~\bibnamefont {Pichler}}, \bibinfo {author} {\bibfnamefont
  {M.}~\bibnamefont {Kalinowski}}, \bibinfo {author} {\bibfnamefont
  {R.}~\bibnamefont {Samajdar}}, \bibinfo {author} {\bibfnamefont
  {A.}~\bibnamefont {Omran}}, \bibinfo {author} {\bibfnamefont
  {S.}~\bibnamefont {Sachdev}}, \bibinfo {author} {\bibfnamefont
  {A.}~\bibnamefont {Vishwanath}}, \bibinfo {author} {\bibfnamefont
  {M.}~\bibnamefont {Greiner}}, \bibinfo {author} {\bibfnamefont
  {V.}~\bibnamefont {Vuleti{\'c}}},\ and\ \bibinfo {author} {\bibfnamefont
  {M.~D.}\ \bibnamefont {Lukin}},\ }\bibfield  {title} {\bibinfo {title}
  {Probing topological spin liquids on a programmable quantum simulator},\
  }\href {https://doi.org/10.1126/science.abi8794} {\bibfield  {journal}
  {\bibinfo  {journal} {Science}\ }\textbf {\bibinfo {volume} {374}},\ \bibinfo
  {pages} {1242} (\bibinfo {year} {2021})}\BibitemShut {NoStop}%
\bibitem [{\citenamefont {Scholl}\ \emph {et~al.}(2022)\citenamefont {Scholl},
  \citenamefont {Williams}, \citenamefont {Bornet}, \citenamefont {Wallner},
  \citenamefont {Barredo}, \citenamefont {Henriet}, \citenamefont {Signoles},
  \citenamefont {Hainaut}, \citenamefont {Franz}, \citenamefont {Geier},
  \citenamefont {Tebben}, \citenamefont {Salzinger}, \citenamefont {Z\"urn},
  \citenamefont {Lahaye}, \citenamefont {Weidem\"uller},\ and\ \citenamefont
  {Browaeys}}]{Scholl_2022}%
  \BibitemOpen
  \bibfield  {author} {\bibinfo {author} {\bibfnamefont {P.}~\bibnamefont
  {Scholl}}, \bibinfo {author} {\bibfnamefont {H.~J.}\ \bibnamefont
  {Williams}}, \bibinfo {author} {\bibfnamefont {G.}~\bibnamefont {Bornet}},
  \bibinfo {author} {\bibfnamefont {F.}~\bibnamefont {Wallner}}, \bibinfo
  {author} {\bibfnamefont {D.}~\bibnamefont {Barredo}}, \bibinfo {author}
  {\bibfnamefont {L.}~\bibnamefont {Henriet}}, \bibinfo {author} {\bibfnamefont
  {A.}~\bibnamefont {Signoles}}, \bibinfo {author} {\bibfnamefont
  {C.}~\bibnamefont {Hainaut}}, \bibinfo {author} {\bibfnamefont
  {T.}~\bibnamefont {Franz}}, \bibinfo {author} {\bibfnamefont
  {S.}~\bibnamefont {Geier}}, \bibinfo {author} {\bibfnamefont
  {A.}~\bibnamefont {Tebben}}, \bibinfo {author} {\bibfnamefont
  {A.}~\bibnamefont {Salzinger}}, \bibinfo {author} {\bibfnamefont
  {G.}~\bibnamefont {Z\"urn}}, \bibinfo {author} {\bibfnamefont
  {T.}~\bibnamefont {Lahaye}}, \bibinfo {author} {\bibfnamefont
  {M.}~\bibnamefont {Weidem\"uller}},\ and\ \bibinfo {author} {\bibfnamefont
  {A.}~\bibnamefont {Browaeys}},\ }\bibfield  {title} {\bibinfo {title}
  {Microwave engineering of programmable $xxz$ hamiltonians in arrays of
  rydberg atoms},\ }\href {https://doi.org/10.1103/PRXQuantum.3.020303}
  {\bibfield  {journal} {\bibinfo  {journal} {PRX Quantum}\ }\textbf {\bibinfo
  {volume} {3}},\ \bibinfo {pages} {020303} (\bibinfo {year}
  {2022})}\BibitemShut {NoStop}%
\bibitem [{\citenamefont {Kaufman}\ \emph {et~al.}(2014)\citenamefont
  {Kaufman}, \citenamefont {Lester}, \citenamefont {Reynolds}, \citenamefont
  {Wall}, \citenamefont {Foss-Feig}, \citenamefont {Hazzard}, \citenamefont
  {Rey},\ and\ \citenamefont {Regal}}]{Kaufman2014}%
  \BibitemOpen
  \bibfield  {author} {\bibinfo {author} {\bibfnamefont {A.~M.}\ \bibnamefont
  {Kaufman}}, \bibinfo {author} {\bibfnamefont {B.~J.}\ \bibnamefont {Lester}},
  \bibinfo {author} {\bibfnamefont {C.~M.}\ \bibnamefont {Reynolds}}, \bibinfo
  {author} {\bibfnamefont {M.~L.}\ \bibnamefont {Wall}}, \bibinfo {author}
  {\bibfnamefont {M.}~\bibnamefont {Foss-Feig}}, \bibinfo {author}
  {\bibfnamefont {K.~R.~A.}\ \bibnamefont {Hazzard}}, \bibinfo {author}
  {\bibfnamefont {A.~M.}\ \bibnamefont {Rey}},\ and\ \bibinfo {author}
  {\bibfnamefont {C.~A.}\ \bibnamefont {Regal}},\ }\bibfield  {title} {\bibinfo
  {title} {Two-particle quantum interference in tunnel-coupled optical
  tweezers},\ }\href {https://doi.org/10.1126/science.1250057} {\bibfield
  {journal} {\bibinfo  {journal} {Science}\ }\textbf {\bibinfo {volume}
  {345}},\ \bibinfo {pages} {306} (\bibinfo {year} {2014})}\BibitemShut
  {NoStop}%
\bibitem [{\citenamefont {Murmann}\ \emph {et~al.}(2015)\citenamefont
  {Murmann}, \citenamefont {Bergschneider}, \citenamefont {Klinkhamer},
  \citenamefont {Z\"urn}, \citenamefont {Lompe},\ and\ \citenamefont
  {Jochim}}]{Murmann_2015}%
  \BibitemOpen
  \bibfield  {author} {\bibinfo {author} {\bibfnamefont {S.}~\bibnamefont
  {Murmann}}, \bibinfo {author} {\bibfnamefont {A.}~\bibnamefont
  {Bergschneider}}, \bibinfo {author} {\bibfnamefont {V.~M.}\ \bibnamefont
  {Klinkhamer}}, \bibinfo {author} {\bibfnamefont {G.}~\bibnamefont {Z\"urn}},
  \bibinfo {author} {\bibfnamefont {T.}~\bibnamefont {Lompe}},\ and\ \bibinfo
  {author} {\bibfnamefont {S.}~\bibnamefont {Jochim}},\ }\bibfield  {title}
  {\bibinfo {title} {Two fermions in a double well: Exploring a fundamental
  building block of the hubbard model},\ }\href
  {https://doi.org/10.1103/PhysRevLett.114.080402} {\bibfield  {journal}
  {\bibinfo  {journal} {Phys. Rev. Lett.}\ }\textbf {\bibinfo {volume} {114}},\
  \bibinfo {pages} {080402} (\bibinfo {year} {2015})}\BibitemShut {NoStop}%
\bibitem [{\citenamefont {Bergschneider}\ \emph {et~al.}(2019)\citenamefont
  {Bergschneider}, \citenamefont {Klinkhamer}, \citenamefont {Becher},
  \citenamefont {Klemt}, \citenamefont {Palm}, \citenamefont {Z{\"u}rn},
  \citenamefont {Jochim},\ and\ \citenamefont {Preiss}}]{Bergschneider_2019}%
  \BibitemOpen
  \bibfield  {author} {\bibinfo {author} {\bibfnamefont {A.}~\bibnamefont
  {Bergschneider}}, \bibinfo {author} {\bibfnamefont {V.~M.}\ \bibnamefont
  {Klinkhamer}}, \bibinfo {author} {\bibfnamefont {J.~H.}\ \bibnamefont
  {Becher}}, \bibinfo {author} {\bibfnamefont {R.}~\bibnamefont {Klemt}},
  \bibinfo {author} {\bibfnamefont {L.}~\bibnamefont {Palm}}, \bibinfo {author}
  {\bibfnamefont {G.}~\bibnamefont {Z{\"u}rn}}, \bibinfo {author}
  {\bibfnamefont {S.}~\bibnamefont {Jochim}},\ and\ \bibinfo {author}
  {\bibfnamefont {P.~M.}\ \bibnamefont {Preiss}},\ }\bibfield  {title}
  {\bibinfo {title} {Experimental characterization of two-particle entanglement
  through position and momentum correlations},\ }\href
  {https://doi.org/10.1038/s41567-019-0508-6} {\bibfield  {journal} {\bibinfo
  {journal} {Nature Physics}\ }\textbf {\bibinfo {volume} {15}},\ \bibinfo
  {pages} {640} (\bibinfo {year} {2019})}\BibitemShut {NoStop}%
\bibitem [{\citenamefont {Becher}\ \emph {et~al.}(2020)\citenamefont {Becher},
  \citenamefont {Sindici}, \citenamefont {Klemt}, \citenamefont {Jochim},
  \citenamefont {Daley},\ and\ \citenamefont {Preiss}}]{Becher_2020}%
  \BibitemOpen
  \bibfield  {author} {\bibinfo {author} {\bibfnamefont {J.~H.}\ \bibnamefont
  {Becher}}, \bibinfo {author} {\bibfnamefont {E.}~\bibnamefont {Sindici}},
  \bibinfo {author} {\bibfnamefont {R.}~\bibnamefont {Klemt}}, \bibinfo
  {author} {\bibfnamefont {S.}~\bibnamefont {Jochim}}, \bibinfo {author}
  {\bibfnamefont {A.~J.}\ \bibnamefont {Daley}},\ and\ \bibinfo {author}
  {\bibfnamefont {P.~M.}\ \bibnamefont {Preiss}},\ }\bibfield  {title}
  {\bibinfo {title} {Measurement of identical particle entanglement and the
  influence of antisymmetrization},\ }\href
  {https://doi.org/10.1103/PhysRevLett.125.180402} {\bibfield  {journal}
  {\bibinfo  {journal} {Phys. Rev. Lett.}\ }\textbf {\bibinfo {volume} {125}},\
  \bibinfo {pages} {180402} (\bibinfo {year} {2020})}\BibitemShut {NoStop}%
\bibitem [{\citenamefont {Spar}\ \emph {et~al.}(2022)\citenamefont {Spar},
  \citenamefont {Guardado-Sanchez}, \citenamefont {Chi}, \citenamefont {Yan},\
  and\ \citenamefont {Bakr}}]{Spar_2022}%
  \BibitemOpen
  \bibfield  {author} {\bibinfo {author} {\bibfnamefont {B.~M.}\ \bibnamefont
  {Spar}}, \bibinfo {author} {\bibfnamefont {E.}~\bibnamefont
  {Guardado-Sanchez}}, \bibinfo {author} {\bibfnamefont {S.}~\bibnamefont
  {Chi}}, \bibinfo {author} {\bibfnamefont {Z.~Z.}\ \bibnamefont {Yan}},\ and\
  \bibinfo {author} {\bibfnamefont {W.~S.}\ \bibnamefont {Bakr}},\ }\bibfield
  {title} {\bibinfo {title} {Realization of a fermi-hubbard optical tweezer
  array},\ }\href {https://doi.org/10.1103/PhysRevLett.128.223202} {\bibfield
  {journal} {\bibinfo  {journal} {Phys. Rev. Lett.}\ }\textbf {\bibinfo
  {volume} {128}},\ \bibinfo {pages} {223202} (\bibinfo {year}
  {2022})}\BibitemShut {NoStop}%
\bibitem [{\citenamefont {Young}\ \emph {et~al.}(2022)\citenamefont {Young},
  \citenamefont {Eckner}, \citenamefont {Schine}, \citenamefont {Childs},\ and\
  \citenamefont {Kaufman}}]{Young_2022}%
  \BibitemOpen
  \bibfield  {author} {\bibinfo {author} {\bibfnamefont {A.~W.}\ \bibnamefont
  {Young}}, \bibinfo {author} {\bibfnamefont {W.~J.}\ \bibnamefont {Eckner}},
  \bibinfo {author} {\bibfnamefont {N.}~\bibnamefont {Schine}}, \bibinfo
  {author} {\bibfnamefont {A.~M.}\ \bibnamefont {Childs}},\ and\ \bibinfo
  {author} {\bibfnamefont {A.~M.}\ \bibnamefont {Kaufman}},\ }\bibfield
  {title} {\bibinfo {title} {Tweezer-programmable 2d quantum walks in a
  hubbard-regime lattice},\ }\href {https://doi.org/10.1126/science.abo0608}
  {\bibfield  {journal} {\bibinfo  {journal} {Science}\ }\textbf {\bibinfo
  {volume} {377}},\ \bibinfo {pages} {885} (\bibinfo {year}
  {2022})}\BibitemShut {NoStop}%
\bibitem [{\citenamefont {Yan}\ \emph {et~al.}(2022)\citenamefont {Yan},
  \citenamefont {Spar}, \citenamefont {Prichard}, \citenamefont {Chi},
  \citenamefont {Wei}, \citenamefont {Ibarra-Garc\'{\i}a-Padilla},
  \citenamefont {Hazzard},\ and\ \citenamefont {Bakr}}]{Yan_2022}%
  \BibitemOpen
  \bibfield  {author} {\bibinfo {author} {\bibfnamefont {Z.~Z.}\ \bibnamefont
  {Yan}}, \bibinfo {author} {\bibfnamefont {B.~M.}\ \bibnamefont {Spar}},
  \bibinfo {author} {\bibfnamefont {M.~L.}\ \bibnamefont {Prichard}}, \bibinfo
  {author} {\bibfnamefont {S.}~\bibnamefont {Chi}}, \bibinfo {author}
  {\bibfnamefont {H.-T.}\ \bibnamefont {Wei}}, \bibinfo {author} {\bibfnamefont
  {E.}~\bibnamefont {Ibarra-Garc\'{\i}a-Padilla}}, \bibinfo {author}
  {\bibfnamefont {K.~R.~A.}\ \bibnamefont {Hazzard}},\ and\ \bibinfo {author}
  {\bibfnamefont {W.~S.}\ \bibnamefont {Bakr}},\ }\bibfield  {title} {\bibinfo
  {title} {Two-dimensional programmable tweezer arrays of fermions},\ }\href
  {https://doi.org/10.1103/PhysRevLett.129.123201} {\bibfield  {journal}
  {\bibinfo  {journal} {Phys. Rev. Lett.}\ }\textbf {\bibinfo {volume} {129}},\
  \bibinfo {pages} {123201} (\bibinfo {year} {2022})}\BibitemShut {NoStop}%
\bibitem [{\citenamefont {Jaksch}\ \emph {et~al.}(1999)\citenamefont {Jaksch},
  \citenamefont {Briegel}, \citenamefont {Cirac}, \citenamefont {Gardiner},\
  and\ \citenamefont {Zoller}}]{Jaksch_1999}%
  \BibitemOpen
  \bibfield  {author} {\bibinfo {author} {\bibfnamefont {D.}~\bibnamefont
  {Jaksch}}, \bibinfo {author} {\bibfnamefont {H.-J.}\ \bibnamefont {Briegel}},
  \bibinfo {author} {\bibfnamefont {J.~I.}\ \bibnamefont {Cirac}}, \bibinfo
  {author} {\bibfnamefont {C.~W.}\ \bibnamefont {Gardiner}},\ and\ \bibinfo
  {author} {\bibfnamefont {P.}~\bibnamefont {Zoller}},\ }\bibfield  {title}
  {\bibinfo {title} {Entanglement of atoms via cold controlled collisions},\
  }\href {https://doi.org/10.1103/PhysRevLett.82.1975} {\bibfield  {journal}
  {\bibinfo  {journal} {Phys. Rev. Lett.}\ }\textbf {\bibinfo {volume} {82}},\
  \bibinfo {pages} {1975} (\bibinfo {year} {1999})}\BibitemShut {NoStop}%
\bibitem [{\citenamefont {Mandel}\ \emph
  {et~al.}(2003{\natexlab{a}})\citenamefont {Mandel}, \citenamefont {Greiner},
  \citenamefont {Widera}, \citenamefont {Rom}, \citenamefont {H\"ansch},\ and\
  \citenamefont {Bloch}}]{Mandel_2003_0}%
  \BibitemOpen
  \bibfield  {author} {\bibinfo {author} {\bibfnamefont {O.}~\bibnamefont
  {Mandel}}, \bibinfo {author} {\bibfnamefont {M.}~\bibnamefont {Greiner}},
  \bibinfo {author} {\bibfnamefont {A.}~\bibnamefont {Widera}}, \bibinfo
  {author} {\bibfnamefont {T.}~\bibnamefont {Rom}}, \bibinfo {author}
  {\bibfnamefont {T.~W.}\ \bibnamefont {H\"ansch}},\ and\ \bibinfo {author}
  {\bibfnamefont {I.}~\bibnamefont {Bloch}},\ }\bibfield  {title} {\bibinfo
  {title} {Coherent transport of neutral atoms in spin-dependent optical
  lattice potentials},\ }\href {https://doi.org/10.1103/PhysRevLett.91.010407}
  {\bibfield  {journal} {\bibinfo  {journal} {Phys. Rev. Lett.}\ }\textbf
  {\bibinfo {volume} {91}},\ \bibinfo {pages} {010407} (\bibinfo {year}
  {2003}{\natexlab{a}})}\BibitemShut {NoStop}%
\bibitem [{\citenamefont {Mandel}\ \emph
  {et~al.}(2003{\natexlab{b}})\citenamefont {Mandel}, \citenamefont {Greiner},
  \citenamefont {Widera}, \citenamefont {Rom}, \citenamefont {H{\"a}nsch},\
  and\ \citenamefont {Bloch}}]{Mandel_2003}%
  \BibitemOpen
  \bibfield  {author} {\bibinfo {author} {\bibfnamefont {O.}~\bibnamefont
  {Mandel}}, \bibinfo {author} {\bibfnamefont {M.}~\bibnamefont {Greiner}},
  \bibinfo {author} {\bibfnamefont {A.}~\bibnamefont {Widera}}, \bibinfo
  {author} {\bibfnamefont {T.}~\bibnamefont {Rom}}, \bibinfo {author}
  {\bibfnamefont {T.~W.}\ \bibnamefont {H{\"a}nsch}},\ and\ \bibinfo {author}
  {\bibfnamefont {I.}~\bibnamefont {Bloch}},\ }\bibfield  {title} {\bibinfo
  {title} {Controlled collisions for multi-particle entanglement of optically
  trapped atoms},\ }\href {https://doi.org/10.1038/nature02008} {\bibfield
  {journal} {\bibinfo  {journal} {Nature}\ }\textbf {\bibinfo {volume} {425}},\
  \bibinfo {pages} {937} (\bibinfo {year} {2003}{\natexlab{b}})}\BibitemShut
  {NoStop}%
\bibitem [{\citenamefont {Daley}\ \emph {et~al.}(2008)\citenamefont {Daley},
  \citenamefont {Boyd}, \citenamefont {Ye},\ and\ \citenamefont
  {Zoller}}]{Daley_2008}%
  \BibitemOpen
  \bibfield  {author} {\bibinfo {author} {\bibfnamefont {A.~J.}\ \bibnamefont
  {Daley}}, \bibinfo {author} {\bibfnamefont {M.~M.}\ \bibnamefont {Boyd}},
  \bibinfo {author} {\bibfnamefont {J.}~\bibnamefont {Ye}},\ and\ \bibinfo
  {author} {\bibfnamefont {P.}~\bibnamefont {Zoller}},\ }\bibfield  {title}
  {\bibinfo {title} {Quantum computing with alkaline-earth-metal atoms},\
  }\href {https://doi.org/10.1103/PhysRevLett.101.170504} {\bibfield  {journal}
  {\bibinfo  {journal} {Phys. Rev. Lett.}\ }\textbf {\bibinfo {volume} {101}},\
  \bibinfo {pages} {170504} (\bibinfo {year} {2008})}\BibitemShut {NoStop}%
\bibitem [{\citenamefont {Belmechri}\ \emph {et~al.}(2013)\citenamefont
  {Belmechri}, \citenamefont {F{\"o}rster}, \citenamefont {Alt}, \citenamefont
  {Widera}, \citenamefont {Meschede},\ and\ \citenamefont
  {Alberti}}]{belmechri2013microwave}%
  \BibitemOpen
  \bibfield  {author} {\bibinfo {author} {\bibfnamefont {N.}~\bibnamefont
  {Belmechri}}, \bibinfo {author} {\bibfnamefont {L.}~\bibnamefont
  {F{\"o}rster}}, \bibinfo {author} {\bibfnamefont {W.}~\bibnamefont {Alt}},
  \bibinfo {author} {\bibfnamefont {A.}~\bibnamefont {Widera}}, \bibinfo
  {author} {\bibfnamefont {D.}~\bibnamefont {Meschede}},\ and\ \bibinfo
  {author} {\bibfnamefont {A.}~\bibnamefont {Alberti}},\ }\bibfield  {title}
  {\bibinfo {title} {Microwave control of atomic motional states in a
  spin-dependent optical lattice},\ }\href@noop {} {\bibfield  {journal}
  {\bibinfo  {journal} {Journal of Physics B: Atomic, Molecular and Optical
  Physics}\ }\textbf {\bibinfo {volume} {46}},\ \bibinfo {pages} {104006}
  (\bibinfo {year} {2013})}\BibitemShut {NoStop}%
\bibitem [{\citenamefont {Robens}\ \emph {et~al.}(2015)\citenamefont {Robens},
  \citenamefont {Alt}, \citenamefont {Meschede}, \citenamefont {Emary},\ and\
  \citenamefont {Alberti}}]{Robens_2015}%
  \BibitemOpen
  \bibfield  {author} {\bibinfo {author} {\bibfnamefont {C.}~\bibnamefont
  {Robens}}, \bibinfo {author} {\bibfnamefont {W.}~\bibnamefont {Alt}},
  \bibinfo {author} {\bibfnamefont {D.}~\bibnamefont {Meschede}}, \bibinfo
  {author} {\bibfnamefont {C.}~\bibnamefont {Emary}},\ and\ \bibinfo {author}
  {\bibfnamefont {A.}~\bibnamefont {Alberti}},\ }\bibfield  {title} {\bibinfo
  {title} {Ideal negative measurements in quantum walks disprove theories based
  on classical trajectories},\ }\href
  {https://doi.org/10.1103/PhysRevX.5.011003} {\bibfield  {journal} {\bibinfo
  {journal} {Phys. Rev. X}\ }\textbf {\bibinfo {volume} {5}},\ \bibinfo {pages}
  {011003} (\bibinfo {year} {2015})}\BibitemShut {NoStop}%
\bibitem [{\citenamefont {Lam}\ \emph {et~al.}(2021{\natexlab{a}})\citenamefont
  {Lam}, \citenamefont {Peter}, \citenamefont {Groh}, \citenamefont {Alt},
  \citenamefont {Robens}, \citenamefont {Meschede}, \citenamefont {Negretti},
  \citenamefont {Montangero}, \citenamefont {Calarco},\ and\ \citenamefont
  {Alberti}}]{Lam_2021}%
  \BibitemOpen
  \bibfield  {author} {\bibinfo {author} {\bibfnamefont {M.~R.}\ \bibnamefont
  {Lam}}, \bibinfo {author} {\bibfnamefont {N.}~\bibnamefont {Peter}}, \bibinfo
  {author} {\bibfnamefont {T.}~\bibnamefont {Groh}}, \bibinfo {author}
  {\bibfnamefont {W.}~\bibnamefont {Alt}}, \bibinfo {author} {\bibfnamefont
  {C.}~\bibnamefont {Robens}}, \bibinfo {author} {\bibfnamefont
  {D.}~\bibnamefont {Meschede}}, \bibinfo {author} {\bibfnamefont
  {A.}~\bibnamefont {Negretti}}, \bibinfo {author} {\bibfnamefont
  {S.}~\bibnamefont {Montangero}}, \bibinfo {author} {\bibfnamefont
  {T.}~\bibnamefont {Calarco}},\ and\ \bibinfo {author} {\bibfnamefont
  {A.}~\bibnamefont {Alberti}},\ }\bibfield  {title} {\bibinfo {title}
  {Demonstration of quantum brachistochrones between distant states of an
  atom},\ }\href {https://doi.org/10.1103/PhysRevX.11.011035} {\bibfield
  {journal} {\bibinfo  {journal} {Phys. Rev. X}\ }\textbf {\bibinfo {volume}
  {11}},\ \bibinfo {pages} {011035} (\bibinfo {year}
  {2021}{\natexlab{a}})}\BibitemShut {NoStop}%
\bibitem [{\citenamefont {Zhang}\ \emph
  {et~al.}(2022{\natexlab{b}})\citenamefont {Zhang}, \citenamefont {He},
  \citenamefont {Sun}, \citenamefont {Zheng}, \citenamefont {Liu},
  \citenamefont {Luo}, \citenamefont {Wang}, \citenamefont {Zhu}, \citenamefont
  {Qiu}, \citenamefont {Shen}, \citenamefont {Wang}, \citenamefont {Lin},
  \citenamefont {Yu}, \citenamefont {Li}, \citenamefont {Xiao}, \citenamefont
  {Li}, \citenamefont {Yang}, \citenamefont {Jiang}, \citenamefont {Dai},
  \citenamefont {Zhou}, \citenamefont {Ma}, \citenamefont {Yuan},\ and\
  \citenamefont {Pan}}]{Zhang_2022}%
  \BibitemOpen
  \bibfield  {author} {\bibinfo {author} {\bibfnamefont {W.-Y.}\ \bibnamefont
  {Zhang}}, \bibinfo {author} {\bibfnamefont {M.-G.}\ \bibnamefont {He}},
  \bibinfo {author} {\bibfnamefont {H.}~\bibnamefont {Sun}}, \bibinfo {author}
  {\bibfnamefont {Y.-G.}\ \bibnamefont {Zheng}}, \bibinfo {author}
  {\bibfnamefont {Y.}~\bibnamefont {Liu}}, \bibinfo {author} {\bibfnamefont
  {A.}~\bibnamefont {Luo}}, \bibinfo {author} {\bibfnamefont {H.-Y.}\
  \bibnamefont {Wang}}, \bibinfo {author} {\bibfnamefont {Z.-H.}\ \bibnamefont
  {Zhu}}, \bibinfo {author} {\bibfnamefont {P.-Y.}\ \bibnamefont {Qiu}},
  \bibinfo {author} {\bibfnamefont {Y.-C.}\ \bibnamefont {Shen}}, \bibinfo
  {author} {\bibfnamefont {X.-K.}\ \bibnamefont {Wang}}, \bibinfo {author}
  {\bibfnamefont {W.}~\bibnamefont {Lin}}, \bibinfo {author} {\bibfnamefont
  {S.-T.}\ \bibnamefont {Yu}}, \bibinfo {author} {\bibfnamefont {B.-C.}\
  \bibnamefont {Li}}, \bibinfo {author} {\bibfnamefont {B.}~\bibnamefont
  {Xiao}}, \bibinfo {author} {\bibfnamefont {M.-D.}\ \bibnamefont {Li}},
  \bibinfo {author} {\bibfnamefont {Y.-M.}\ \bibnamefont {Yang}}, \bibinfo
  {author} {\bibfnamefont {X.}~\bibnamefont {Jiang}}, \bibinfo {author}
  {\bibfnamefont {H.-N.}\ \bibnamefont {Dai}}, \bibinfo {author} {\bibfnamefont
  {Y.}~\bibnamefont {Zhou}}, \bibinfo {author} {\bibfnamefont {X.}~\bibnamefont
  {Ma}}, \bibinfo {author} {\bibfnamefont {Z.-S.}\ \bibnamefont {Yuan}},\ and\
  \bibinfo {author} {\bibfnamefont {J.-W.}\ \bibnamefont {Pan}},\ }\href
  {https://doi.org/10.48550/ARXIV.2210.02936} {\bibinfo {title} {Functional
  building blocks for scalable multipartite entanglement in optical lattices}}
  (\bibinfo {year} {2022}{\natexlab{b}})\BibitemShut {NoStop}%
\bibitem [{\citenamefont {Ringbauer}\ \emph {et~al.}(2022)\citenamefont
  {Ringbauer}, \citenamefont {Meth}, \citenamefont {Postler}, \citenamefont
  {Stricker}, \citenamefont {Blatt}, \citenamefont {Schindler},\ and\
  \citenamefont {Monz}}]{Ringbauer_2022}%
  \BibitemOpen
  \bibfield  {author} {\bibinfo {author} {\bibfnamefont {M.}~\bibnamefont
  {Ringbauer}}, \bibinfo {author} {\bibfnamefont {M.}~\bibnamefont {Meth}},
  \bibinfo {author} {\bibfnamefont {L.}~\bibnamefont {Postler}}, \bibinfo
  {author} {\bibfnamefont {R.}~\bibnamefont {Stricker}}, \bibinfo {author}
  {\bibfnamefont {R.}~\bibnamefont {Blatt}}, \bibinfo {author} {\bibfnamefont
  {P.}~\bibnamefont {Schindler}},\ and\ \bibinfo {author} {\bibfnamefont
  {T.}~\bibnamefont {Monz}},\ }\bibfield  {title} {\bibinfo {title} {A
  universal qudit quantum processor with trapped ions},\ }\href
  {https://doi.org/10.1038/s41567-022-01658-0} {\bibfield  {journal} {\bibinfo
  {journal} {Nature Physics}\ }\textbf {\bibinfo {volume} {18}},\ \bibinfo
  {pages} {1053} (\bibinfo {year} {2022})}\BibitemShut {NoStop}%
\bibitem [{\citenamefont {Lloyd}(1996)}]{Lloyd_1996}%
  \BibitemOpen
  \bibfield  {author} {\bibinfo {author} {\bibfnamefont {S.}~\bibnamefont
  {Lloyd}},\ }\bibfield  {title} {\bibinfo {title} {Universal quantum
  simulators},\ }\href {https://doi.org/10.1126/science.273.5278.1073}
  {\bibfield  {journal} {\bibinfo  {journal} {Science}\ }\textbf {\bibinfo
  {volume} {273}},\ \bibinfo {pages} {1073} (\bibinfo {year}
  {1996})}\BibitemShut {NoStop}%
\bibitem [{\citenamefont {Lloyd}(1995)}]{Lloyd_1995}%
  \BibitemOpen
  \bibfield  {author} {\bibinfo {author} {\bibfnamefont {S.}~\bibnamefont
  {Lloyd}},\ }\bibfield  {title} {\bibinfo {title} {Almost any quantum logic
  gate is universal},\ }\href {https://doi.org/10.1103/PhysRevLett.75.346}
  {\bibfield  {journal} {\bibinfo  {journal} {Phys. Rev. Lett.}\ }\textbf
  {\bibinfo {volume} {75}},\ \bibinfo {pages} {346} (\bibinfo {year}
  {1995})}\BibitemShut {NoStop}%
\bibitem [{\citenamefont {McArdle}\ \emph {et~al.}(2020)\citenamefont
  {McArdle}, \citenamefont {Endo}, \citenamefont {Aspuru-Guzik}, \citenamefont
  {Benjamin},\ and\ \citenamefont {Yuan}}]{McArdle_2020}%
  \BibitemOpen
  \bibfield  {author} {\bibinfo {author} {\bibfnamefont {S.}~\bibnamefont
  {McArdle}}, \bibinfo {author} {\bibfnamefont {S.}~\bibnamefont {Endo}},
  \bibinfo {author} {\bibfnamefont {A.}~\bibnamefont {Aspuru-Guzik}}, \bibinfo
  {author} {\bibfnamefont {S.~C.}\ \bibnamefont {Benjamin}},\ and\ \bibinfo
  {author} {\bibfnamefont {X.}~\bibnamefont {Yuan}},\ }\bibfield  {title}
  {\bibinfo {title} {Quantum computational chemistry},\ }\href
  {https://doi.org/10.1103/RevModPhys.92.015003} {\bibfield  {journal}
  {\bibinfo  {journal} {Rev. Mod. Phys.}\ }\textbf {\bibinfo {volume} {92}},\
  \bibinfo {pages} {015003} (\bibinfo {year} {2020})}\BibitemShut {NoStop}%
\bibitem [{\citenamefont {Ye}\ \emph {et~al.}(2008)\citenamefont {Ye},
  \citenamefont {Kimble},\ and\ \citenamefont {Katori}}]{ye2008quantum}%
  \BibitemOpen
  \bibfield  {author} {\bibinfo {author} {\bibfnamefont {J.}~\bibnamefont
  {Ye}}, \bibinfo {author} {\bibfnamefont {H.}~\bibnamefont {Kimble}},\ and\
  \bibinfo {author} {\bibfnamefont {H.}~\bibnamefont {Katori}},\ }\bibfield
  {title} {\bibinfo {title} {Quantum state engineering and precision metrology
  using state-insensitive light traps},\ }\href@noop {} {\bibfield  {journal}
  {\bibinfo  {journal} {science}\ }\textbf {\bibinfo {volume} {320}},\ \bibinfo
  {pages} {1734} (\bibinfo {year} {2008})}\BibitemShut {NoStop}%
\bibitem [{\citenamefont {Cooper}\ \emph {et~al.}(2018)\citenamefont {Cooper},
  \citenamefont {Covey}, \citenamefont {Madjarov}, \citenamefont {Porsev},
  \citenamefont {Safronova},\ and\ \citenamefont {Endres}}]{Cooper_2018}%
  \BibitemOpen
  \bibfield  {author} {\bibinfo {author} {\bibfnamefont {A.}~\bibnamefont
  {Cooper}}, \bibinfo {author} {\bibfnamefont {J.~P.}\ \bibnamefont {Covey}},
  \bibinfo {author} {\bibfnamefont {I.~S.}\ \bibnamefont {Madjarov}}, \bibinfo
  {author} {\bibfnamefont {S.~G.}\ \bibnamefont {Porsev}}, \bibinfo {author}
  {\bibfnamefont {M.~S.}\ \bibnamefont {Safronova}},\ and\ \bibinfo {author}
  {\bibfnamefont {M.}~\bibnamefont {Endres}},\ }\bibfield  {title} {\bibinfo
  {title} {Alkaline-earth atoms in optical tweezers},\ }\href
  {https://doi.org/10.1103/PhysRevX.8.041055} {\bibfield  {journal} {\bibinfo
  {journal} {Phys. Rev. X}\ }\textbf {\bibinfo {volume} {8}},\ \bibinfo {pages}
  {041055} (\bibinfo {year} {2018})}\BibitemShut {NoStop}%
\bibitem [{\citenamefont {Heinz}\ \emph {et~al.}(2020)\citenamefont {Heinz},
  \citenamefont {Park}, \citenamefont {\ifmmode \check{S}\else
  \v{S}\fi{}anti\ifmmode~\acute{c}\else \'{c}\fi{}}, \citenamefont {Trautmann},
  \citenamefont {Porsev}, \citenamefont {Safronova}, \citenamefont {Bloch},\
  and\ \citenamefont {Blatt}}]{Heinz_2020}%
  \BibitemOpen
  \bibfield  {author} {\bibinfo {author} {\bibfnamefont {A.}~\bibnamefont
  {Heinz}}, \bibinfo {author} {\bibfnamefont {A.~J.}\ \bibnamefont {Park}},
  \bibinfo {author} {\bibfnamefont {N.}~\bibnamefont {\ifmmode \check{S}\else
  \v{S}\fi{}anti\ifmmode~\acute{c}\else \'{c}\fi{}}}, \bibinfo {author}
  {\bibfnamefont {J.}~\bibnamefont {Trautmann}}, \bibinfo {author}
  {\bibfnamefont {S.~G.}\ \bibnamefont {Porsev}}, \bibinfo {author}
  {\bibfnamefont {M.~S.}\ \bibnamefont {Safronova}}, \bibinfo {author}
  {\bibfnamefont {I.}~\bibnamefont {Bloch}},\ and\ \bibinfo {author}
  {\bibfnamefont {S.}~\bibnamefont {Blatt}},\ }\bibfield  {title} {\bibinfo
  {title} {State-dependent optical lattices for the strontium optical qubit},\
  }\href {https://doi.org/10.1103/PhysRevLett.124.203201} {\bibfield  {journal}
  {\bibinfo  {journal} {Phys. Rev. Lett.}\ }\textbf {\bibinfo {volume} {124}},\
  \bibinfo {pages} {203201} (\bibinfo {year} {2020})}\BibitemShut {NoStop}%
\bibitem [{\citenamefont {Schine}\ \emph {et~al.}(2022)\citenamefont {Schine},
  \citenamefont {Young}, \citenamefont {Eckner}, \citenamefont {Martin},\ and\
  \citenamefont {Kaufman}}]{Schine2022}%
  \BibitemOpen
  \bibfield  {author} {\bibinfo {author} {\bibfnamefont {N.}~\bibnamefont
  {Schine}}, \bibinfo {author} {\bibfnamefont {A.~W.}\ \bibnamefont {Young}},
  \bibinfo {author} {\bibfnamefont {W.~J.}\ \bibnamefont {Eckner}}, \bibinfo
  {author} {\bibfnamefont {M.~J.}\ \bibnamefont {Martin}},\ and\ \bibinfo
  {author} {\bibfnamefont {A.~M.}\ \bibnamefont {Kaufman}},\ }\bibfield
  {title} {\bibinfo {title} {Long-lived bell states in an array of optical
  clock qubits},\ }\bibfield  {journal} {\bibinfo  {journal} {Nature Physics}\
  }\href {https://doi.org/10.1038/s41567-022-01678-w}
  {10.1038/s41567-022-01678-w} (\bibinfo {year} {2022})\BibitemShut {NoStop}%
\bibitem [{\citenamefont {Wilson}\ \emph {et~al.}(2022)\citenamefont {Wilson},
  \citenamefont {Saskin}, \citenamefont {Meng}, \citenamefont {Ma},
  \citenamefont {Dilip}, \citenamefont {Burgers},\ and\ \citenamefont
  {Thompson}}]{Wilson_2022}%
  \BibitemOpen
  \bibfield  {author} {\bibinfo {author} {\bibfnamefont {J.~T.}\ \bibnamefont
  {Wilson}}, \bibinfo {author} {\bibfnamefont {S.}~\bibnamefont {Saskin}},
  \bibinfo {author} {\bibfnamefont {Y.}~\bibnamefont {Meng}}, \bibinfo {author}
  {\bibfnamefont {S.}~\bibnamefont {Ma}}, \bibinfo {author} {\bibfnamefont
  {R.}~\bibnamefont {Dilip}}, \bibinfo {author} {\bibfnamefont {A.~P.}\
  \bibnamefont {Burgers}},\ and\ \bibinfo {author} {\bibfnamefont {J.~D.}\
  \bibnamefont {Thompson}},\ }\bibfield  {title} {\bibinfo {title} {Trapping
  alkaline earth rydberg atoms optical tweezer arrays},\ }\href
  {https://doi.org/10.1103/PhysRevLett.128.033201} {\bibfield  {journal}
  {\bibinfo  {journal} {Phys. Rev. Lett.}\ }\textbf {\bibinfo {volume} {128}},\
  \bibinfo {pages} {033201} (\bibinfo {year} {2022})}\BibitemShut {NoStop}%
\bibitem [{\citenamefont {Kandala}\ \emph
  {et~al.}(2017{\natexlab{b}})\citenamefont {Kandala}, \citenamefont
  {Mezzacapo}, \citenamefont {Temme}, \citenamefont {Takita}, \citenamefont
  {Brink}, \citenamefont {Chow},\ and\ \citenamefont
  {Gambetta}}]{Kandala_2017}%
  \BibitemOpen
  \bibfield  {author} {\bibinfo {author} {\bibfnamefont {A.}~\bibnamefont
  {Kandala}}, \bibinfo {author} {\bibfnamefont {A.}~\bibnamefont {Mezzacapo}},
  \bibinfo {author} {\bibfnamefont {K.}~\bibnamefont {Temme}}, \bibinfo
  {author} {\bibfnamefont {M.}~\bibnamefont {Takita}}, \bibinfo {author}
  {\bibfnamefont {M.}~\bibnamefont {Brink}}, \bibinfo {author} {\bibfnamefont
  {J.~M.}\ \bibnamefont {Chow}},\ and\ \bibinfo {author} {\bibfnamefont
  {J.~M.}\ \bibnamefont {Gambetta}},\ }\bibfield  {title} {\bibinfo {title}
  {Hardware-efficient variational quantum eigensolver for small molecules and
  quantum magnets},\ }\href {https://doi.org/10.1038/nature23879} {\bibfield
  {journal} {\bibinfo  {journal} {Nature}\ }\textbf {\bibinfo {volume} {549}},\
  \bibinfo {pages} {242} (\bibinfo {year} {2017}{\natexlab{b}})}\BibitemShut
  {NoStop}%
\bibitem [{\citenamefont {Trotter}(1959)}]{Trotter_1959}%
  \BibitemOpen
  \bibfield  {author} {\bibinfo {author} {\bibfnamefont {H.~F.}\ \bibnamefont
  {Trotter}},\ }\bibfield  {title} {\bibinfo {title} {On the product of
  semi-groups of operators},\ }\href {http://www.jstor.org/stable/2033649}
  {\bibfield  {journal} {\bibinfo  {journal} {Proceedings of the American
  Mathematical Society}\ }\textbf {\bibinfo {volume} {10}},\ \bibinfo {pages}
  {545} (\bibinfo {year} {1959})}\BibitemShut {NoStop}%
\bibitem [{\citenamefont {McClean}\ \emph {et~al.}(2016)\citenamefont
  {McClean}, \citenamefont {Romero}, \citenamefont {Babbush},\ and\
  \citenamefont {Aspuru-Guzik}}]{McClean_2016}%
  \BibitemOpen
  \bibfield  {author} {\bibinfo {author} {\bibfnamefont {J.~R.}\ \bibnamefont
  {McClean}}, \bibinfo {author} {\bibfnamefont {J.}~\bibnamefont {Romero}},
  \bibinfo {author} {\bibfnamefont {R.}~\bibnamefont {Babbush}},\ and\ \bibinfo
  {author} {\bibfnamefont {A.}~\bibnamefont {Aspuru-Guzik}},\ }\bibfield
  {title} {\bibinfo {title} {The theory of variational hybrid quantum-classical
  algorithms},\ }\href {https://doi.org/10.1088/1367-2630/18/2/023023}
  {\bibfield  {journal} {\bibinfo  {journal} {New Journal of Physics}\ }\textbf
  {\bibinfo {volume} {18}},\ \bibinfo {pages} {023023} (\bibinfo {year}
  {2016})}\BibitemShut {NoStop}%
\bibitem [{\citenamefont {Anand}\ \emph {et~al.}(2022)\citenamefont {Anand},
  \citenamefont {Schleich}, \citenamefont {Alperin-Lea}, \citenamefont
  {Jensen}, \citenamefont {Sim}, \citenamefont {D{\'\i}az-Tinoco},
  \citenamefont {Kottmann}, \citenamefont {Degroote}, \citenamefont
  {Izmaylov},\ and\ \citenamefont {Aspuru-Guzik}}]{Anand_2022}%
  \BibitemOpen
  \bibfield  {author} {\bibinfo {author} {\bibfnamefont {A.}~\bibnamefont
  {Anand}}, \bibinfo {author} {\bibfnamefont {P.}~\bibnamefont {Schleich}},
  \bibinfo {author} {\bibfnamefont {S.}~\bibnamefont {Alperin-Lea}}, \bibinfo
  {author} {\bibfnamefont {P.~W.~K.}\ \bibnamefont {Jensen}}, \bibinfo {author}
  {\bibfnamefont {S.}~\bibnamefont {Sim}}, \bibinfo {author} {\bibfnamefont
  {M.}~\bibnamefont {D{\'\i}az-Tinoco}}, \bibinfo {author} {\bibfnamefont
  {J.~S.}\ \bibnamefont {Kottmann}}, \bibinfo {author} {\bibfnamefont
  {M.}~\bibnamefont {Degroote}}, \bibinfo {author} {\bibfnamefont {A.~F.}\
  \bibnamefont {Izmaylov}},\ and\ \bibinfo {author} {\bibfnamefont
  {A.}~\bibnamefont {Aspuru-Guzik}},\ }\bibfield  {title} {\bibinfo {title} {A
  quantum computing view on unitary coupled cluster theory},\ }\href
  {https://doi.org/10.1039/D1CS00932J} {\bibfield  {journal} {\bibinfo
  {journal} {Chem. Soc. Rev.}\ }\textbf {\bibinfo {volume} {51}},\ \bibinfo
  {pages} {1659} (\bibinfo {year} {2022})}\BibitemShut {NoStop}%
\bibitem [{\citenamefont {{Naldesi}}\ \emph {et~al.}(2022)\citenamefont
  {{Naldesi}}, \citenamefont {{Elben}}, \citenamefont {{Minguzzi}},
  \citenamefont {{Cl{\'e}ment}}, \citenamefont {{Zoller}},\ and\ \citenamefont
  {{Vermersch}}}]{Naldesi2022}%
  \BibitemOpen
  \bibfield  {author} {\bibinfo {author} {\bibfnamefont {P.}~\bibnamefont
  {{Naldesi}}}, \bibinfo {author} {\bibfnamefont {A.}~\bibnamefont {{Elben}}},
  \bibinfo {author} {\bibfnamefont {A.}~\bibnamefont {{Minguzzi}}}, \bibinfo
  {author} {\bibfnamefont {D.}~\bibnamefont {{Cl{\'e}ment}}}, \bibinfo {author}
  {\bibfnamefont {P.}~\bibnamefont {{Zoller}}},\ and\ \bibinfo {author}
  {\bibfnamefont {B.}~\bibnamefont {{Vermersch}}},\ }\bibfield  {title}
  {\bibinfo {title} {{Fermionic correlation functions from randomized
  measurements in programmable atomic quantum devices}},\ }\href
  {https://doi.org/10.48550/arXiv.2205.00981} {\bibfield  {journal} {\bibinfo
  {journal} {arXiv e-prints}\ ,\ \bibinfo {eid} {arXiv:2205.00981}} (\bibinfo
  {year} {2022})},\ \Eprint {https://arxiv.org/abs/2205.00981}
  {arXiv:2205.00981 [cond-mat.quant-gas]} \BibitemShut {NoStop}%
\bibitem [{\citenamefont {Low}(2022)}]{Low_2022}%
  \BibitemOpen
  \bibfield  {author} {\bibinfo {author} {\bibfnamefont {G.~H.}\ \bibnamefont
  {Low}},\ }\bibfield  {title} {\bibinfo {title} {Classical shadows of fermions
  with particle number symmetry},\ }\href {https://arxiv.org/abs/2208.08964}
  {\bibfield  {journal} {\bibinfo  {journal} {arXiv:2208.08964}\ } (\bibinfo
  {year} {2022})}\BibitemShut {NoStop}%
\bibitem [{Note1()}]{Note1}%
  \BibitemOpen
  \bibinfo {note} {We consider the LiH molecule at a fixed interatomic distance
  of $1.45$ {\r A} in the Born-Oppenheimer approximation, with two electrons
  and four active orbitals. The Hamiltonian parameters $h^{(1)}_{i,j}$ and
  $h^{(2)}_{i,j,k,l}$ are calculated using OpenFermion~\cite
  {McClean_2020}.}\BibitemShut {Stop}%
\bibitem [{\citenamefont {De~Chiara}\ \emph {et~al.}(2008)\citenamefont
  {De~Chiara}, \citenamefont {Calarco}, \citenamefont {Anderlini},
  \citenamefont {Montangero}, \citenamefont {Lee}, \citenamefont {Brown},
  \citenamefont {Phillips},\ and\ \citenamefont {Porto}}]{DeChiara_2008}%
  \BibitemOpen
  \bibfield  {author} {\bibinfo {author} {\bibfnamefont {G.}~\bibnamefont
  {De~Chiara}}, \bibinfo {author} {\bibfnamefont {T.}~\bibnamefont {Calarco}},
  \bibinfo {author} {\bibfnamefont {M.}~\bibnamefont {Anderlini}}, \bibinfo
  {author} {\bibfnamefont {S.}~\bibnamefont {Montangero}}, \bibinfo {author}
  {\bibfnamefont {P.~J.}\ \bibnamefont {Lee}}, \bibinfo {author} {\bibfnamefont
  {B.~L.}\ \bibnamefont {Brown}}, \bibinfo {author} {\bibfnamefont {W.~D.}\
  \bibnamefont {Phillips}},\ and\ \bibinfo {author} {\bibfnamefont {J.~V.}\
  \bibnamefont {Porto}},\ }\bibfield  {title} {\bibinfo {title} {Optimal
  control of atom transport for quantum gates in optical lattices},\ }\href
  {https://doi.org/10.1103/PhysRevA.77.052333} {\bibfield  {journal} {\bibinfo
  {journal} {Phys. Rev. A}\ }\textbf {\bibinfo {volume} {77}},\ \bibinfo
  {pages} {052333} (\bibinfo {year} {2008})}\BibitemShut {NoStop}%
\bibitem [{\citenamefont {Pagano}\ \emph {et~al.}(2022)\citenamefont {Pagano},
  \citenamefont {Weber}, \citenamefont {Jaschke}, \citenamefont {Pfau},
  \citenamefont {Meinert}, \citenamefont {Montangero},\ and\ \citenamefont
  {B\"uchler}}]{Pagano_2022}%
  \BibitemOpen
  \bibfield  {author} {\bibinfo {author} {\bibfnamefont {A.}~\bibnamefont
  {Pagano}}, \bibinfo {author} {\bibfnamefont {S.}~\bibnamefont {Weber}},
  \bibinfo {author} {\bibfnamefont {D.}~\bibnamefont {Jaschke}}, \bibinfo
  {author} {\bibfnamefont {T.}~\bibnamefont {Pfau}}, \bibinfo {author}
  {\bibfnamefont {F.}~\bibnamefont {Meinert}}, \bibinfo {author} {\bibfnamefont
  {S.}~\bibnamefont {Montangero}},\ and\ \bibinfo {author} {\bibfnamefont
  {H.~P.}\ \bibnamefont {B\"uchler}},\ }\bibfield  {title} {\bibinfo {title}
  {Error budgeting for a controlled-phase gate with strontium-88 rydberg
  atoms},\ }\href {https://doi.org/10.1103/PhysRevResearch.4.033019} {\bibfield
   {journal} {\bibinfo  {journal} {Phys. Rev. Research}\ }\textbf {\bibinfo
  {volume} {4}},\ \bibinfo {pages} {033019} (\bibinfo {year}
  {2022})}\BibitemShut {NoStop}%
\bibitem [{\citenamefont {Grimsley}\ \emph {et~al.}(2019)\citenamefont
  {Grimsley}, \citenamefont {Economou}, \citenamefont {Barnes},\ and\
  \citenamefont {Mayhall}}]{Grimsley_2019}%
  \BibitemOpen
  \bibfield  {author} {\bibinfo {author} {\bibfnamefont {H.~R.}\ \bibnamefont
  {Grimsley}}, \bibinfo {author} {\bibfnamefont {S.~E.}\ \bibnamefont
  {Economou}}, \bibinfo {author} {\bibfnamefont {E.}~\bibnamefont {Barnes}},\
  and\ \bibinfo {author} {\bibfnamefont {N.~J.}\ \bibnamefont {Mayhall}},\
  }\bibfield  {title} {\bibinfo {title} {An adaptive variational algorithm for
  exact molecular simulations on a quantum computer},\ }\href
  {https://doi.org/10.1038/s41467-019-10988-2} {\bibfield  {journal} {\bibinfo
  {journal} {Nature Communications}\ }\textbf {\bibinfo {volume} {10}},\
  \bibinfo {pages} {3007} (\bibinfo {year} {2019})}\BibitemShut {NoStop}%
\bibitem [{\citenamefont {Tang}\ \emph {et~al.}(2021)\citenamefont {Tang},
  \citenamefont {Shkolnikov}, \citenamefont {Barron}, \citenamefont {Grimsley},
  \citenamefont {Mayhall}, \citenamefont {Barnes},\ and\ \citenamefont
  {Economou}}]{qubitadapt_vqe}%
  \BibitemOpen
  \bibfield  {author} {\bibinfo {author} {\bibfnamefont {H.~L.}\ \bibnamefont
  {Tang}}, \bibinfo {author} {\bibfnamefont {V.}~\bibnamefont {Shkolnikov}},
  \bibinfo {author} {\bibfnamefont {G.~S.}\ \bibnamefont {Barron}}, \bibinfo
  {author} {\bibfnamefont {H.~R.}\ \bibnamefont {Grimsley}}, \bibinfo {author}
  {\bibfnamefont {N.~J.}\ \bibnamefont {Mayhall}}, \bibinfo {author}
  {\bibfnamefont {E.}~\bibnamefont {Barnes}},\ and\ \bibinfo {author}
  {\bibfnamefont {S.~E.}\ \bibnamefont {Economou}},\ }\bibfield  {title}
  {\bibinfo {title} {Qubit-adapt-vqe: An adaptive algorithm for constructing
  hardware-efficient ans\"atze on a quantum processor},\ }\href
  {https://doi.org/10.1103/PRXQuantum.2.020310} {\bibfield  {journal} {\bibinfo
   {journal} {PRX Quantum}\ }\textbf {\bibinfo {volume} {2}},\ \bibinfo {pages}
  {020310} (\bibinfo {year} {2021})}\BibitemShut {NoStop}%
\bibitem [{\citenamefont {Anastasiou}\ \emph {et~al.}(2022)\citenamefont
  {Anastasiou}, \citenamefont {Chen}, \citenamefont {Mayhall}, \citenamefont
  {Barnes},\ and\ \citenamefont {Economou}}]{tetris_vqe}%
  \BibitemOpen
  \bibfield  {author} {\bibinfo {author} {\bibfnamefont {P.~G.}\ \bibnamefont
  {Anastasiou}}, \bibinfo {author} {\bibfnamefont {Y.}~\bibnamefont {Chen}},
  \bibinfo {author} {\bibfnamefont {N.~J.}\ \bibnamefont {Mayhall}}, \bibinfo
  {author} {\bibfnamefont {E.}~\bibnamefont {Barnes}},\ and\ \bibinfo {author}
  {\bibfnamefont {S.~E.}\ \bibnamefont {Economou}},\ }\bibfield  {title}
  {\bibinfo {title} {Tetris-adapt-vqe: An adaptive algorithm that yields
  shallower, denser circuit ansätze}\ }\href
  {https://doi.org/10.48550/ARXIV.2209.10562} {10.48550/ARXIV.2209.10562}
  (\bibinfo {year} {2022})\BibitemShut {NoStop}%
\bibitem [{Nal(2023)}]{Naldesi_2023}%
  \BibitemOpen
  \href@noop {} {\bibinfo {title} {in preparation}} (\bibinfo {year} {2023}),\
  \bibinfo {note} {p. Naldesi, \textit{et al.}}\BibitemShut {Stop}%
\bibitem [{\citenamefont {Kitaev}(1995)}]{Kitaev_1995}%
  \BibitemOpen
  \bibfield  {author} {\bibinfo {author} {\bibfnamefont {A.~Y.}\ \bibnamefont
  {Kitaev}},\ }\href@noop {} {\bibinfo {title} {Quantum measurements and the
  abelian stabilizer problem}} (\bibinfo {year} {1995}),\ \Eprint
  {https://arxiv.org/abs/arXiv:quant-ph/9511026} {arXiv:quant-ph/9511026}
  \BibitemShut {NoStop}%
\bibitem [{\citenamefont {Borla}\ \emph {et~al.}(2020)\citenamefont {Borla},
  \citenamefont {Verresen}, \citenamefont {Grusdt},\ and\ \citenamefont
  {Moroz}}]{Borla_2020}%
  \BibitemOpen
  \bibfield  {author} {\bibinfo {author} {\bibfnamefont {U.}~\bibnamefont
  {Borla}}, \bibinfo {author} {\bibfnamefont {R.}~\bibnamefont {Verresen}},
  \bibinfo {author} {\bibfnamefont {F.}~\bibnamefont {Grusdt}},\ and\ \bibinfo
  {author} {\bibfnamefont {S.}~\bibnamefont {Moroz}},\ }\bibfield  {title}
  {\bibinfo {title} {Confined phases of one-dimensional spinless fermions
  coupled to ${Z}_{2}$ gauge theory},\ }\href
  {https://doi.org/10.1103/PhysRevLett.124.120503} {\bibfield  {journal}
  {\bibinfo  {journal} {Phys. Rev. Lett.}\ }\textbf {\bibinfo {volume} {124}},\
  \bibinfo {pages} {120503} (\bibinfo {year} {2020})}\BibitemShut {NoStop}%
\bibitem [{\citenamefont {Gonz\'alez-Cuadra}\ \emph {et~al.}(2020)\citenamefont
  {Gonz\'alez-Cuadra}, \citenamefont {Tagliacozzo}, \citenamefont
  {Lewenstein},\ and\ \citenamefont {Bermudez}}]{Gonzalez_2020}%
  \BibitemOpen
  \bibfield  {author} {\bibinfo {author} {\bibfnamefont {D.}~\bibnamefont
  {Gonz\'alez-Cuadra}}, \bibinfo {author} {\bibfnamefont {L.}~\bibnamefont
  {Tagliacozzo}}, \bibinfo {author} {\bibfnamefont {M.}~\bibnamefont
  {Lewenstein}},\ and\ \bibinfo {author} {\bibfnamefont {A.}~\bibnamefont
  {Bermudez}},\ }\bibfield  {title} {\bibinfo {title} {Robust topological order
  in fermionic ${\mathbb{z}}_{2}$ gauge theories: From aharonov-bohm
  instability to soliton-induced deconfinement},\ }\href
  {https://doi.org/10.1103/PhysRevX.10.041007} {\bibfield  {journal} {\bibinfo
  {journal} {Phys. Rev. X}\ }\textbf {\bibinfo {volume} {10}},\ \bibinfo
  {pages} {041007} (\bibinfo {year} {2020})}\BibitemShut {NoStop}%
\bibitem [{\citenamefont {Fradkin}(2013)}]{Fradkin_2013}%
  \BibitemOpen
  \bibfield  {author} {\bibinfo {author} {\bibfnamefont {E.}~\bibnamefont
  {Fradkin}},\ }\href {https://doi.org/10.1017/CBO9781139015509} {\emph
  {\bibinfo {title} {Field Theories of Condensed Matter Physics}}},\ \bibinfo
  {edition} {2nd}\ ed.\ (\bibinfo  {publisher} {Cambridge University Press},\
  \bibinfo {year} {2013})\BibitemShut {NoStop}%
\bibitem [{\citenamefont {Baskaran}\ and\ \citenamefont
  {Anderson}(1988)}]{Baskaran_1988}%
  \BibitemOpen
  \bibfield  {author} {\bibinfo {author} {\bibfnamefont {G.}~\bibnamefont
  {Baskaran}}\ and\ \bibinfo {author} {\bibfnamefont {P.~W.}\ \bibnamefont
  {Anderson}},\ }\bibfield  {title} {\bibinfo {title} {Gauge theory of
  high-temperature superconductors and strongly correlated fermi systems},\
  }\href {https://doi.org/10.1103/PhysRevB.37.580} {\bibfield  {journal}
  {\bibinfo  {journal} {Phys. Rev. B}\ }\textbf {\bibinfo {volume} {37}},\
  \bibinfo {pages} {580} (\bibinfo {year} {1988})}\BibitemShut {NoStop}%
\bibitem [{\citenamefont {Wen}(2017)}]{Wen_2017}%
  \BibitemOpen
  \bibfield  {author} {\bibinfo {author} {\bibfnamefont {X.-G.}\ \bibnamefont
  {Wen}},\ }\bibfield  {title} {\bibinfo {title} {Colloquium: Zoo of
  quantum-topological phases of matter},\ }\href
  {https://doi.org/10.1103/RevModPhys.89.041004} {\bibfield  {journal}
  {\bibinfo  {journal} {Rev. Mod. Phys.}\ }\textbf {\bibinfo {volume} {89}},\
  \bibinfo {pages} {041004} (\bibinfo {year} {2017})}\BibitemShut {NoStop}%
\bibitem [{\citenamefont {Assaad}\ and\ \citenamefont
  {Grover}(2016)}]{Assaad_2016}%
  \BibitemOpen
  \bibfield  {author} {\bibinfo {author} {\bibfnamefont {F.~F.}\ \bibnamefont
  {Assaad}}\ and\ \bibinfo {author} {\bibfnamefont {T.}~\bibnamefont
  {Grover}},\ }\bibfield  {title} {\bibinfo {title} {Simple fermionic model of
  deconfined phases and phase transitions},\ }\href
  {https://doi.org/10.1103/PhysRevX.6.041049} {\bibfield  {journal} {\bibinfo
  {journal} {Phys. Rev. X}\ }\textbf {\bibinfo {volume} {6}},\ \bibinfo {pages}
  {041049} (\bibinfo {year} {2016})}\BibitemShut {NoStop}%
\bibitem [{\citenamefont {Borla}\ \emph {et~al.}(2022)\citenamefont {Borla},
  \citenamefont {Jeevanesan}, \citenamefont {Pollmann},\ and\ \citenamefont
  {Moroz}}]{Borla_2022}%
  \BibitemOpen
  \bibfield  {author} {\bibinfo {author} {\bibfnamefont {U.}~\bibnamefont
  {Borla}}, \bibinfo {author} {\bibfnamefont {B.}~\bibnamefont {Jeevanesan}},
  \bibinfo {author} {\bibfnamefont {F.}~\bibnamefont {Pollmann}},\ and\
  \bibinfo {author} {\bibfnamefont {S.}~\bibnamefont {Moroz}},\ }\bibfield
  {title} {\bibinfo {title} {Quantum phases of two-dimensional
  ${\mathbb{z}}_{2}$ gauge theory coupled to single-component fermion matter},\
  }\href {https://doi.org/10.1103/PhysRevB.105.075132} {\bibfield  {journal}
  {\bibinfo  {journal} {Phys. Rev. B}\ }\textbf {\bibinfo {volume} {105}},\
  \bibinfo {pages} {075132} (\bibinfo {year} {2022})}\BibitemShut {NoStop}%
\bibitem [{\citenamefont {Iadecola}\ and\ \citenamefont
  {Schecter}(2020)}]{Iadecola_2020}%
  \BibitemOpen
  \bibfield  {author} {\bibinfo {author} {\bibfnamefont {T.}~\bibnamefont
  {Iadecola}}\ and\ \bibinfo {author} {\bibfnamefont {M.}~\bibnamefont
  {Schecter}},\ }\bibfield  {title} {\bibinfo {title} {Quantum many-body scar
  states with emergent kinetic constraints and finite-entanglement revivals},\
  }\href {https://doi.org/10.1103/PhysRevB.101.024306} {\bibfield  {journal}
  {\bibinfo  {journal} {Phys. Rev. B}\ }\textbf {\bibinfo {volume} {101}},\
  \bibinfo {pages} {024306} (\bibinfo {year} {2020})}\BibitemShut {NoStop}%
\bibitem [{\citenamefont {Aramthottil}\ \emph {et~al.}(2022)\citenamefont
  {Aramthottil}, \citenamefont {Bhattacharya}, \citenamefont
  {Gonz\'alez-Cuadra}, \citenamefont {Lewenstein}, \citenamefont {Barbiero},\
  and\ \citenamefont {Zakrzewski}}]{Sai_2022}%
  \BibitemOpen
  \bibfield  {author} {\bibinfo {author} {\bibfnamefont {A.~S.}\ \bibnamefont
  {Aramthottil}}, \bibinfo {author} {\bibfnamefont {U.}~\bibnamefont
  {Bhattacharya}}, \bibinfo {author} {\bibfnamefont {D.}~\bibnamefont
  {Gonz\'alez-Cuadra}}, \bibinfo {author} {\bibfnamefont {M.}~\bibnamefont
  {Lewenstein}}, \bibinfo {author} {\bibfnamefont {L.}~\bibnamefont
  {Barbiero}},\ and\ \bibinfo {author} {\bibfnamefont {J.}~\bibnamefont
  {Zakrzewski}},\ }\bibfield  {title} {\bibinfo {title} {Scar states in
  deconfined ${\mathbb{z}}_{2}$ lattice gauge theories},\ }\href
  {https://doi.org/10.1103/PhysRevB.106.L041101} {\bibfield  {journal}
  {\bibinfo  {journal} {Phys. Rev. B}\ }\textbf {\bibinfo {volume} {106}},\
  \bibinfo {pages} {L041101} (\bibinfo {year} {2022})}\BibitemShut {NoStop}%
\bibitem [{\citenamefont {{Halimeh}}\ \emph {et~al.}(2022)\citenamefont
  {{Halimeh}}, \citenamefont {{Barbiero}}, \citenamefont {{Hauke}},
  \citenamefont {{Grusdt}},\ and\ \citenamefont {{Bohrdt}}}]{Halimeh_2022}%
  \BibitemOpen
  \bibfield  {author} {\bibinfo {author} {\bibfnamefont {J.~C.}\ \bibnamefont
  {{Halimeh}}}, \bibinfo {author} {\bibfnamefont {L.}~\bibnamefont
  {{Barbiero}}}, \bibinfo {author} {\bibfnamefont {P.}~\bibnamefont {{Hauke}}},
  \bibinfo {author} {\bibfnamefont {F.}~\bibnamefont {{Grusdt}}},\ and\
  \bibinfo {author} {\bibfnamefont {A.}~\bibnamefont {{Bohrdt}}},\ }\bibfield
  {title} {\bibinfo {title} {{Robust quantum many-body scars in lattice gauge
  theories}},\ }\href@noop {} {\bibfield  {journal} {\bibinfo  {journal} {arXiv
  e-prints}\ ,\ \bibinfo {eid} {arXiv:2203.08828}} (\bibinfo {year} {2022})},\
  \Eprint {https://arxiv.org/abs/2203.08828} {arXiv:2203.08828
  [cond-mat.quant-gas]} \BibitemShut {NoStop}%
\bibitem [{\citenamefont {Wiese}(2013)}]{Wiese_2013}%
  \BibitemOpen
  \bibfield  {author} {\bibinfo {author} {\bibfnamefont {U.-J.}\ \bibnamefont
  {Wiese}},\ }\bibfield  {title} {\bibinfo {title} {Ultracold quantum gases and
  lattice systems: quantum simulation of lattice gauge theories},\ }\href
  {https://doi.org/https://doi.org/10.1002/andp.201300104} {\bibfield
  {journal} {\bibinfo  {journal} {Annalen der Physik}\ }\textbf {\bibinfo
  {volume} {525}},\ \bibinfo {pages} {777} (\bibinfo {year}
  {2013})}\BibitemShut {NoStop}%
\bibitem [{\citenamefont {Zohar}\ \emph {et~al.}(2015)\citenamefont {Zohar},
  \citenamefont {Cirac},\ and\ \citenamefont {Reznik}}]{Zohar_2015}%
  \BibitemOpen
  \bibfield  {author} {\bibinfo {author} {\bibfnamefont {E.}~\bibnamefont
  {Zohar}}, \bibinfo {author} {\bibfnamefont {J.~I.}\ \bibnamefont {Cirac}},\
  and\ \bibinfo {author} {\bibfnamefont {B.}~\bibnamefont {Reznik}},\
  }\bibfield  {title} {\bibinfo {title} {Quantum simulations of lattice gauge
  theories using ultracold atoms in optical lattices},\ }\href
  {https://doi.org/10.1088/0034-4885/79/1/014401} {\bibfield  {journal}
  {\bibinfo  {journal} {Reports on Progress in Physics}\ }\textbf {\bibinfo
  {volume} {79}},\ \bibinfo {pages} {014401} (\bibinfo {year}
  {2015})}\BibitemShut {NoStop}%
\bibitem [{\citenamefont {Dalmonte}\ and\ \citenamefont
  {Montangero}(2016)}]{Dalmonte_2016}%
  \BibitemOpen
  \bibfield  {author} {\bibinfo {author} {\bibfnamefont {M.}~\bibnamefont
  {Dalmonte}}\ and\ \bibinfo {author} {\bibfnamefont {S.}~\bibnamefont
  {Montangero}},\ }\bibfield  {title} {\bibinfo {title} {Lattice gauge theory
  simulations in the quantum information era},\ }\href
  {https://doi.org/10.1080/00107514.2016.1151199} {\bibfield  {journal}
  {\bibinfo  {journal} {Contemporary Physics}\ }\textbf {\bibinfo {volume}
  {57}},\ \bibinfo {pages} {388} (\bibinfo {year} {2016})}\BibitemShut
  {NoStop}%
\bibitem [{\citenamefont {Ba{\~n}uls}\ \emph {et~al.}(2020)\citenamefont
  {Ba{\~n}uls}, \citenamefont {Blatt}, \citenamefont {Catani}, \citenamefont
  {Celi}, \citenamefont {Cirac}, \citenamefont {Dalmonte}, \citenamefont
  {Fallani}, \citenamefont {Jansen}, \citenamefont {Lewenstein}, \citenamefont
  {Montangero}, \citenamefont {Muschik}, \citenamefont {Reznik}, \citenamefont
  {Rico}, \citenamefont {Tagliacozzo}, \citenamefont {Van~Acoleyen},
  \citenamefont {Verstraete}, \citenamefont {Wiese}, \citenamefont {Wingate},
  \citenamefont {Zakrzewski},\ and\ \citenamefont {Zoller}}]{Banuls_2020}%
  \BibitemOpen
  \bibfield  {author} {\bibinfo {author} {\bibfnamefont {M.~C.}\ \bibnamefont
  {Ba{\~n}uls}}, \bibinfo {author} {\bibfnamefont {R.}~\bibnamefont {Blatt}},
  \bibinfo {author} {\bibfnamefont {J.}~\bibnamefont {Catani}}, \bibinfo
  {author} {\bibfnamefont {A.}~\bibnamefont {Celi}}, \bibinfo {author}
  {\bibfnamefont {J.~I.}\ \bibnamefont {Cirac}}, \bibinfo {author}
  {\bibfnamefont {M.}~\bibnamefont {Dalmonte}}, \bibinfo {author}
  {\bibfnamefont {L.}~\bibnamefont {Fallani}}, \bibinfo {author} {\bibfnamefont
  {K.}~\bibnamefont {Jansen}}, \bibinfo {author} {\bibfnamefont
  {M.}~\bibnamefont {Lewenstein}}, \bibinfo {author} {\bibfnamefont
  {S.}~\bibnamefont {Montangero}}, \bibinfo {author} {\bibfnamefont {C.~A.}\
  \bibnamefont {Muschik}}, \bibinfo {author} {\bibfnamefont {B.}~\bibnamefont
  {Reznik}}, \bibinfo {author} {\bibfnamefont {E.}~\bibnamefont {Rico}},
  \bibinfo {author} {\bibfnamefont {L.}~\bibnamefont {Tagliacozzo}}, \bibinfo
  {author} {\bibfnamefont {K.}~\bibnamefont {Van~Acoleyen}}, \bibinfo {author}
  {\bibfnamefont {F.}~\bibnamefont {Verstraete}}, \bibinfo {author}
  {\bibfnamefont {U.-J.}\ \bibnamefont {Wiese}}, \bibinfo {author}
  {\bibfnamefont {M.}~\bibnamefont {Wingate}}, \bibinfo {author} {\bibfnamefont
  {J.}~\bibnamefont {Zakrzewski}},\ and\ \bibinfo {author} {\bibfnamefont
  {P.}~\bibnamefont {Zoller}},\ }\bibfield  {title} {\bibinfo {title}
  {Simulating lattice gauge theories within quantum technologies},\ }\href
  {https://doi.org/10.1140/epjd/e2020-100571-8} {\bibfield  {journal} {\bibinfo
   {journal} {The European Physical Journal D}\ }\textbf {\bibinfo {volume}
  {74}},\ \bibinfo {pages} {165} (\bibinfo {year} {2020})}\BibitemShut
  {NoStop}%
\bibitem [{\citenamefont {Aidelsburger}\ \emph {et~al.}(2022)\citenamefont
  {Aidelsburger}, \citenamefont {Barbiero}, \citenamefont {Bermudez},
  \citenamefont {Chanda}, \citenamefont {Dauphin}, \citenamefont
  {Gonz{\'a}lez-Cuadra}, \citenamefont {Grzybowski}, \citenamefont {Hands},
  \citenamefont {Jendrzejewski}, \citenamefont {J{\"u}nemann}, \citenamefont
  {Juzeli{\=u}nas}, \citenamefont {Kasper}, \citenamefont {Piga}, \citenamefont
  {Ran}, \citenamefont {Rizzi}, \citenamefont {Sierra}, \citenamefont
  {Tagliacozzo}, \citenamefont {Tirrito}, \citenamefont {Zache}, \citenamefont
  {Zakrzewski}, \citenamefont {Zohar},\ and\ \citenamefont
  {Lewenstein}}]{Aidelsburger_2022}%
  \BibitemOpen
  \bibfield  {author} {\bibinfo {author} {\bibfnamefont {M.}~\bibnamefont
  {Aidelsburger}}, \bibinfo {author} {\bibfnamefont {L.}~\bibnamefont
  {Barbiero}}, \bibinfo {author} {\bibfnamefont {A.}~\bibnamefont {Bermudez}},
  \bibinfo {author} {\bibfnamefont {T.}~\bibnamefont {Chanda}}, \bibinfo
  {author} {\bibfnamefont {A.}~\bibnamefont {Dauphin}}, \bibinfo {author}
  {\bibfnamefont {D.}~\bibnamefont {Gonz{\'a}lez-Cuadra}}, \bibinfo {author}
  {\bibfnamefont {P.~R.}\ \bibnamefont {Grzybowski}}, \bibinfo {author}
  {\bibfnamefont {S.}~\bibnamefont {Hands}}, \bibinfo {author} {\bibfnamefont
  {F.}~\bibnamefont {Jendrzejewski}}, \bibinfo {author} {\bibfnamefont
  {J.}~\bibnamefont {J{\"u}nemann}}, \bibinfo {author} {\bibfnamefont
  {G.}~\bibnamefont {Juzeli{\=u}nas}}, \bibinfo {author} {\bibfnamefont
  {V.}~\bibnamefont {Kasper}}, \bibinfo {author} {\bibfnamefont
  {A.}~\bibnamefont {Piga}}, \bibinfo {author} {\bibfnamefont {S.-J.}\
  \bibnamefont {Ran}}, \bibinfo {author} {\bibfnamefont {M.}~\bibnamefont
  {Rizzi}}, \bibinfo {author} {\bibfnamefont {G.}~\bibnamefont {Sierra}},
  \bibinfo {author} {\bibfnamefont {L.}~\bibnamefont {Tagliacozzo}}, \bibinfo
  {author} {\bibfnamefont {E.}~\bibnamefont {Tirrito}}, \bibinfo {author}
  {\bibfnamefont {T.~V.}\ \bibnamefont {Zache}}, \bibinfo {author}
  {\bibfnamefont {J.}~\bibnamefont {Zakrzewski}}, \bibinfo {author}
  {\bibfnamefont {E.}~\bibnamefont {Zohar}},\ and\ \bibinfo {author}
  {\bibfnamefont {M.}~\bibnamefont {Lewenstein}},\ }\bibfield  {title}
  {\bibinfo {title} {Cold atoms meet lattice gauge theory},\ }\href
  {https://doi.org/10.1098/rsta.2021.0064} {\bibfield  {journal} {\bibinfo
  {journal} {Philosophical Transactions of the Royal Society A: Mathematical,
  Physical and Engineering Sciences}\ }\textbf {\bibinfo {volume} {380}},\
  \bibinfo {pages} {20210064} (\bibinfo {year} {2022})}\BibitemShut {NoStop}%
\bibitem [{\citenamefont {Klco}\ \emph {et~al.}(2022)\citenamefont {Klco},
  \citenamefont {Roggero},\ and\ \citenamefont {Savage}}]{Klco_2022}%
  \BibitemOpen
  \bibfield  {author} {\bibinfo {author} {\bibfnamefont {N.}~\bibnamefont
  {Klco}}, \bibinfo {author} {\bibfnamefont {A.}~\bibnamefont {Roggero}},\ and\
  \bibinfo {author} {\bibfnamefont {M.~J.}\ \bibnamefont {Savage}},\ }\bibfield
   {title} {\bibinfo {title} {Standard model physics and the digital quantum
  revolution: thoughts about the interface},\ }\href
  {https://doi.org/10.1088/1361-6633/ac58a4} {\bibfield  {journal} {\bibinfo
  {journal} {Reports on Progress in Physics}\ }\textbf {\bibinfo {volume}
  {85}},\ \bibinfo {pages} {064301} (\bibinfo {year} {2022})}\BibitemShut
  {NoStop}%
\bibitem [{\citenamefont {Zohar}\ \emph {et~al.}(2017)\citenamefont {Zohar},
  \citenamefont {Farace}, \citenamefont {Reznik},\ and\ \citenamefont
  {Cirac}}]{Zohar_2017}%
  \BibitemOpen
  \bibfield  {author} {\bibinfo {author} {\bibfnamefont {E.}~\bibnamefont
  {Zohar}}, \bibinfo {author} {\bibfnamefont {A.}~\bibnamefont {Farace}},
  \bibinfo {author} {\bibfnamefont {B.}~\bibnamefont {Reznik}},\ and\ \bibinfo
  {author} {\bibfnamefont {J.~I.}\ \bibnamefont {Cirac}},\ }\bibfield  {title}
  {\bibinfo {title} {Digital quantum simulation of ${\mathbb{z}}_{2}$ lattice
  gauge theories with dynamical fermionic matter},\ }\href
  {https://doi.org/10.1103/PhysRevLett.118.070501} {\bibfield  {journal}
  {\bibinfo  {journal} {Phys. Rev. Lett.}\ }\textbf {\bibinfo {volume} {118}},\
  \bibinfo {pages} {070501} (\bibinfo {year} {2017})}\BibitemShut {NoStop}%
\bibitem [{\citenamefont {Lamm}\ \emph {et~al.}(2019)\citenamefont {Lamm},
  \citenamefont {Lawrence},\ and\ \citenamefont {Yamauchi}}]{Lamm_2019}%
  \BibitemOpen
  \bibfield  {author} {\bibinfo {author} {\bibfnamefont {H.}~\bibnamefont
  {Lamm}}, \bibinfo {author} {\bibfnamefont {S.}~\bibnamefont {Lawrence}},\
  and\ \bibinfo {author} {\bibfnamefont {Y.}~\bibnamefont {Yamauchi}} (\bibinfo
  {collaboration} {NuQS Collaboration}),\ }\bibfield  {title} {\bibinfo {title}
  {General methods for digital quantum simulation of gauge theories},\ }\href
  {https://doi.org/10.1103/PhysRevD.100.034518} {\bibfield  {journal} {\bibinfo
   {journal} {Phys. Rev. D}\ }\textbf {\bibinfo {volume} {100}},\ \bibinfo
  {pages} {034518} (\bibinfo {year} {2019})}\BibitemShut {NoStop}%
\bibitem [{\citenamefont {Gonz\'alez-Cuadra}\ \emph {et~al.}(2022)\citenamefont
  {Gonz\'alez-Cuadra}, \citenamefont {Zache}, \citenamefont {Carrasco},
  \citenamefont {Kraus},\ and\ \citenamefont {Zoller}}]{Gonzalez_2022}%
  \BibitemOpen
  \bibfield  {author} {\bibinfo {author} {\bibfnamefont {D.}~\bibnamefont
  {Gonz\'alez-Cuadra}}, \bibinfo {author} {\bibfnamefont {T.~V.}\ \bibnamefont
  {Zache}}, \bibinfo {author} {\bibfnamefont {J.}~\bibnamefont {Carrasco}},
  \bibinfo {author} {\bibfnamefont {B.}~\bibnamefont {Kraus}},\ and\ \bibinfo
  {author} {\bibfnamefont {P.}~\bibnamefont {Zoller}},\ }\bibfield  {title}
  {\bibinfo {title} {Hardware efficient quantum simulation of non-abelian gauge
  theories with qudits on rydberg platforms},\ }\href
  {https://doi.org/10.1103/PhysRevLett.129.160501} {\bibfield  {journal}
  {\bibinfo  {journal} {Phys. Rev. Lett.}\ }\textbf {\bibinfo {volume} {129}},\
  \bibinfo {pages} {160501} (\bibinfo {year} {2022})}\BibitemShut {NoStop}%
\bibitem [{Gon(2023)}]{Gonzalez_2023}%
  \BibitemOpen
  \href@noop {} {\bibinfo {title} {in preparation}} (\bibinfo {year} {2023}),\
  \bibinfo {note} {t. V. Zache \textit{et al.}}\BibitemShut {Stop}%
\bibitem [{\citenamefont {Serwane}\ \emph {et~al.}(2011)\citenamefont
  {Serwane}, \citenamefont {Z{\"u}rn}, \citenamefont {Lompe}, \citenamefont
  {Ottenstein}, \citenamefont {Wenz},\ and\ \citenamefont
  {Jochim}}]{serwane2011deterministic}%
  \BibitemOpen
  \bibfield  {author} {\bibinfo {author} {\bibfnamefont {F.}~\bibnamefont
  {Serwane}}, \bibinfo {author} {\bibfnamefont {G.}~\bibnamefont {Z{\"u}rn}},
  \bibinfo {author} {\bibfnamefont {T.}~\bibnamefont {Lompe}}, \bibinfo
  {author} {\bibfnamefont {T.}~\bibnamefont {Ottenstein}}, \bibinfo {author}
  {\bibfnamefont {A.}~\bibnamefont {Wenz}},\ and\ \bibinfo {author}
  {\bibfnamefont {S.}~\bibnamefont {Jochim}},\ }\bibfield  {title} {\bibinfo
  {title} {Deterministic preparation of a tunable few-fermion system},\
  }\href@noop {} {\bibfield  {journal} {\bibinfo  {journal} {Science}\ }\textbf
  {\bibinfo {volume} {332}},\ \bibinfo {pages} {336} (\bibinfo {year}
  {2011})}\BibitemShut {NoStop}%
\bibitem [{\citenamefont {Cong}\ \emph {et~al.}(2022)\citenamefont {Cong},
  \citenamefont {Levine}, \citenamefont {Keesling}, \citenamefont {Bluvstein},
  \citenamefont {Wang},\ and\ \citenamefont {Lukin}}]{Cong_2022}%
  \BibitemOpen
  \bibfield  {author} {\bibinfo {author} {\bibfnamefont {I.}~\bibnamefont
  {Cong}}, \bibinfo {author} {\bibfnamefont {H.}~\bibnamefont {Levine}},
  \bibinfo {author} {\bibfnamefont {A.}~\bibnamefont {Keesling}}, \bibinfo
  {author} {\bibfnamefont {D.}~\bibnamefont {Bluvstein}}, \bibinfo {author}
  {\bibfnamefont {S.-T.}\ \bibnamefont {Wang}},\ and\ \bibinfo {author}
  {\bibfnamefont {M.~D.}\ \bibnamefont {Lukin}},\ }\bibfield  {title} {\bibinfo
  {title} {{Hardware-Efficient, Fault-Tolerant Quantum Computation with Rydberg
  Atoms}},\ }\href {https://doi.org/10.1103/physrevx.12.021049} {\bibfield
  {journal} {\bibinfo  {journal} {Phys. Rev. X}\ }\textbf {\bibinfo {volume}
  {12}},\ \bibinfo {pages} {021049} (\bibinfo {year} {2022})}\BibitemShut
  {NoStop}%
\bibitem [{\citenamefont {Lam}\ \emph {et~al.}(2021{\natexlab{b}})\citenamefont
  {Lam}, \citenamefont {Peter}, \citenamefont {Groh}, \citenamefont {Alt},
  \citenamefont {Robens}, \citenamefont {Meschede}, \citenamefont {Negretti},
  \citenamefont {Montangero}, \citenamefont {Calarco},\ and\ \citenamefont
  {Alberti}}]{Lam.2021}%
  \BibitemOpen
  \bibfield  {author} {\bibinfo {author} {\bibfnamefont {M.~R.}\ \bibnamefont
  {Lam}}, \bibinfo {author} {\bibfnamefont {N.}~\bibnamefont {Peter}}, \bibinfo
  {author} {\bibfnamefont {T.}~\bibnamefont {Groh}}, \bibinfo {author}
  {\bibfnamefont {W.}~\bibnamefont {Alt}}, \bibinfo {author} {\bibfnamefont
  {C.}~\bibnamefont {Robens}}, \bibinfo {author} {\bibfnamefont
  {D.}~\bibnamefont {Meschede}}, \bibinfo {author} {\bibfnamefont
  {A.}~\bibnamefont {Negretti}}, \bibinfo {author} {\bibfnamefont
  {S.}~\bibnamefont {Montangero}}, \bibinfo {author} {\bibfnamefont
  {T.}~\bibnamefont {Calarco}},\ and\ \bibinfo {author} {\bibfnamefont
  {A.}~\bibnamefont {Alberti}},\ }\bibfield  {title} {\bibinfo {title}
  {{Demonstration of Quantum Brachistochrones between Distant States of an
  Atom}},\ }\href {https://doi.org/10.1103/physrevx.11.011035} {\bibfield
  {journal} {\bibinfo  {journal} {Phys. Rev. X}\ }\textbf {\bibinfo {volume}
  {11}},\ \bibinfo {pages} {011035} (\bibinfo {year}
  {2021}{\natexlab{b}})}\BibitemShut {NoStop}%
\bibitem [{\citenamefont {Hahn}(1950)}]{Hahn_1950}%
  \BibitemOpen
  \bibfield  {author} {\bibinfo {author} {\bibfnamefont {E.~L.}\ \bibnamefont
  {Hahn}},\ }\bibfield  {title} {\bibinfo {title} {Spin echoes},\ }\href
  {https://doi.org/10.1103/PhysRev.80.580} {\bibfield  {journal} {\bibinfo
  {journal} {Phys. Rev.}\ }\textbf {\bibinfo {volume} {80}},\ \bibinfo {pages}
  {580} (\bibinfo {year} {1950})}\BibitemShut {NoStop}%
\bibitem [{\citenamefont {Nascimb\`ene}\ \emph {et~al.}(2012)\citenamefont
  {Nascimb\`ene}, \citenamefont {Chen}, \citenamefont {Atala}, \citenamefont
  {Aidelsburger}, \citenamefont {Trotzky}, \citenamefont {Paredes},\ and\
  \citenamefont {Bloch}}]{Nascimbene_2012}%
  \BibitemOpen
  \bibfield  {author} {\bibinfo {author} {\bibfnamefont {S.}~\bibnamefont
  {Nascimb\`ene}}, \bibinfo {author} {\bibfnamefont {Y.-A.}\ \bibnamefont
  {Chen}}, \bibinfo {author} {\bibfnamefont {M.}~\bibnamefont {Atala}},
  \bibinfo {author} {\bibfnamefont {M.}~\bibnamefont {Aidelsburger}}, \bibinfo
  {author} {\bibfnamefont {S.}~\bibnamefont {Trotzky}}, \bibinfo {author}
  {\bibfnamefont {B.}~\bibnamefont {Paredes}},\ and\ \bibinfo {author}
  {\bibfnamefont {I.}~\bibnamefont {Bloch}},\ }\bibfield  {title} {\bibinfo
  {title} {Experimental realization of plaquette resonating valence-bond states
  with ultracold atoms in optical superlattices},\ }\href
  {https://doi.org/10.1103/PhysRevLett.108.205301} {\bibfield  {journal}
  {\bibinfo  {journal} {Phys. Rev. Lett.}\ }\textbf {\bibinfo {volume} {108}},\
  \bibinfo {pages} {205301} (\bibinfo {year} {2012})}\BibitemShut {NoStop}%
\bibitem [{\citenamefont {Zhang}\ \emph {et~al.}(2014)\citenamefont {Zhang},
  \citenamefont {Bishof}, \citenamefont {Bromley}, \citenamefont {Kraus},
  \citenamefont {Safronova}, \citenamefont {Zoller}, \citenamefont {Rey},\ and\
  \citenamefont {Ye}}]{Zhang_2014}%
  \BibitemOpen
  \bibfield  {author} {\bibinfo {author} {\bibfnamefont {X.}~\bibnamefont
  {Zhang}}, \bibinfo {author} {\bibfnamefont {M.}~\bibnamefont {Bishof}},
  \bibinfo {author} {\bibfnamefont {S.~L.}\ \bibnamefont {Bromley}}, \bibinfo
  {author} {\bibfnamefont {C.~V.}\ \bibnamefont {Kraus}}, \bibinfo {author}
  {\bibfnamefont {M.~S.}\ \bibnamefont {Safronova}}, \bibinfo {author}
  {\bibfnamefont {P.}~\bibnamefont {Zoller}}, \bibinfo {author} {\bibfnamefont
  {A.~M.}\ \bibnamefont {Rey}},\ and\ \bibinfo {author} {\bibfnamefont
  {J.}~\bibnamefont {Ye}},\ }\bibfield  {title} {\bibinfo {title}
  {Spectroscopic observation of su(<i>n</i>)-symmetric interactions in sr
  orbital magnetism},\ }\href {https://doi.org/10.1126/science.1254978}
  {\bibfield  {journal} {\bibinfo  {journal} {Science}\ }\textbf {\bibinfo
  {volume} {345}},\ \bibinfo {pages} {1467} (\bibinfo {year}
  {2014})}\BibitemShut {NoStop}%
\bibitem [{\citenamefont {Goban}\ \emph {et~al.}(2018)\citenamefont {Goban},
  \citenamefont {Hutson}, \citenamefont {Marti}, \citenamefont {Campbell},
  \citenamefont {Perlin}, \citenamefont {Julienne}, \citenamefont {D'Incao},
  \citenamefont {Rey},\ and\ \citenamefont {Ye}}]{Goban_2018}%
  \BibitemOpen
  \bibfield  {author} {\bibinfo {author} {\bibfnamefont {A.}~\bibnamefont
  {Goban}}, \bibinfo {author} {\bibfnamefont {R.~B.}\ \bibnamefont {Hutson}},
  \bibinfo {author} {\bibfnamefont {G.~E.}\ \bibnamefont {Marti}}, \bibinfo
  {author} {\bibfnamefont {S.~L.}\ \bibnamefont {Campbell}}, \bibinfo {author}
  {\bibfnamefont {M.~A.}\ \bibnamefont {Perlin}}, \bibinfo {author}
  {\bibfnamefont {P.~S.}\ \bibnamefont {Julienne}}, \bibinfo {author}
  {\bibfnamefont {J.~P.}\ \bibnamefont {D'Incao}}, \bibinfo {author}
  {\bibfnamefont {A.~M.}\ \bibnamefont {Rey}},\ and\ \bibinfo {author}
  {\bibfnamefont {J.}~\bibnamefont {Ye}},\ }\bibfield  {title} {\bibinfo
  {title} {Emergence of multi-body interactions in a fermionic lattice clock},\
  }\href {https://doi.org/10.1038/s41586-018-0661-6} {\bibfield  {journal}
  {\bibinfo  {journal} {Nature}\ }\textbf {\bibinfo {volume} {563}},\ \bibinfo
  {pages} {369} (\bibinfo {year} {2018})}\BibitemShut {NoStop}%
\bibitem [{\citenamefont {Bishof}\ \emph {et~al.}(2011)\citenamefont {Bishof},
  \citenamefont {Martin}, \citenamefont {Swallows}, \citenamefont {Benko},
  \citenamefont {Lin}, \citenamefont {Qu\'em\'ener}, \citenamefont {Rey},\ and\
  \citenamefont {Ye}}]{Bishof_2011}%
  \BibitemOpen
  \bibfield  {author} {\bibinfo {author} {\bibfnamefont {M.}~\bibnamefont
  {Bishof}}, \bibinfo {author} {\bibfnamefont {M.~J.}\ \bibnamefont {Martin}},
  \bibinfo {author} {\bibfnamefont {M.~D.}\ \bibnamefont {Swallows}}, \bibinfo
  {author} {\bibfnamefont {C.}~\bibnamefont {Benko}}, \bibinfo {author}
  {\bibfnamefont {Y.}~\bibnamefont {Lin}}, \bibinfo {author} {\bibfnamefont
  {G.}~\bibnamefont {Qu\'em\'ener}}, \bibinfo {author} {\bibfnamefont {A.~M.}\
  \bibnamefont {Rey}},\ and\ \bibinfo {author} {\bibfnamefont {J.}~\bibnamefont
  {Ye}},\ }\bibfield  {title} {\bibinfo {title} {Inelastic collisions and
  density-dependent excitation suppression in a ${}^{87}$sr optical lattice
  clock},\ }\href {https://doi.org/10.1103/PhysRevA.84.052716} {\bibfield
  {journal} {\bibinfo  {journal} {Phys. Rev. A}\ }\textbf {\bibinfo {volume}
  {84}},\ \bibinfo {pages} {052716} (\bibinfo {year} {2011})}\BibitemShut
  {NoStop}%
\bibitem [{\citenamefont {McClean}\ \emph {et~al.}(2020)\citenamefont
  {McClean}, \citenamefont {Rubin}, \citenamefont {Sung}, \citenamefont
  {Kivlichan}, \citenamefont {Bonet-Monroig}, \citenamefont {Cao},
  \citenamefont {Dai}, \citenamefont {Fried}, \citenamefont {Gidney},
  \citenamefont {Gimby}, \citenamefont {Gokhale}, \citenamefont {H{\"a}ner},
  \citenamefont {Hardikar}, \citenamefont {Havl{\'\i}{\v c}ek}, \citenamefont
  {Higgott}, \citenamefont {Huang}, \citenamefont {Izaac}, \citenamefont
  {Jiang}, \citenamefont {Liu}, \citenamefont {McArdle}, \citenamefont
  {Neeley}, \citenamefont {O'Brien}, \citenamefont {O'Gorman}, \citenamefont
  {Ozfidan}, \citenamefont {Radin}, \citenamefont {Romero}, \citenamefont
  {Sawaya}, \citenamefont {Senjean}, \citenamefont {Setia}, \citenamefont
  {Sim}, \citenamefont {Steiger}, \citenamefont {Steudtner}, \citenamefont
  {Sun}, \citenamefont {Sun}, \citenamefont {Wang}, \citenamefont {Zhang},\
  and\ \citenamefont {Babbush}}]{McClean_2020}%
  \BibitemOpen
  \bibfield  {author} {\bibinfo {author} {\bibfnamefont {J.~R.}\ \bibnamefont
  {McClean}}, \bibinfo {author} {\bibfnamefont {N.~C.}\ \bibnamefont {Rubin}},
  \bibinfo {author} {\bibfnamefont {K.~J.}\ \bibnamefont {Sung}}, \bibinfo
  {author} {\bibfnamefont {I.~D.}\ \bibnamefont {Kivlichan}}, \bibinfo {author}
  {\bibfnamefont {X.}~\bibnamefont {Bonet-Monroig}}, \bibinfo {author}
  {\bibfnamefont {Y.}~\bibnamefont {Cao}}, \bibinfo {author} {\bibfnamefont
  {C.}~\bibnamefont {Dai}}, \bibinfo {author} {\bibfnamefont {E.~S.}\
  \bibnamefont {Fried}}, \bibinfo {author} {\bibfnamefont {C.}~\bibnamefont
  {Gidney}}, \bibinfo {author} {\bibfnamefont {B.}~\bibnamefont {Gimby}},
  \bibinfo {author} {\bibfnamefont {P.}~\bibnamefont {Gokhale}}, \bibinfo
  {author} {\bibfnamefont {T.}~\bibnamefont {H{\"a}ner}}, \bibinfo {author}
  {\bibfnamefont {T.}~\bibnamefont {Hardikar}}, \bibinfo {author}
  {\bibfnamefont {V.}~\bibnamefont {Havl{\'\i}{\v c}ek}}, \bibinfo {author}
  {\bibfnamefont {O.}~\bibnamefont {Higgott}}, \bibinfo {author} {\bibfnamefont
  {C.}~\bibnamefont {Huang}}, \bibinfo {author} {\bibfnamefont
  {J.}~\bibnamefont {Izaac}}, \bibinfo {author} {\bibfnamefont
  {Z.}~\bibnamefont {Jiang}}, \bibinfo {author} {\bibfnamefont
  {X.}~\bibnamefont {Liu}}, \bibinfo {author} {\bibfnamefont {S.}~\bibnamefont
  {McArdle}}, \bibinfo {author} {\bibfnamefont {M.}~\bibnamefont {Neeley}},
  \bibinfo {author} {\bibfnamefont {T.}~\bibnamefont {O'Brien}}, \bibinfo
  {author} {\bibfnamefont {B.}~\bibnamefont {O'Gorman}}, \bibinfo {author}
  {\bibfnamefont {I.}~\bibnamefont {Ozfidan}}, \bibinfo {author} {\bibfnamefont
  {M.~D.}\ \bibnamefont {Radin}}, \bibinfo {author} {\bibfnamefont
  {J.}~\bibnamefont {Romero}}, \bibinfo {author} {\bibfnamefont {N.~P.~D.}\
  \bibnamefont {Sawaya}}, \bibinfo {author} {\bibfnamefont {B.}~\bibnamefont
  {Senjean}}, \bibinfo {author} {\bibfnamefont {K.}~\bibnamefont {Setia}},
  \bibinfo {author} {\bibfnamefont {S.}~\bibnamefont {Sim}}, \bibinfo {author}
  {\bibfnamefont {D.~S.}\ \bibnamefont {Steiger}}, \bibinfo {author}
  {\bibfnamefont {M.}~\bibnamefont {Steudtner}}, \bibinfo {author}
  {\bibfnamefont {Q.}~\bibnamefont {Sun}}, \bibinfo {author} {\bibfnamefont
  {W.}~\bibnamefont {Sun}}, \bibinfo {author} {\bibfnamefont {D.}~\bibnamefont
  {Wang}}, \bibinfo {author} {\bibfnamefont {F.}~\bibnamefont {Zhang}},\ and\
  \bibinfo {author} {\bibfnamefont {R.}~\bibnamefont {Babbush}},\ }\bibfield
  {title} {\bibinfo {title} {Openfermion: the electronic structure package for
  quantum computers},\ }\href {https://doi.org/10.1088/2058-9565/ab8ebc}
  {\bibfield  {journal} {\bibinfo  {journal} {Quantum Science and Technology}\
  }\textbf {\bibinfo {volume} {5}},\ \bibinfo {pages} {034014} (\bibinfo {year}
  {2020})}\BibitemShut {NoStop}%
\end{thebibliography}%

\end{document}